\documentclass{article} 
\usepackage{iclr2025_conference,times}

\usepackage{amsmath,amsfonts,bm}

\def\eqref#1{equation~\ref{#1}}

\def\1{\bm{1}}

\def\rmB{{\mathbf{B}}}

\def\rmX{{\mathbf{X}}}

\def\vx{{\bm{x}}}

\def\vz{{\bm{z}}}

\def\mD{{\bm{D}}}

\def\mH{{\bm{H}}}

\def\mV{{\bm{V}}}

\def\mX{{\bm{X}}}

\def\mZ{{\bm{Z}}}

\DeclareMathAlphabet{\mathsfit}{\encodingdefault}{\sfdefault}{m}{sl}
\SetMathAlphabet{\mathsfit}{bold}{\encodingdefault}{\sfdefault}{bx}{n}

\def\gD{{\mathcal{D}}}
\def\gE{{\mathcal{E}}}

\def\gG{{\mathcal{G}}}

\def\gL{{\mathcal{L}}}

\def\gN{{\mathcal{N}}}

\def\gS{{\mathcal{S}}}

\def\gV{{\mathcal{V}}}

\def\sR{{\mathbb{R}}}

\newcommand*{\dif}{\mathop{}\!\mathrm{d}}

\newcommand{\E}{\mathbb{E}}

\usepackage{hyperref}
\usepackage{url}
\usepackage{amsthm}
\usepackage{amsmath}
\usepackage{mathtools}
\usepackage{multirow}
\usepackage[normalem]{ulem}
\usepackage{graphicx}
\usepackage{wrapfig}
\usepackage{algorithm}
\usepackage{algorithmic}
\usepackage{nicefrac}
\usepackage{tikz}
\usepackage{subcaption}
\usepackage{caption}
\usepackage{float}
\useunder{\uline}{\ul}{}

\theoremstyle{plain}
\newtheorem{theorem}{Theorem}[section]
\newtheorem{proposition}[theorem]{Proposition}

\theoremstyle{definition}

\theoremstyle{remark}

\RequirePackage{hyperref}
\definecolor{darkblue}{rgb}{0,0.08,0.45}
\hypersetup{
    colorlinks=true,
    linkcolor=darkblue,
    citecolor=darkblue,
    filecolor=darkblue,
    urlcolor=darkblue,
}
\usepackage[capitalize,noabbrev]{cleveref}
\crefname{section}{\textsection}{\textsection}
\crefname{equation}{Eq.}{Eqs.}
\crefname{figure}{Fig.}{Figs.}
\crefname{theorem}{Theorem}{Theorems}
\crefname{proposition}{Proposition}{Propositions}
\crefname{lemma}{Lemma}{Lemmas}
\crefname{corollary}{Corollary}{Corollaries}
\crefname{definition}{Definition}{Definitions}
\crefname{assumption}{Assumption}{Assumptions}
\crefname{remark}{Remark}{Remarks}

\title{Force-Guided Bridge Matching for Full-Atom Time-Coarsened Molecular Dynamics}

\author{%
  Ziyang Yu$^1$~~Wenbing Huang$^{3,4,}$\thanks{Corresponding authors: Wenbing Huang, Yang Liu.}~~Yang Liu$^{1,2,*}$ \\
  \small $^1$Department of Computer Science and Technology, Tsinghua University \\
  \small $^2$Institute for AI Industry Research (AIR), Tsinghua University \\
  \small $^3$Gaoling School of Artificial Intelligence, Renmin University of China\\
  \small $^4$Beijing Key Laboratory of Big Data Management and Analysis Methods, Beijing, China\\
  \small{\texttt{yu-zy24@mails.tsinghua.edu.cn},~\texttt{hwenbing@126.com},~\texttt{liuyang2011@tsinghua.edu.cn}}
}

\iclrfinalcopy
\begin{document}

\maketitle

\begin{abstract}
Molecular Dynamics (MD) is crucial in various fields such as materials science, chemistry, and pharmacology to name a few. Conventional MD software struggles with the balance between time cost and prediction accuracy, which restricts its wider application. Recently, data-driven approaches based on deep generative models have been devised for time-coarsened dynamics, which aim at learning dynamics of diverse molecular systems over a long timestep, enjoying both universality and efficiency. Nevertheless, most current methods are designed solely to learn from the data distribution regardless of the underlying Boltzmann distribution, and the physics priors such as energies and forces are constantly overlooked. In this work, we propose a conditional generative model called Force-guided Bridge Matching (FBM), which learns full-atom time-coarsened dynamics and targets the Boltzmann-constrained distribution. With the guidance of our delicately-designed intermediate force field, FBM leverages favourable physics priors into the generation process, giving rise to enhanced simulations. Experiments on two datasets consisting of peptides verify our superiority in terms of comprehensive metrics and demonstrate transferability to unseen systems.
\end{abstract}

\section{Introduction}
\label{sec:intro}

\begin{wrapfigure}{r}{0.5\linewidth}
  \centering
  \includegraphics[width=\linewidth]{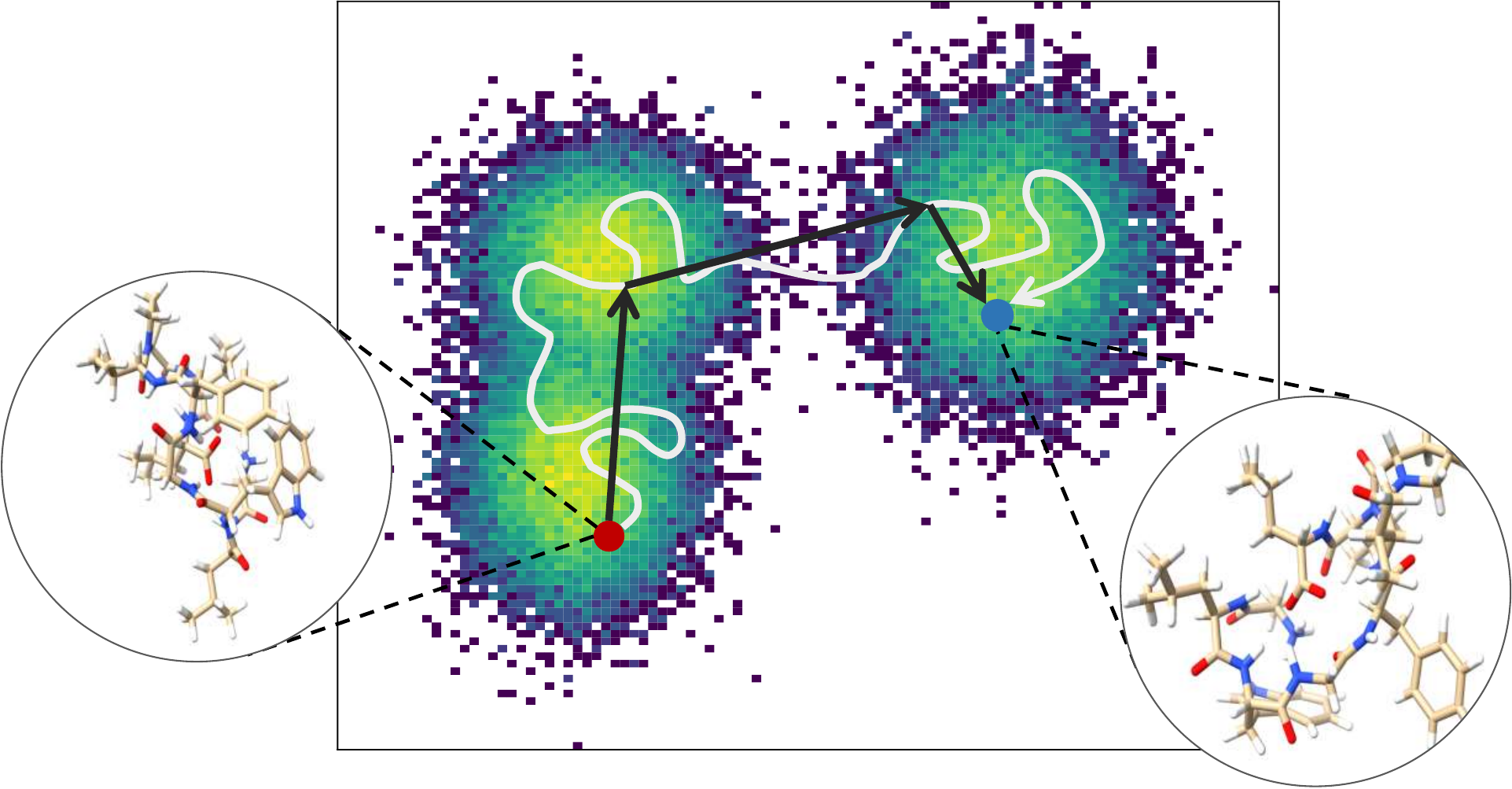}
  \caption{Illustration of how molecular conformations transfer from one state to another by MD (path in white) and time-coarsened dynamics (path in black).}
  \label{fig:time-coarsened-dynamics}
\end{wrapfigure}

Molecular Dynamics (MD), which simulates the physical movements of molecular systems at the atomic level via \emph{in silico} methods, are widely applied in the fields of materials science, physics, chemistry, and pharmacology~\citep{wolf2005deformation, durrant2011molecular, salo2020molecular}. 
Accurate MD simulations enable the researcher to comprehend the equilibrium thermodynamics and kinetics of different molecular phases without the need for expensive wet-lab experiments.

Conventional MD software, like AMBER~\citep{pearlman1995amber} and CHARMM~\citep{vanommeslaeghe2010charmm}, mostly pre-defines an empirical force field of molecular systems and performs MD based on the numerical integration of Newtonian equations over the timestep $\Delta{t}$. To minimize discretization errors, $\Delta{t}$ should be chosen small enough, typically on the order of femtoseconds ($10^{-15}$s).  Consequently, simulating the full duration of critical phase transitions that occur on the microsecond ($10^{-6}$s) or even millisecond ($10^{-3}$s) timescale becomes virtually impossible within a reasonable wall-clock time. To overcome the efficiency limitation in long-timescale simulation, a surge of approaches have been devised, including MD-like methods~\citep{voter2000temperature, pang2017gaussian, laio2002escaping} and Monte Carlo-based methods~\citep{sadigh2012scalable, sugita1999replica, neyts2014combining}. However, all these methods share a common drawback: different molecular systems require  customized simulations, despite the fact that many atomistic systems should, in principle, exhibit similar dynamic mechanisms.

Recently, data-driven approaches~\citep{klein2024timewarp, schreiner2024implicit, hsu2024score, li2024f, jing2024generative} leveraging deep generative models have been used to enhance MD simulations. Unlike traditional numerical integration, these methods directly learn MD from diverse observed trajectories, offering superior transferability across different molecular systems. Additionally, they can accelerate simulations through the use of \textit{time-coarsened dynamics}, where models learn to generate new states after a significantly larger timestep ($\tau \gg \Delta t$) starting from an initial state, as illustrated in Figure~\ref{fig:time-coarsened-dynamics}. Nevertheless, existing learning-based methods still face two significant issues. The first issue is that most current methods are designed solely to learn from the data distribution, which may be biased compared to the underlying Boltzmann distribution~\citep{boltzmann1877beziehung} --- a fundamental concept for describing the thermal equilibrium of molecular systems in relation to atomic positions and velocities. Although Timewarp~\citep{klein2024timewarp} alleviates this issues by employing the Metropolis-Hastings algorithm~\citep{metropolis1953equation} to resample from the equilibrium, the unbearably low acceptance rate still hinders its applicability.
The second issue is that physics priors (\emph{e.g.}, energies and forces) are constantly overlooked during the learning process, which, yet, are critical for providing insights into the dynamics of molecular systems.

In this work, we delicately design a generative model to learn time-coarsened dynamics, and target the Boltzmann-constrained distribution $p(\Vec{\rmX})$. This distribution incorporates a potential energy term $\exp(-k\varepsilon(\Vec{\rmX}))$, into the original data distribution $q(\Vec{\rmX})$, resulting in $p(\Vec{\rmX})=\frac{1}{Z}q(\Vec{\rmX})\exp(-k\varepsilon(\Vec{\rmX}))$. Here, $\Vec{\rmX}$ stands for the molecular conformation, and $Z$ is the normalization factor. The definition of $p(\Vec{\rmX})$ ensures that regions of high density in the generated distribution correspond to low potentials $\varepsilon(\Vec{\rmX})$, thereby aligning more closely with thermodynamic principles.
To learn $p(\Vec{\rmX})$, we utilize the \textit{bridge matching} generative framework~\citep{shi2024diffusion} in order to make the generation to be conditioned on the starting molecular conformations rather than Gaussian priors. 
The most challenging part is that, the vector field to generate $p(\Vec{\rmX})$ is strongly associated with the virtual ``energies'' and ``forces'' during the generation process,  which, yet, are nontrivial to obtain. To address the issue, we derive an effective and rigorous way to interpolate a well-designed \textit{intermediate force field} into the bridge matching framework based on reasonable assumptions. The proposed framework is termed as Force-guided Bridge Matching (FBM). It is also worth noting that FBM employs \texttt{TorchMD-NET}~\citep{pelaez2024torchmdnet} as the backbone model, which enables full-atom modeling while preserving $\mathrm{SO}(3)$-equivariance. Our main contributions are summarized as follows:
\begin{enumerate}
    \item To our best knowledge, FBM is the first full-atom generative model that directly targets the Boltzmann-constrained distribution without extra resampling steps, which is tailored for time-coarsened dynamics.
    \item Under rigorous theoretical derivation, we integrate a well-defined intermediate force field into the bridge matching framework, which effectively involves the guidance of physics priors for better MD simulations.
    \item We evaluate FBM on two datasets consisting of small peptides, where FBM exhibits transferability to unseen peptide systems and consistently showcases state-of-the-art results across various peptides.
\end{enumerate}

\section{Related Work}
\label{sec:related}

\paragraph{Boltzmann Generator}
An important objective of MD research is to quickly sample from the Boltzmann distribution, thereby revealing the free energy landscape and collective variables of matter of molecular systems. \textit{Boltzmann generators}~\citep{noe2019boltzmann, kohler2021smooth, kohler2023rigid, falkner2023conditioning, klein2024timewarp, klein2024transferable} employ generative models to produce samples asymptotically from the Boltzmann distribution mainly by: (i) Apply reweighting techniques to i.i.d. generated samples. (ii) Inference in a Markov Chain Monte Carlo (MCMC) procedure. These approaches heavily rely on MC resampling techniques, which are the bottleneck of the sampling efficiency due to costly energy calculation and low acceptance rates. Most similar to our work, \citet{wang2024protein} propose ConfDiff that incorporate the energy and force guidance directly into score diffusion to target the Boltzmann-constrained distribution, yet it only works well on protein backbones and generate conformations in an unconditional way that fails to capture temporal dynamics. In contrast, FBM introduces the force guidance into the bridge matching framework, making it possible to sample straightforwardly from the Boltzmann-constrained distribution.

\paragraph{Time-Coarsened Dynamics}
To overcome the instability of numerical integration of conventional MD simulations, many deep learning methods have adopted the fashion of \textit{time-coarsened dynamics}, where models learn dynamics of diverse molecular systems over a long timestep. \citet{fu2023simulate} proposes a multi-scale graph network to learn dynamics of polymers, which fails to operate in the all atom system. ITO~\citep{schreiner2024implicit} is devised to learn the transition probability over multiple time resolutions, yet its transferability across chemical space remains unknown. Recently, Timewarp~\citep{klein2024timewarp} and TBG~\citep{klein2024transferable} utilize augmented normalizing flows and flow matching respectively for transferable time-coarsened dynamics of small peptides, while both need additional reweighting procedures to debias expectations to the Boltzmann distribution. In addition, Score Dynamics~\citep{hsu2024score} applies score diffusion to capture dynamic patterns merely from the data distribution, while F$^3$low~\citep{li2024f} and MDGen~\citep{jing2024generative} learn transformations of proteins under coarse-grained representations. On the contrary, FBM is designed to learn time-coarsened dynamics at the atomic level and target the Boltzmann-constrained distribution by directly inference without extra steps, which aligns well to thermodynamic principles and exhibits transferability to unseen peptide systems as well.

\section{Method}
\label{sec:method}
In this section, we will present the overall workflow of our method. Specifically, in \S~\ref{sec:repr}, we will define our task formulation of time-coarsened dynamics, with providing necessary notations. Then in \S~\ref{sec:bbm}, we introduce how to learn the time-coarsened dynamics from the data distribution, via a conditional generative model based on bridge matching, which is dubbed as FBM-{\small BASE}. Built upon this baseline, in \S~\ref{sec:fbm}, we further propose a novel generative framework that targets the Boltzmann-constrained distribution by incorporating an intermediate force field into the FBM-{\small BASE} model, leading to our eventual model FBM. 
All proofs of propositions are provided in \S~\ref{sec:appendix-proof}.


\subsection{Molecular Representation}
\label{sec:repr}
We represent each molecule (namely peptide in our experiments) as a graph $\gG=(\gV,\gE)$ consisting of the node set $\gV$ and the edge set $\gE$. For a molecule with $N$ atoms (including hydrogens), $\gV=\{v_0,\cdots,v_{N-1}\}$ where $v_i$ ($0\leq{i}<N$) represents the $i$-th atom of the molecule. Each node $v_i$ is further attributed with Cartesian coordinates $\Vec{\vx}_i\in\sR^{3}$ from the structural information and node features $\vz_i\in\sR^{H}$ from the embedding of atom types, where $H$ represents the hidden dimension. Particularly for peptides that are composed of 20 natural amino acids and exhibit similar features, we construct the atom type vocabulary based on the atom nomenclature of Protein Data Bank~\citep{berman2000protein} to obtain more refined atomic representations. Formally, the features and Cartesian coordinates of all nodes are concatenated as node representations:
\begin{equation}
    \Vec{\mX}=[\Vec{\vx}_0,\cdots,\Vec{\vx}_{N-1}]^{\top}\in\sR^{N\times{3}},~\mZ=[\vz_0,\cdots,\vz_{N-1}]^{\top}\in\sR^{N\times{H}}.
\end{equation}
The edges are constructed with the cutoff radius $r_{\mathrm{cut}}$. For any node pair $(v_i, v_j)$, the connection is established iff $||\Vec{\vx}_i-\Vec{\vx}_j||<r_{\mathrm{cut}}$. Constructing cutoff graphs is a favorable choice for modelling molecular systems since forces and chemical bonds are highly related to interatomic distances.
With the aforementioned notations, we provide the task formulation below.

\paragraph{Task Formulation}
Given each MD trajectory, we extract molecule pairs $\gD = \{(\gG_s, \gG_{s+\tau}) \mid s \in \gS\}$  to create a training dataset, where $\gG_s$ denotes the starting state at time $s$ and $\gG_{s+\tau}$ represents the future state after a temporal interval \(\tau\).
Our goals are: (i) Train a \textit{baseline} model that fits the conditional data distribution $\mu(\Vec{\mX}_{s+\tau}|\Vec{\mX}_s)$ from $\gD$, denoted as $q$. (ii) Based on the trained baseline model, we further train a new model that admits the Boltzmann constraint: $p(\Vec{\rmX})=\frac{1}{Z}q(\Vec{\rmX})\exp(-k\varepsilon(\Vec{\rmX}))$, where $k$, $\varepsilon$, and $Z$ denote the inverse temperature, the potential of the molecular system, and the partition function, respectively. During the training process, we require the queries of the MD energies and forces for each pair $(\gG_s, \gG_{s+\tau})$, namely, $(\varepsilon(\Vec{\mX}_s), \varepsilon(\Vec{\mX}_{s+\tau}))$ and $(\nabla\varepsilon(\Vec{\mX}_s), \nabla\varepsilon(\Vec{\mX}_{s+\tau}))$. In the following context, we will train three neural networks $v_\theta$, $u_\theta$, and $w_\theta$ to approximate the vector fields that are used to generate the distributions $q$ and $p$.

\subsection{Baseline Bridge Matching}
\label{sec:bbm}
We will present the baseline BM model (\emph{i.e.}, FBM-{\small BASE}) to fit the conditional data distribution $\mu(\Vec{\mX}_{s+\tau}|\Vec{\mX_s})$ from dataset $\gD$. For the sake of convenience and consistency with the continuous diffusion time between $[0,1]$, we refer to the prior graph as $\gG_0$ and the target graph as $\gG_1$ in what follows, and denote their corresponding coordinates as $\Vec{\mX}_0$ and $\Vec{\mX}_1$ accordingly.

\paragraph{Bridge Matching Framework} We leverage the generative framework of \textit{bridge matching}~\citep{shi2024diffusion}, which learns a mimicking diffusion process between two arbitrary distributions, allowing for more flexible choices of priors. Let $q_0, q_1$ denote the prior and target data distributions, respectively, and $t\in[0,1]$ be the continuous diffusion time. We define the forward process:
\begin{equation}
    \label{eq:diffusion}
    \dif\Vec{\rmX}_t=f_t(\Vec{\rmX}_t)\dif t+\sigma_t\dif\rmB_t,~\Vec{\rmX}_0\sim{q_0},
\end{equation}
where $\Vec{\rmX}_t$ and $\rmB_t$  represent the random variable and the Brownian motion of the diffusion process at time $t$, respectively. With the process pinned down at an initial point $\Vec{\mX}_0$ and terminal point $ \Vec{\mX}_1$, the conditional distribution $q_t(\cdot|\Vec{\mX}_0,\Vec{\mX}_1)$ will be realized as a \textit{diffusion bridge} in the form of $\dif\Vec{\rmX}_t=\{f_t(\Vec{\rmX}_t)+\sigma_t^2\nabla\log{q_t(\Vec{\mX}_1|\Vec{\rmX}_t)}\}\dif t+\sigma_t\dif\rmB_t$ with $\Vec{\rmX}_0=\Vec{\mX}_0$, where Doob $h$-transform theory~\citep{rogers2000diffusions} guarantees $\Vec{\rmX}_1=\Vec{\mX}_1$. For simplification, by considering $f_t=0$ and $\sigma_t=\sigma$, the process will degenerate to the following \textit{Brownian bridge}:
\begin{equation}
    \label{eq:bridge}
    \dif\Vec{\rmX}_t=\frac{\Vec{\mX}_1-\Vec{\rmX}_t}{1-t}\dif t+\sigma{\dif\rmB_t},~\Vec{\rmX}_0=\Vec{\mX}_0,
\end{equation}
which yields the conditional distribution $q_t(\Vec{\rmX}_t|\Vec{\mX}_0,\Vec{\mX}_1)$ at time $t\in[0,1]$, with its marginal defined as $q_t(\Vec{\rmX}_t)$. The core of bridge matching is to find a Markov diffusion governed by a \textit{vector field} $v$:
\begin{equation}
    \label{eq:markov-diffusion}
    \dif\Vec{\rmX}_t=v(\Vec{\rmX}_t, t)\dif t+\sigma{\dif\rmB_t},
\end{equation}
which admits the same marginal $\Vec{\rmX}_t\sim{q_t}$, such that $\Vec{\rmX}_1\sim{q_1}$ holds. To achieve this, we can learn a parametric vector field $v_{\theta}$ via the following regression loss:
\begin{equation}
    \label{eq:loss-fwd}
    \gL_{\mathrm{fwd}}=\E_{t\sim\mathrm{Uni}(0,1),(\gG_0,\gG_1)\sim\gD,\Vec{\mX}_t\sim{q_t(\cdot|\Vec{\mX}_0,\Vec{\mX}_1)}}[||\frac{\Vec{\mX}_1-\Vec{\mX}_t}{1-t}-v_{\theta}(\Vec{\mX}_t, t)||^2],
\end{equation}
where $\mathrm{Uni}(0,1)$ denotes the uniform distribution of $[0,1]$, and $v_{\theta}$ is implemented by a neural network parameterized by $\theta$. The calculation of \cref{eq:loss-fwd} requires sampling from $q_t(\Vec{\rmX}_t|\Vec{\mX}_0,\Vec{\mX}_1)$, which usually needs additional SDE simulations. Fortunately, \cref{eq:bridge} enables a closed-form solution as follows:
\begin{equation}
    \label{eq:conditional}
    q_t(\Vec{\rmX}_t|\Vec{\mX}_0,\Vec{\mX}_1)=\gN(t\Vec{\mX}_1+(1-t)\Vec{\mX}_0, t(1-t)\sigma^2\bm{I}),
\end{equation}
where we can sample $\Vec{\mX}_t$ efficiently during each training step with any given $t$.

The optimal vector field $v_{\theta}$ to minimize the loss in \cref{eq:loss-fwd} is actually equal to the expectation $v^*(\Vec{\rmX}_t,t)=\E_{q_t(\cdot,\cdot|\Vec{\rmX}_t)}[\frac{\Vec{\mX}_1-\Vec{\rmX}_t}{1-t}]$. Then the distribution $q_t$ can be estimated by performing SDE sampling in \cref{eq:markov-diffusion}, using $v_{\theta}$ as the vector field $v$. In our experiments,  $v_{\theta}$ is built upon \texttt{TorchMD-NET}~\citep{pelaez2024torchmdnet} for full-atom modeling while satisfying $\mathrm{SO}(3)$-equivariance. Details of the architecture of neural networks used in our model are further elucidated in \S~\ref{sec:arch}.  

\subsection{Force-guided Bridge Matching}
\label{sec:fbm}
\begin{figure}[t]
  \centering
  \vspace{-3em}
  \includegraphics[width=\linewidth]{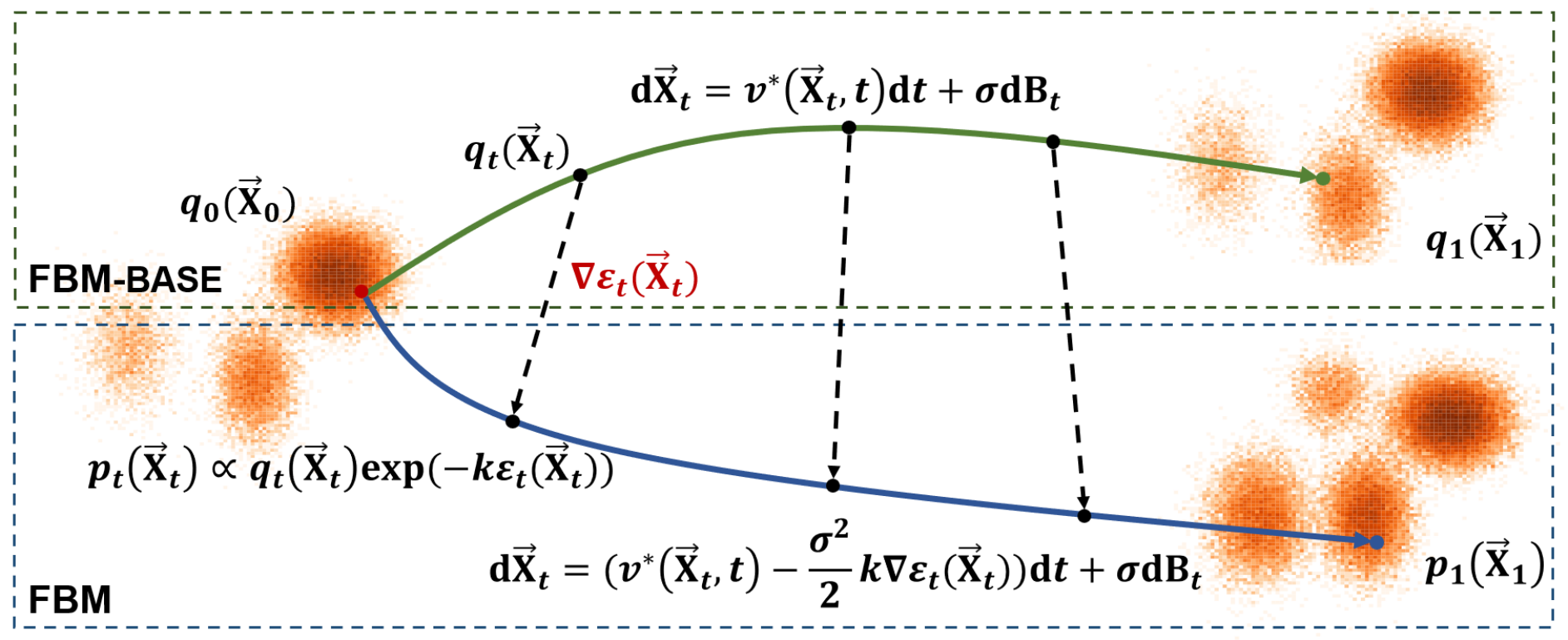}
  \caption{The overall framework of FBM-{\small BASE} and FBM. \textbf{A.} Firstly, FBM-{\small BASE} leverages the bridge matching framework to learn time-coarsened dynamics from the data distributions $q_0$ and $q_1$. \textbf{B.} With the guidance of the intermediate force field $\nabla\varepsilon_t$ at diffusion time $t$, the marginal distribution admits $p_t(\Vec{\rmX}_t)\propto{q_t(\Vec{\rmX}_t)\exp(-k\varepsilon_t(\Vec{\rmX}_t))}$, thereby the target distribution of FBM is debiased to the Boltzmann-constrained distribution $p_1$.}
  \label{fig:fbm-process}
\end{figure}

In most cases, the training datasets are biased from the underlying Boltzmann distribution, leading to defective prediction even for superior generative models. In this section, we will introduce how to train a force-guided generative model FBM, which admits the Boltzmann-constrained distribution denoted as $p_1(\Vec{\rmX}_1)=\frac{1}{Z}q_1(\Vec{\rmX}_1)\exp(-k\varepsilon(\Vec{\rmX}_1))$. Here, the exponential term $\exp(-k\varepsilon(\Vec{\rmX}_1))$ serves as the physics prior from thermodynamic principles. The overall framework of FBM as well as FBM-{\small BASE} is illustrated in Figure~\ref{fig:fbm-process}.

\paragraph{Force-guided Bridge Matching Framework}
In order to learn $p_1$ under the bridge matching framework, our key idea is to construct a new probabilistic path $p_t$ based on the existing probabilistic path $q_t$ from \cref{eq:bridge}, such that the following condition is satisfied for $t\in[0,1]$:
\begin{equation}
    \label{eq:Boltzmann-constrained-dis}
    p_t(\Vec{\rmX}_t)=\frac{1}{Z_t}q_t(\Vec{\rmX}_t)\exp(-k\varepsilon_t(\Vec{\rmX}_t)),~\varepsilon_0(\cdot)=\varepsilon_1(\cdot)=\varepsilon(\cdot).
\end{equation}
Here $Z_t$ is the partition function and $\varepsilon_t$ is an artificially-designed \textit{intermediate potential} of the process, which should converge to the real MD potential $\varepsilon$ when $t\to0^+$ and $t\to1^-$ for consistency. Further, we assume $p_t(\Vec{\rmX}_t|\Vec{\mX}_0,\Vec{\mX}_1)=q_t(\Vec{\rmX}_t|\Vec{\mX}_0,\Vec{\mX}_1)$ as in \cref{eq:conditional}, thereby the stochastic process governed by $p_t$ shares the same form with \cref{eq:bridge}, which can be modeled by the following Markov diffusion process associated with a vector field $v'$:
\begin{equation}
    \label{eq:markov-diffusion-p}
    \dif\Vec{\rmX}_t=v'(\Vec{\rmX}_t, t)\dif t+\sigma{\dif\rmB_t}.
\end{equation}
Therefore, we are able to inference $p_t$ if we know how to learn $v'$ from the dataset. Interestingly, we prove that $v'(\Vec{\rmX}_t,t)$ can be expressed in terms of the vector field $v^*(\Vec{\rmX}_t, t)$ generating $q_t$ and the intermediate force field $\nabla\varepsilon_t(\Vec{\rmX}_t)$ allowing for the Boltzmann constraint, which will be formally demonstrated in Proposition~\ref{prop:vector-field}. In prior to showing this proposition, we first derive the form of the intermediate force field below:
\begin{proposition}
    \label{prop:force-field}
    Assume that the joint distributions $q(\Vec{\rmX}_0,\Vec{\rmX}_1)$ and $p(\Vec{\rmX}_0,\Vec{\rmX}_1)$ satisfy $p(\Vec{\rmX}_0,\Vec{\rmX}_1)\propto{q(\Vec{\rmX}_0,\Vec{\rmX}_1)\exp(-k(\varepsilon(\Vec{\rmX}_0)+\varepsilon(\Vec{\rmX}_1)))}$, the intermediate force field $\nabla\varepsilon_t$ is given by:
\begin{align}
    \label{eq:force-field}
    \nabla\varepsilon_t(\Vec{\rmX}_t)&=\frac{\E_{q(\Vec{\rmX}_0,\Vec{\rmX}_1)}[q_t(\Vec{\rmX}_t|\Vec{\mX}_0,\Vec{\mX}_1)\exp(-k(\varepsilon(\Vec{\mX}_0)+\varepsilon(\Vec{\mX}_1)))\zeta(\Vec{\mX}_0,\Vec{\mX}_1,\Vec{\rmX}_t)]}{k\E_{q(\Vec{\rmX}_0,\Vec{\rmX}_1)}[q_t(\Vec{\rmX}_t|\Vec{\mX}_0,\Vec{\mX}_1)\exp(-k(\varepsilon(\Vec{\mX}_0)+\varepsilon(\Vec{\mX}_1)))]},
\end{align}
where we denote $\zeta(\Vec{\mX}_0,\Vec{\mX}_1,\Vec{\rmX}_t)=\nabla\log{q_t}(\Vec{\rmX}_t)-\nabla\log{q_t}(\Vec{\rmX}_t|\Vec{\mX}_0,\Vec{\mX}_1)$ for brevity.
\end{proposition}
\begin{proposition}
    \label{prop:vector-field}
    Given the prerequisites in Proposition~\ref{prop:force-field}, we have $v'(\Vec{\rmX}_t, t)=v^*(\Vec{\rmX}_t, t)-\frac{\sigma^2}{2}k\nabla\varepsilon_t(\Vec{\rmX}_t)$ under some mild assumptions.
\end{proposition}

\paragraph{Estimation of Immediate Force Field}
However, it is challenging to calculate the intermediate force field $\nabla\varepsilon_t(\Vec{\rmX}_t)$ owing to its nontrivial form in \cref{eq:force-field}. Note that $q_t(\Vec{\rmX}_t|\Vec{\mX}_0,\Vec{\mX}_1)$ can be directly calculated by \cref{eq:conditional}, $\varepsilon(\Vec{\mX}_0),\varepsilon(\Vec{\mX}_1)$ are both known as MD potentials, and the the expectation in the denominator can be estimated with samples during a training mini-batch instead of the entire data distribution~\citep{lu2023contrastive}. The most challenging part is the computation of the term $\nabla\log{q_t}(\Vec{\rmX}_t)$ in $\zeta(\Vec{\mX}_0,\Vec{\mX}_1,\Vec{\rmX}_t)$. To provide an unbiased estimation of this score, we investigate its relation with the vector fields. 
Given $\gG_0, \gG_1$, we first define the conditional score as $s_t(\Vec{\rmX}_t|\Vec{\mX}_0,\Vec{\mX}_1)=\nabla\log{q_t(\Vec{\rmX}_t|\Vec{\mX}_0,\Vec{\mX}_1)}$. Based on \cref{eq:conditional}, the closed-form of $s_t$ is given by:
\begin{equation}
    \label{eq:score}
    s_t(\Vec{\rmX}_t|\Vec{\mX}_0,\Vec{\mX}_1)=-\frac{\Vec{\rmX}_t-[t\Vec{\mX}_1+(1-t)\Vec{\mX}_0]}{t(1-t)\sigma^2}=\frac{1}{\sigma^2}[\frac{\Vec{\mX}_1-\Vec{\rmX}_t}{1-t}-\frac{\Vec{\rmX}_t-\Vec{\mX}_0}{t}].
\end{equation}
Note that the first term of \cref{eq:score} has the same form as in the training objective of \cref{eq:loss-fwd}. Similarly, we train another network $u_{\theta}$ to imitate the second term:
\begin{equation}
    \label{eq:loss-rev}
    \gL_{\mathrm{rev}}=\E_{t\sim\mathrm{Uni}(0,1),(\gG_0,\gG_1)\sim\gD,\Vec{\mX}_t\sim{q_t(\cdot|\Vec{\mX}_0,\Vec{\mX}_1)}}[||\frac{\Vec{\mX}_t-\Vec{\mX}_0}{t}-u_{\theta}(\Vec{\mX}_t, t)||^2],
\end{equation}
where the expectation of $u_{\theta}$ can be expressed as $u^*(\Vec{\rmX}_t,t)=\E_{q_t(\cdot,\cdot|\Vec{\rmX}_t)}[\frac{\Vec{\rmX}_t-\Vec{\mX}_0}{t}]$. Then we take the expectation over $\Vec{\rmX}_0,\Vec{\rmX}_1$ conditioned on $\Vec{\rmX}_t$ in \cref{eq:score}, yielding:
\begin{equation}
    \label{eq:marginal-score}
    s_t^*(\Vec{\rmX}_t)=\E_{q_t(\cdot,\cdot|\Vec{\rmX}_t)}[s_t(\Vec{\rmX}_t|\Vec{\rmX}_0,\Vec{\rmX}_1)]=\frac{v^*(\Vec{\rmX}_t,t)-u^*(\Vec{\rmX}_t,t)}{\sigma^2}.
\end{equation}
We present Proposition~\ref{prop:score} to reveal that $s_t^*$ is identical to the marginal score $\nabla\log{q_t}$ of interest:
\begin{proposition}
    \label{prop:score}
    We have $\nabla\log{q_t}(\Vec{\rmX}_t)=s_t^*(\Vec{\rmX}_t)$, where $\nabla\log{q_t}$ is the score of the Brownian bridge defined in \cref{eq:bridge} and $s_t^*(\Vec{\rmX}_t)$ is the expectation of the conditional score given by \cref{eq:marginal-score}.
\end{proposition}
In practice, $\nabla\log{q_t}(\Vec{\rmX}_t)$ is estimated as $\nicefrac{(v_\theta(\Vec{\rmX}_t,t)-u_\theta(\Vec{\rmX}_t,t))}{\sigma^2}$, by replacing the vector fields with the learned neural networks $v_\theta, u_\theta$ in \cref{eq:marginal-score}.
Now, since all quantities related to the intermediate force field $\nabla\varepsilon_t(\Vec{\rmX}_t)$ are commutable, we then train a neural network $w_{\theta}(\Vec{\rmX}_t, t)$ as its unbiased estimator:
\begin{equation}
    \label{eq:loss-iff}
    \gL_{\mathrm{iff}}=\E_{t,(\gG_0,\gG_1),q_t(\cdot|\Vec{\mX}_0,\Vec{\mX}_1)}[||\frac{\exp(-k(\varepsilon(\Vec{\mX}_0)+\varepsilon(\Vec{\mX}_1)))\zeta(\Vec{\mX}_0,\Vec{\mX}_1,\Vec{\mX}_t)}{k\E_{q(\Vec{\mX}_0,\Vec{\mX}_1)^B}[q_t(\Vec{\mX}_t|\Vec{\mX}_0,\Vec{\mX}_1)\exp(-k(\varepsilon(\Vec{\mX}_0)+\varepsilon(\Vec{\mX}_1)))]}-w_{\theta}(\Vec{\mX}_t,t)||^2],
\end{equation}
where $B$ denotes the mini-batch size of each training step. 

\subsection{Full Training Processes and Inference}

In addition to the aforementioned regression losses, we introduce an auxiliary loss from \citet{yim2023se}, which promotes predictions of pairwise atomic relations. The loss is defined as:
\begin{equation}
    \label{eq:loss-aux}
    \gL_{\mathrm{aux}}=(1-t)\cdot\frac{||\bm{1}_{\mD_0<6\mathring{\mathrm{A}}}(\mD_0-\hat{\mD}_0)||^2}{\sum{\bm{1}_{\mD_0<6\mathring{\mathrm{A}}}}-N}+t\cdot\frac{||\bm{1}_{\mD_1<6\mathring{\mathrm{A}}}(\mD_1-\hat{\mD}_1)||^2}{\sum{\bm{1}_{\mD_1<6\mathring{\mathrm{A}}}}-N},
\end{equation}
where $D_0, D_1\in\sR^{N\times{N}}$ denote pairwise distances between all atoms of $\gG_0$ and $\gG_1$, and $\hat{D}_0, \hat{D}_1$ are defined in the same way based on the estimated starting point $\hat{\Vec{\mX}}_0=\Vec{\mX}_t-tu_{\theta}(\Vec{\mX}_t,t)$ and terminal point $\hat{\Vec{\mX}}_1=\Vec{\mX}_t+(1-t)v_{\theta}(\Vec{\mX}_t,t)$. The full loss of FBM-{\small BASE} is given by:
\begin{equation}
    \label{eq:loss-full}
    \gL_{\mathrm{base}}=\gL_{\mathrm{fwd}}+\gL_{\mathrm{rev}}+\lambda_{\mathrm{aux}}\cdot\gL_{\mathrm{aux}},
\end{equation}
where $\lambda_{\mathrm{aux}}$ is a hyper-parameter to balance the weight of different training objectives.

Empirically, large variances are noticed during training FBM with \cref{eq:loss-iff} when $t$ is close to 0 and 1. To address the issue, we find that the intermediate force field converges to the MD force field $\nabla\varepsilon$ at $t=0,1$, which is guaranteed by Proposition~\ref{prop:boundary}:
\begin{proposition}
    \label{prop:boundary}
    Given $\varepsilon_0=\varepsilon_1=\varepsilon$ and the intermediate force field described in \cref{eq:force-field}, the continuity condition $\lim_{t\to{0^+}}\nabla\varepsilon_t(\Vec{\rmX}_t)=\nabla\varepsilon(\Vec{\rmX}_0),~\lim_{t\to{1^-}}\nabla\varepsilon_t(\Vec{\rmX}_t)=\nabla\varepsilon(\Vec{\rmX}_1)$ holds.
\end{proposition}
Therefore, we leverage two separate networks $w^{(1)}_{\theta}, w^{(2)}_{\theta}$ to learn the boundary force fields:
\begin{equation}
    \label{eq:loss-bound}
    \gL_{\mathrm{bnd}}=\E_{t,(\gG_0,\gG_1),q_t(\cdot|\Vec{\mX}_0,\Vec{\mX}_1)}[||\nabla\varepsilon(\Vec{\mX}_0)-w^{(1)}_{\theta}(\Vec{\mX}_t,t)||^2+||\nabla\varepsilon(\Vec{\mX}_1)-w^{(2)}_{\theta}(\Vec{\mX}_t,t)||^2].
\end{equation}
We construct the network $w_{\theta}$ in the interpolation form with another network $w^{(3)}_{\theta}$, similar to~\citet{mate2023learning}: $w_{\theta}(\Vec{\rmX}_t,t)=(1-t)w^{(1)}_{\theta}(\Vec{\rmX}_t,t)+tw^{(2)}_{\theta}(\Vec{\rmX}_t,t)+t(1-t)w^{(3)}_{\theta}(\Vec{\rmX}_t,t)$. The ultimate loss for training FBM is given by:
\begin{equation}
    \label{eq:loss-fbm}
    \gL_{\mathrm{FBM}}=\gL_{\mathrm{iff}}+\gL_{\mathrm{bnd}}.
\end{equation}
\paragraph{Full Training Processes}
We first perform training to derive $v_\theta$ and $u_\theta$ under the base loss in \cref{eq:loss-full}, and then continue the training process to attain $w_\theta$ under the FBM loss in \cref{eq:loss-fbm}.
The pseudo codes for training FBM-{\small BASE} and FBM are in Alg.~\ref{alg:base-train} and Alg.~\ref{alg:fbm-train}, respectively.

\paragraph{Force-guided Inference}
For inference with FBM, we estimate the vector field $v'$ based on Proposition~\ref{prop:vector-field}, by replacing $v^*$ and $\nabla\varepsilon_t$ with the trained neural networks $v_{\theta}$ and $w_{\theta}$. Force-guided inference is then performed following the SDE process of \cref{eq:markov-diffusion-p}, where the diffusion time $t$ is discretized equidistantly. Additional details and pseudo codes for inference are provided in \S~\ref{sec:appendix-train-infer}.

\section{Experiments}
\label{sec:experiment}

\label{sec:experiment-setup}

\paragraph{Dataset Generation}
We evaluate FBM on two datasets consisting of small peptides: \emph{Alanine-Dipeptide (AD)} that is commonly studied previously and \emph{PepMD} which is created by us. 
The AD dataset contains a simple peptide with only 22 atoms. The initial structure and reference MD trajectories of AD are all obtained from \texttt{mdshare}\footnote{https://github.com/markovmodel/mdshare} without post-processing. 
As for PepMD, we first screen valid peptides between 3-10 residues from the sequence data provided by PDB~\citep{berman2000protein}. Next, we perform data cleaning according to the following criterion: each peptide must contain only the 20 natural amino acids, and the number of any type of residue should not exceed 50\% of the sequence length. We then cluster the data with a sequence identity threshold of 60\% by \texttt{MMseq2}~\citep{steinegger2017mmseqs2}, and randomly sample one peptide from each cluster to obtain a non-redundant dataset. Considering the computing resource constraints, we select 136/14 peptides for constructing the training-validation/test set, respectively. The structures of all 150 peptides are predicted by open-source tools \texttt{RDKit}~\citep{rdkit} and \texttt{PDBfixer}\footnote{https://github.com/openmm/pdbfixer}, which are sent as initial states to generate MD trajectories using \texttt{OpenMM}~\citep{eastman2017openmm} afterwards. Finally, the peptide pairs for training are then sampled from trajectories in the way depicted in \S~\ref{sec:repr}. The MD simulation setups and the statistical details of our curated dataset are both illustrated in \S~\ref{sec:appendix-dataset-detail}.


\paragraph{Baselines}
We compare FBM with the following generative models that learn time-coarsened dynamics: (i) Timewarp~\citep{klein2024timewarp}, the current state-of-the-art model that targets the Boltzmann distribution by MCMC resampling, which exhibits superior transferability to unseen peptide systems. (ii) ITO~\citep{schreiner2024implicit}, a conditional diffusion model that learns dynamics at multiple time resolutions. (iii) Score Dynamics~\citep{hsu2024score}, a score-based diffusion model that learns transitions of collective variables. All models were trained from scratch for fair comparison.

\paragraph{Metrics}
Following \citet{wang2024protein}, we evaluate generated conformation ensembles against the full MD trajectories as to their validity, flexibility, and distributional similarity. We provide brief descriptions of the metrics in this part and further details are illustrated in \S~\ref{sec:appendix-metric-detail}:
\begin{itemize}
    \item \textbf{Validity}. We consider a molecular conformation as \textit{valid} when it is governed by certain physics constraints. Following \citet{lu2023str2str}, we judge whether a conformation is valid by the criterion: no bond clashes between any residue pairs and no bond breaks between adjacent residues, based on coordinates of $\alpha$-carbons. This metric, named as V{\small AL}-CA, represents the fraction of valid conformations in the full generated conformation ensembles.
    \item \textbf{Flexibility}. The generated structures are also required to exhibit flexibility to capture dynamic characteristics. Following \citet{janson2023direct}, we report the root mean square error of contact maps between the generated conformation ensembles and the reference MD trajectories as a measure of \textit{flexibility}, termed as C{\small ONTACT}.
    \item \textbf{Distributional similarity}. We estimate the similarity between the generated distribution and the Boltzmann distribution by projecting conformations onto following low-dimensional feature spaces~\citep{lu2023str2str}: (i) pairwise distances between $\alpha$-carbons of residues (P{\small W}D); (ii) radius-of-gyration (R{\small G}) that measures the distribution of $\alpha$-carbons to the center-of-mass; (iii) the \textit{time-lagged independent components}~\citep{perez2013identification} (TIC), where only the slowest two components are taken into consideration. For each metric, the mean Jensen-Shannon (JS) distance along the feature dimensions is reported.
\end{itemize}

\subsection{Metastable States Exploration for AD}
We first investigate how well the generated conformations can travel across different metastable states of AD. Due to the simple structure of AD with only one peptide bond, some metrics are not applicable except for TIC and TIC-2D (\emph{i.e.}, the joint distribution of TIC 0 and TIC 1). In particular, the backbone dihedrals of AD, psi and phi, are commonly considered as two challenging variables for state transitions. Therefore, we include the similarity measurement of the joint distribution of psi and phi, \emph{i.e.} the \textit{Ramachandran plot}~\citep{1963stereochemistry}, denoted as R{\small AM}.

In Table \ref{tab:result-ad} we show evaluation results on AD, where models sample in the time-coarsened manner from the same initial state for a chain length of $10^3$. According to Table \ref{tab:result-ad}, FBM outperforms existing baselines on both R{\small AM} and TIC metrics, and with the introduction of physics priors, it shows considerable improvements in distribution similarity across various feature spaces compared to FBM-{\small BASE}. Although Timewarp surpasses FBM in the TIC-2D metric, we will explain later that it comes at the cost of generating invalid conformations.

\begin{figure}[H]
    \centering
    \resizebox{\textwidth}{!}{
    \begin{tabular}{ccc}
        \begin{subfigure}{0.32\textwidth}
            \centering
            \includegraphics[width=\linewidth]{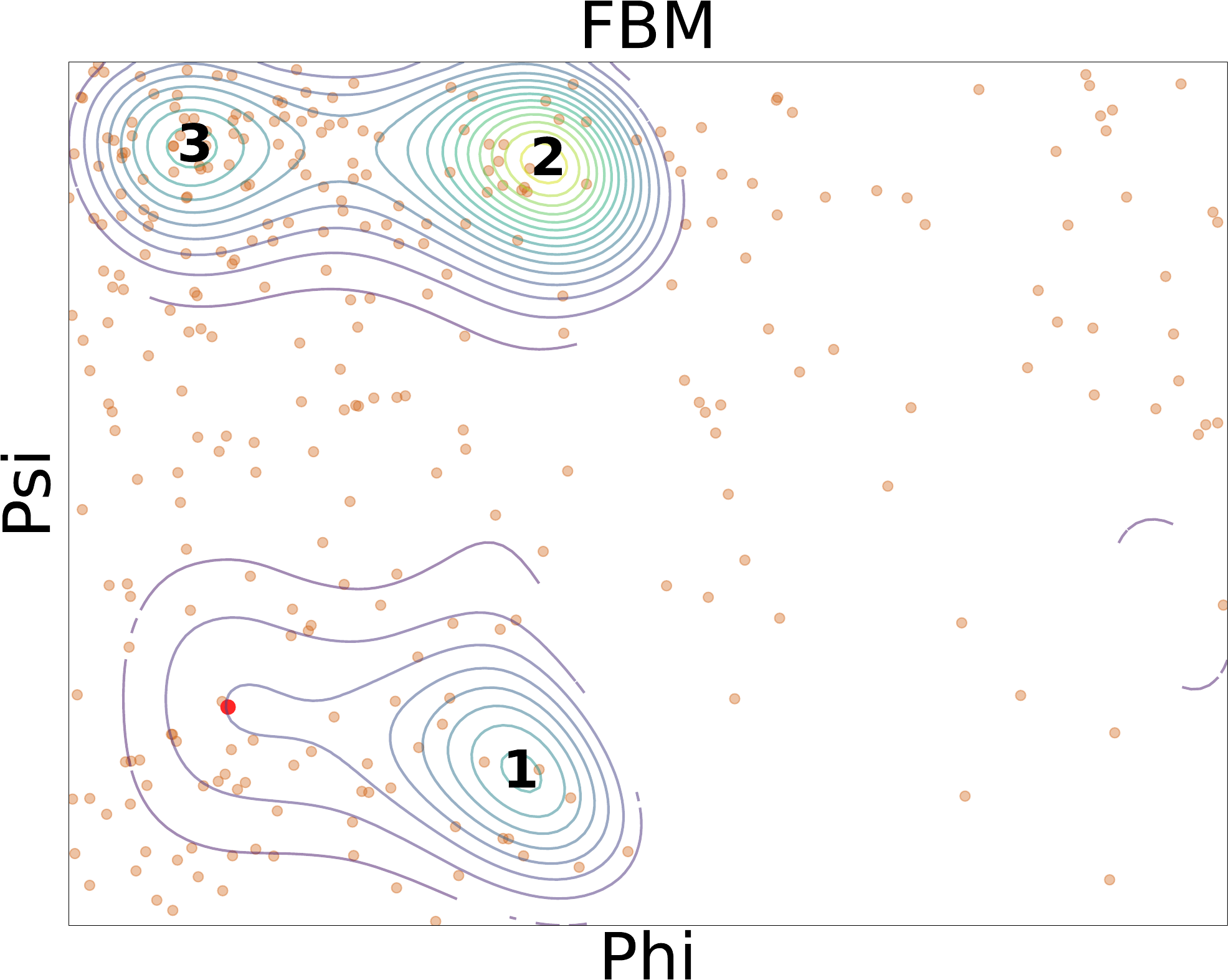} 
        \end{subfigure} &
        \begin{subfigure}{0.32\textwidth}
            \centering
            \includegraphics[width=\linewidth]{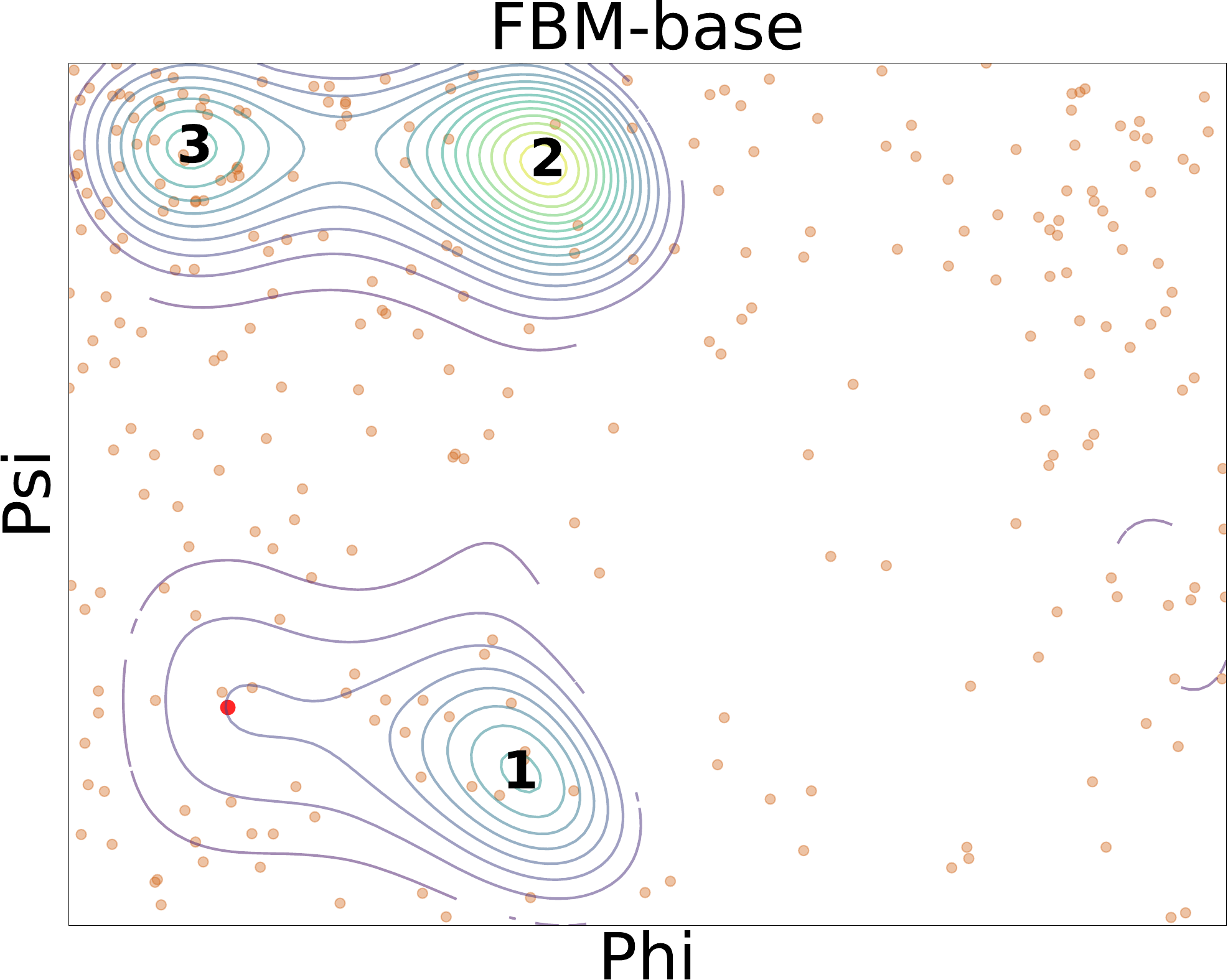} 
        \end{subfigure} &
        \begin{subfigure}{0.32\textwidth}
            \centering
            \includegraphics[width=\linewidth]{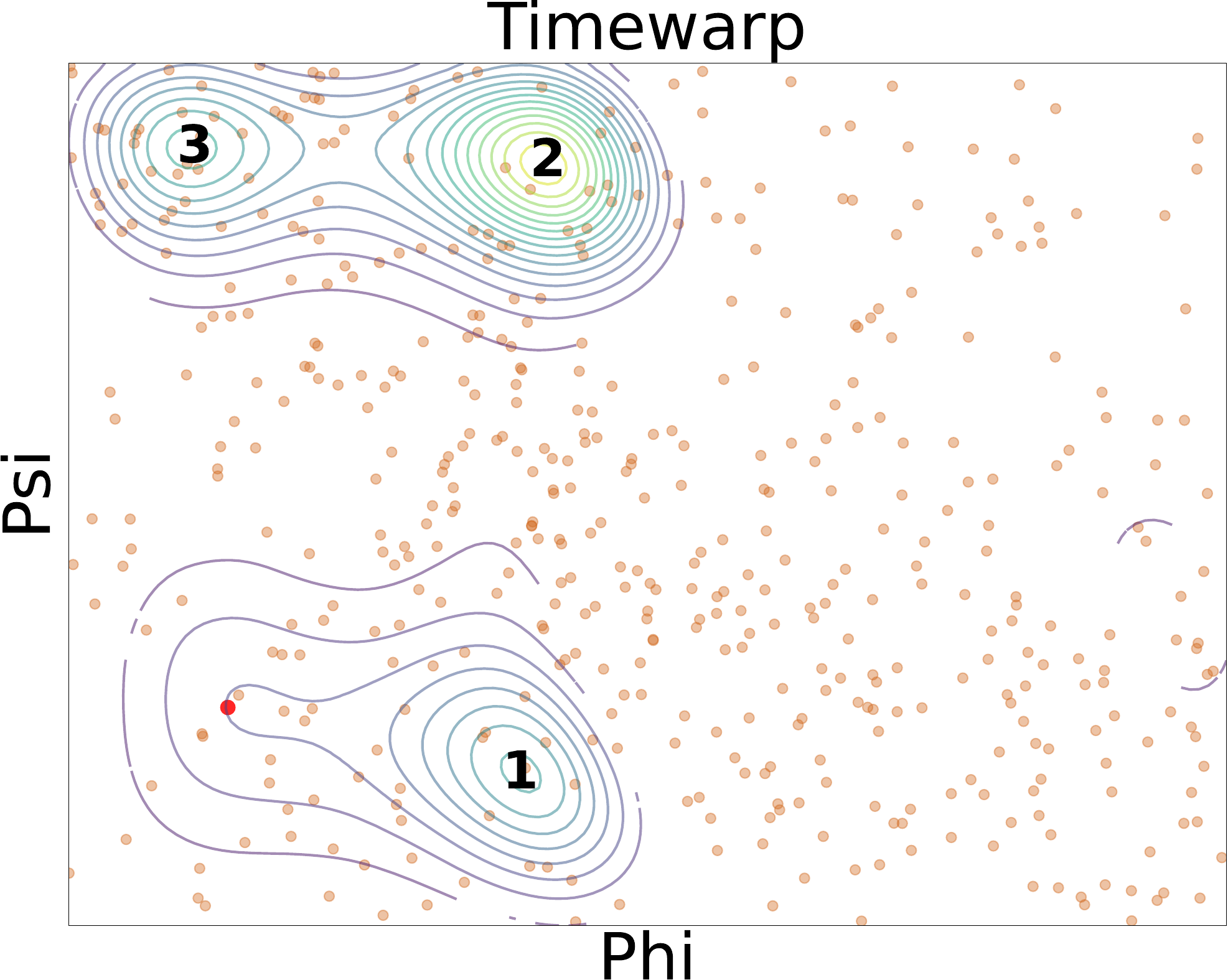} 
        \end{subfigure} \\
        \begin{subfigure}{0.32\textwidth}
            \centering
            \includegraphics[width=\linewidth]{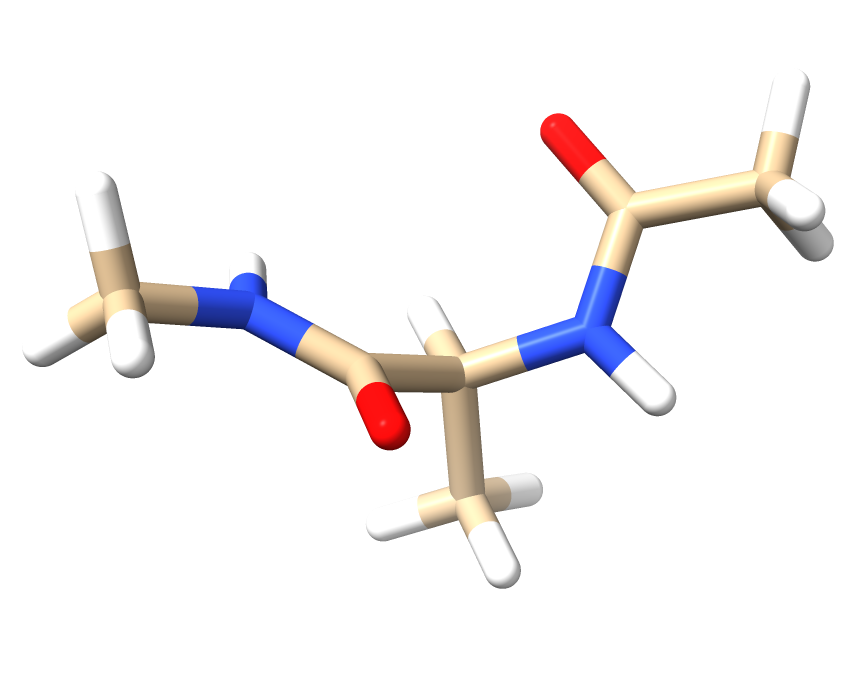}
        \end{subfigure} &
        \begin{subfigure}{0.32\textwidth}
            \centering
            \includegraphics[width=\linewidth]{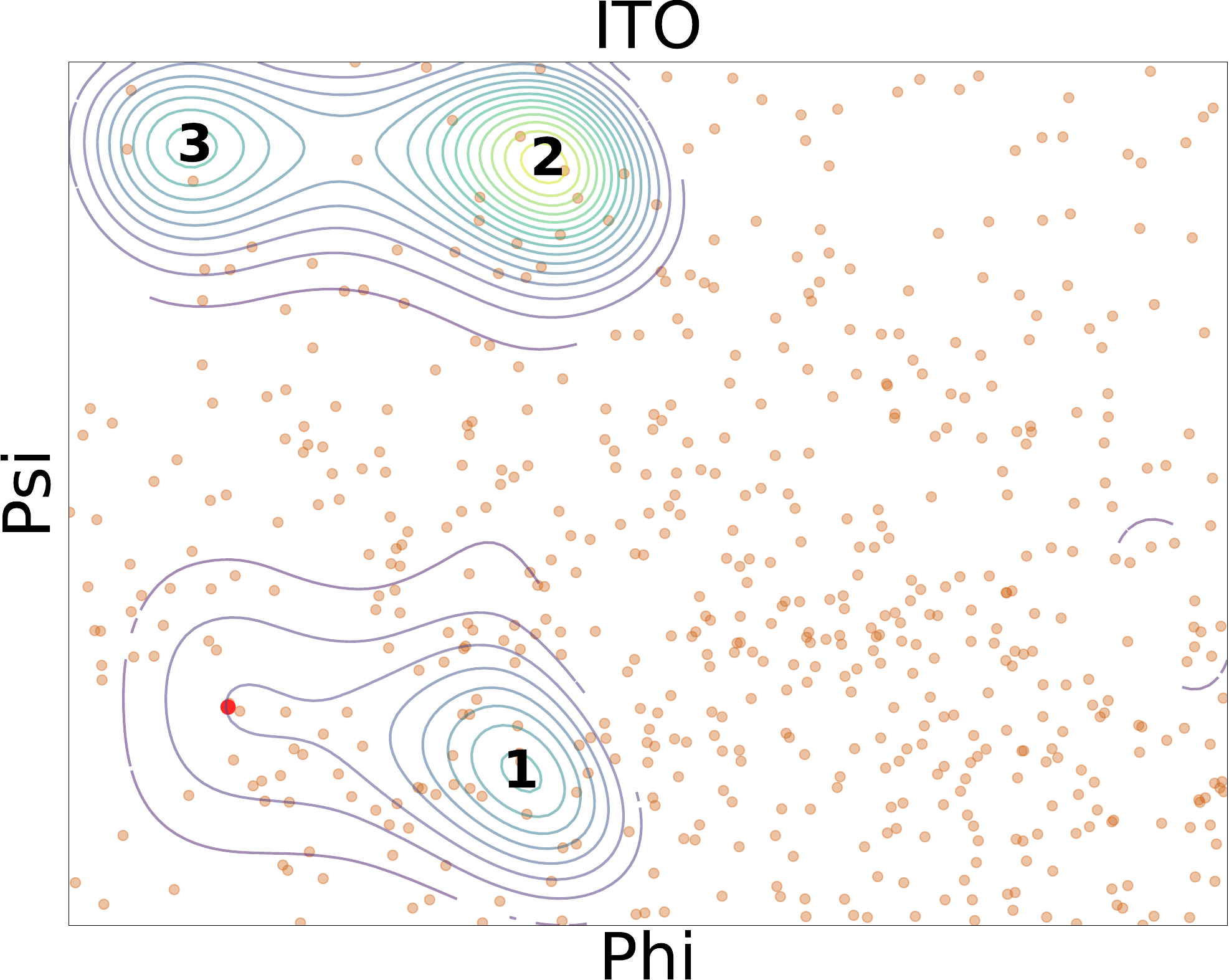} 
        \end{subfigure} &
        \begin{subfigure}{0.32\textwidth}
            \centering
            \includegraphics[width=\linewidth]{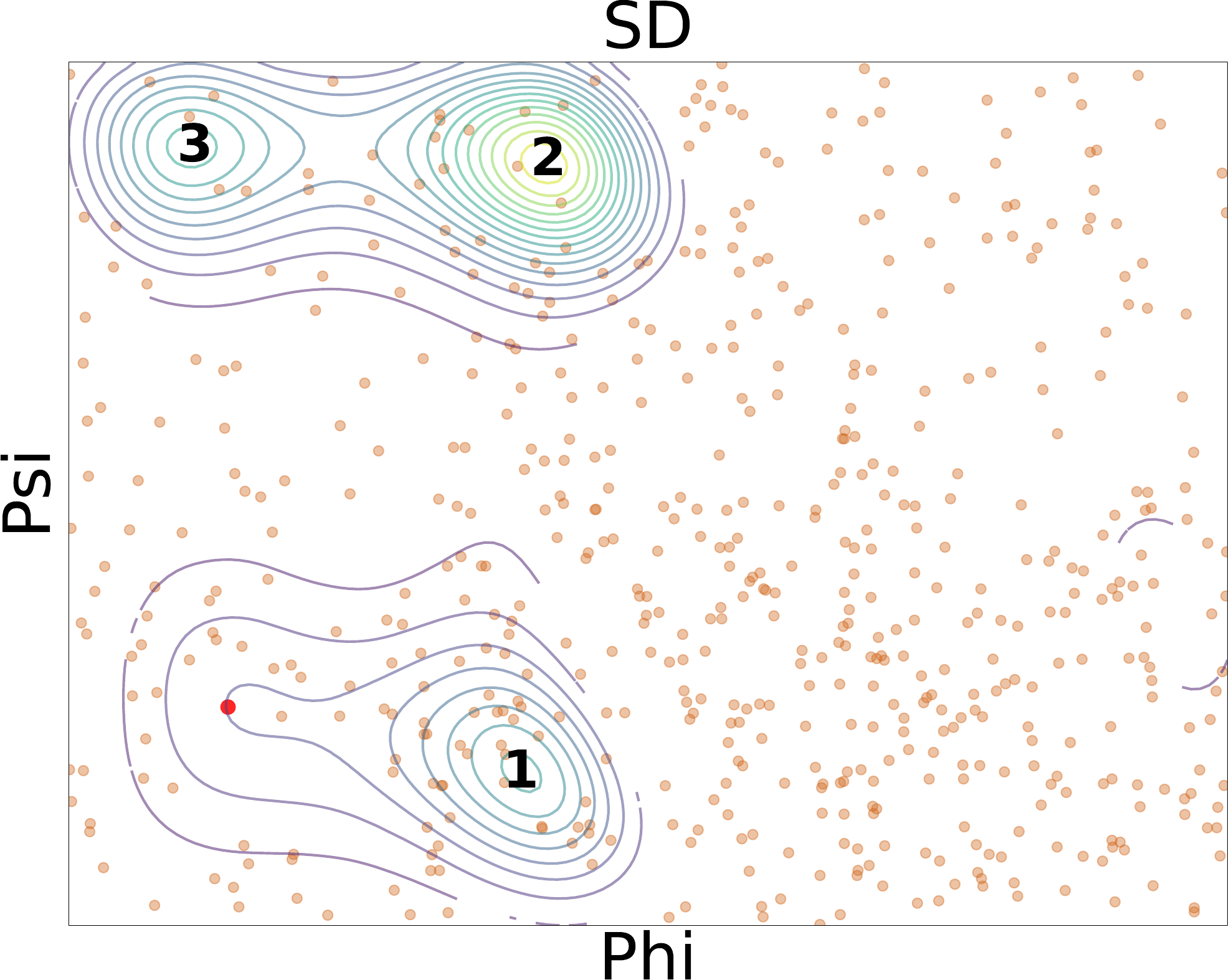} 
        \end{subfigure}
    \end{tabular}
    }
    \caption{Ramachandran plots of alanine dipeptide with conformation ensembles generated by models. The initial state is indicated with the red cross. Contours represent the kernel densities estimated by the MD trajectory and the generated conformations are shown in scatter.}
    \label{fig:ram-ad}
\end{figure}

Further, Ramachandran plots of generated ensembles are illustrated in Figure \ref{fig:ram-ad}, where three known metastable states are recognized based on MD trajectories and labeled in order. Apparently, ITO and Score Dynamics fail to capture the dynamics of AD with samples randomly allocated. Moreover, Timewarp cannot rapidly traverse metastable states, resulting in a great portion of invalid samples. Despite both FBM-{\small BASE} and FBM showing relatively ``clean'' plots with fewer unreasonable conformations, samples of FBM are more concentrated in high-density regions, confirming a strong guidance of the intermediate force field to align closely with thermodynamic principles.

\begin{table}[H]
\centering
\caption{Results on alanine dipeptide. Values of each metric are averaged over three independent runs. The best result for each metric is shown in \textbf{bold} and the second best is {\ul underlined}.}
\label{tab:result-ad}
\begin{tabular}{lccc}
\hline
\multirow{2}{*}{M{\small ODELS}} & \multicolumn{3}{c}{JS {\small DISTANCE} ($\downarrow$)} \\ \cline{2-4} 
                                 & R{\small AM}      & TIC              & TIC-2D           \\ \hline
T{\small IMEWARP}                & {\ul 0.722}       & 0.546            & \textbf{0.719}   \\
ITO                              & 0.740             & 0.696            & 0.833            \\
SD                               & 0.731             & 0.673            & 0.807            \\ \hline
FBM-{\small BASE}                & 0.727             & {\ul 0.533}      & 0.749            \\
FBM                              & \textbf{0.711}    & \textbf{0.525}   & {\ul 0.733}      \\ \hline
\end{tabular}
\end{table}

\subsection{Transferability to Unseen Peptides of PepMD}
We then explore the transferability of models to unseen peptides with various sequence lengths of PepMD. We use all metrics in \S~\ref{sec:experiment-setup} for evaluation and the results on 14 test peptides of PepMD are demonstrated in Table \ref{tab:result-pepmd}, where all samples are generated for a chain length of $10^3$. We find Timewarp achieves a good performance on distributional similarity and flexibility, yet at the cost of only a small portion of valid samples. In contrast, our FBM showcases superiority across all metrics and achieves significant improvement on the validity of generated conformations in particular. It indicate that by introducing the force guidance, the generated ensembles of FBM better comply with the underlying Boltzmann distribution. Additional experimental results can be found in \S~\ref{sec:appendix-additional-experimental-result}.

\begin{table}[H]
\caption{Results on the test set of PepMD. Values of each metric are first averaged over 3 independent runs for each peptide and then shown in mean/std of all 14 test peptides. The best result for each metric is shown in \textbf{bold} and the second best is {\ul underlined}.}
\label{tab:result-pepmd}
\resizebox{\textwidth}{!}{%
\begin{tabular}{lcccccc}
\hline
\multirow{2}{*}{M{\small ODELS}} & \multicolumn{4}{c}{JS {\small DISTANCE} ($\downarrow$)}                                   & \multirow{2}{*}{V{\small AL}-CA ($\uparrow$)} & \multirow{2}{*}{C{\small ONTACT} ($\downarrow$)} \\ \cline{2-5}
                                 & P{\small W}D         & R{\small G}          & TIC                  & TIC-2D               &                                               &                                                  \\ \hline
T{\small IMEWARP}                & {\ul 0.575}/0.082    & 0.561/0.124          & {\ul 0.633}/0.069    & {\ul 0.804}/0.025    & 0.115/0.121                                   & {\ul 0.197}/0.128                                \\
ITO                              & 0.833/0.000          & 0.829/0.012          & 0.789/0.067          & 0.833/0.000          & 0.001/0.000                                   & 0.940/0.081                                      \\
SD                               & 0.823/0.030          & 0.818/0.041          & 0.773/0.032          & 0.832/0.001          & 0.006/0.016                                   & 0.824/0.095                                      \\ \hline
FBM-{\small BASE}                & 0.576/0.066          & {\ul 0.560}/0.153    & 0.639/0.061          & 0.807/0.020          & {\ul 0.367}/0.173                             & 0.208/0.142                                      \\
FBM                              & \textbf{0.573}/0.064 & \textbf{0.542}/0.140 & \textbf{0.631}/0.077 & \textbf{0.801}/0.032 & \textbf{0.616}/0.188                          & \textbf{0.188}/0.127                             \\ \hline
\end{tabular}%
}
\end{table}

\begin{figure}[H]
    \centering
    \includegraphics[width=\linewidth]{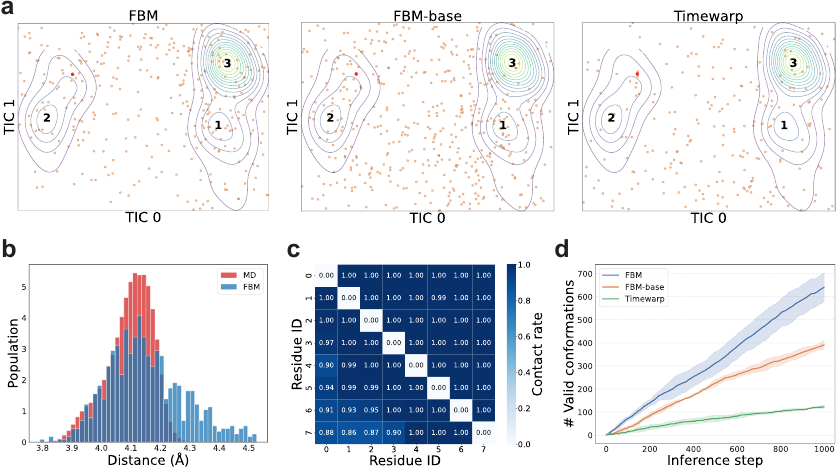}
    \caption{The visualization of comprehensive metrics on peptide 1e28:C. \textbf{a.} Plots of the slowest two TIC components analyzed by feature projections. \textbf{b.} The distribution of the radius-of-gyration. \textbf{c.} The residue contact map, where the data in the lower and upper triangle are obtained from FBM and MD, respectively. \textbf{d.} Cumulated valid conformations during inference over 3 independent runs.}
    \label{fig:1e28-all}
\end{figure}

For better understanding, we provide the visualization of comprehensive metrics on the test peptide 1e28:C (TAFTIPSI) in Figure \ref{fig:1e28-all}. In Figure \ref{fig:1e28-all}(a), we show that the samples generated by FBM exhibit a more pronounced clustering in regions with high reference densities, though all compared methods inevitably generate samples in low-density regions. Figure \ref{fig:1e28-all}(b) demonstrates that FBM accurately captures the peak of the distribution of the radius-of-gyration, with a discrepancy of less than 0.3$\mathring{\mathrm{A}}$ in the right tail of the distribution. FBM and MD also show a close match in terms of the contact rate from Figure \ref{fig:1e28-all}(c). Finally, according to Figure \ref{fig:1e28-all}(d), we emphasize that FBM generates dominantly more valid conformations during the inference step compared to all baselines.


\begin{figure}[H]
  \centering
  \vspace{-1em}
  \includegraphics[width=\linewidth]{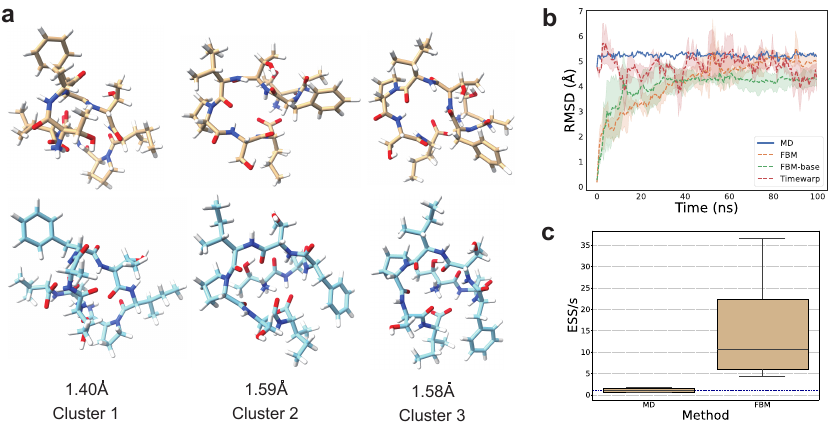}
  \caption{Comparisons between FBM and MD on conformation transitions over time. \textbf{a.} Comparison between the reference equilibrium conformations (blue) and the selected samples of FBM (yellow) of peptide 1e28:C. $C_{\alpha}$-RMSD values are reported below each cluster. \textbf{b.} $C_{\alpha}$-RMSD values along trajectories compared with the initial state of peptide 1e28:C over 3 independent runs. \textbf{c.} The effective sample size per second measured on the test set. All specific values are converted to multiples of the median value of MD, which is shown as the blue dashed line for reference.}
  \label{fig:1e28-cluster}
\end{figure}

In Figure \ref{fig:1e28-cluster}, we conduct further comparisons with MD on conformation transitions over time. We first explore the ability of FBM to recover equilibrium conformations, which is measured by the lowest sample $C_{\alpha}$-RMSD to each cluster center~\citep{wang2024protein}. Reference structures and selected samples of FBM with the lowest $C_{\alpha}$-RMSD for 3 clusters of peptide 1e28:C are provided in Figure \ref{fig:1e28-cluster}(a). The RMSD values of all pairs are below 2$\mathring{\mathrm{A}}$, showing a good recovery of representative conformations. In Figure \ref{fig:1e28-cluster}(b), we provide the $C_{\alpha}$-RMSD values along trajectories compared with the initial state of peptide 1e28:C. Note that, since MD performs local energy minimization on the initial state before simulation, the starting point of its curve is not at the origin. We show that FBM gradually guides the peptide toward equilibrium, reaching a stable RMSD level similar to MD at around 70 ns. In contrast, FBM-{\small BASE} reaches a biased equilibrium at an early stage, while Timewarp exhibits excessive fluctuations over time. In Figure \ref{fig:1e28-cluster}(c), we report the effective-sample-size per second of wall-clock time (ESS/s)~\citep{klein2024timewarp} over the entire test set, where FBM achieves an efficiency improvement of around 10 times relative to MD based on the median values.

\section{Conclusion and Future Work}
\label{sec:conclusion}
In this work, we present a novel generative model called FBM for time-coarsened dynamics in a full-atom fashion. We first leverage the bridge matching framework to construct the baseline model FBM-{\small BASE} for learning dynamics from the data distribution. Based on FBM-{\small BASE}, we further introduce physics priors and interpolate a well-designed intermediate force field, which is theoretically guaranteed to target the Boltzmann-constrained distribution via directly inference without extra steps. Experiments on alanine dipeptide and our curated dataset PepMD showcase superiority of FBM on comprehensive metrics and demonstrate transferability to unseen peptide systems.

As the first attempt to incorporate the intermediate force field to bridge matching for full-atom time-coarsened dynamics, our method has considerable room for improvement. Firstly, our experiments have been conducted on small peptides with fewer than 10 residues. Further exploration on more complex molecular systems (\emph{e.g.}, proteins) is warranted. Secondly, since the training labels for FBM depend on the marginal score calculations provided by FBM-{\small BASE}, we have to adopt a two-stage training process rather than an end-to-end one, which increases the training overhead. Lastly, the transitions between metastable states are still not fast enough, resulting in unreasonable conformations along the generated paths. Therefore, methods for rapid and jump-like state transition are of great importance.

\bibliography{iclr2025_conference}
\bibliographystyle{iclr2025_conference}

\appendix
\section*{Appendix}

\section{Reproducibility}
Our source code and the curated dataset PepMD are all available at \url{https://github.com/yaledeus/FBM}.

\section{Proofs of Propositions}
\label{sec:appendix-proof}

\textit{Proof of Proposition \ref{prop:force-field}.} Given the joint distributions of random variables $\Vec{\rmX}_0$ and $\Vec{\rmX}_1$ following densities $q_0, q_1$ and $p_0, p_1$ satisfy:
\begin{equation}
    p(\Vec{\rmX}_0,\Vec{\rmX}_1)=\frac{1}{Z}q(\Vec{\rmX}_0, \Vec{\rmX}_1)\exp(-k(\varepsilon(\Vec{\rmX}_0)+\varepsilon(\Vec{\rmX}_1))),
\end{equation}
where $Z$ is the partition function to ensure $\iint{p(\Vec{\rmX}_0,\Vec{\rmX}_1)}\dif\Vec{\rmX}_0\dif\Vec{\rmX}_1=1$. It is easy to verify that the condition holds when $\Vec{\rmX}_0$ and $\Vec{\rmX}_1$ are independent variables.

Therefore, the marginal density $p_t$ is given by:
\begin{align}
    p_t(\Vec{\rmX}_t)&=\iint{p_t(\Vec{\rmX}_t|\Vec{\rmX}_0,\Vec{\rmX}_1)p(\Vec{\rmX}_0,\Vec{\rmX}_1)\dif\Vec{\rmX}_0\dif\Vec{\rmX}_1}\\
    &=\iint{q_t(\Vec{\rmX}_t|\Vec{\rmX}_0,\Vec{\rmX}_1)q(\Vec{\rmX}_0,\Vec{\rmX}_1)\frac{\exp(-k(\varepsilon(\Vec{\rmX}_0)+\varepsilon(\Vec{\rmX}_1)))}{Z}\dif\Vec{\rmX}_0\dif\Vec{\rmX}_1}\\
    &=\iint{q_t(\Vec{\rmX}_0,\Vec{\rmX}_1|\Vec{\rmX}_t)q_t(\Vec{\rmX}_t)\frac{\exp(-k(\varepsilon(\Vec{\rmX}_0)+\varepsilon(\Vec{\rmX}_1)))}{Z}\dif\Vec{\rmX}_0\dif\Vec{\rmX}_1}\\
    &=q_t(\Vec{\rmX}_t)\E_{q_t(\cdot,\cdot|\Vec{\rmX}_t)}[\frac{\exp(-k(\varepsilon(\Vec{\rmX}_0)+\varepsilon(\Vec{\rmX}_1)))}{Z}],
\end{align}
where in the first equality we use the assumption that probability paths $p_t$ and $q_t$ share the same conditional distribution given $\Vec{\rmX}_0,\Vec{\rmX}_1$. Considering the assumption that $p_t$ admits the Boltzmann-constrained form as in \cref{eq:Boltzmann-constrained-dis}, we can easily derive the formula for the intermediate potential:

\begin{equation}
    \label{eq:potential}
    \varepsilon_t(\Vec{\rmX}_t)=-\frac{1}{k}\log\E_{q_t(\Vec{\rmX}_0,\Vec{\rmX}_1|\Vec{\mX}_t)}[\exp(-k(\varepsilon(\Vec{\mX}_0)+\varepsilon(\Vec{\mX}_1)))]+\frac{1}{k}\log\frac{Z}{Z_t}.
\end{equation}

Take the gradient of \cref{eq:potential} with regard to $\Vec{\rmX}_t$, the intermediate force field is given by:
\begin{equation}
    \label{eq:force-field-raw}
    \nabla\varepsilon_t(\Vec{\rmX}_t)=-\frac{\iint\exp(-k(\varepsilon(\Vec{\rmX}_0)+\varepsilon(\Vec{\rmX}_1)))\nabla{q_t(\Vec{\rmX}_0,\Vec{\rmX}_1|\Vec{\rmX}_t)}\dif\Vec{\rmX}_0\dif\Vec{\rmX}_1}{k\E_{q_t(\cdot,\cdot|\Vec{\rmX}_t)}[\exp(-k(\varepsilon(\Vec{\rmX}_0)+\varepsilon(\Vec{\rmX}_1)))]},
\end{equation}
where we assume that the integrals and gradients can be commuted. The numerator of \cref{eq:force-field-raw} can be further expanded by:
\begin{align}
    &\iint\exp(-k(\varepsilon(\Vec{\rmX}_0)+\varepsilon(\Vec{\rmX}_1)))\nabla{q_t(\Vec{\rmX}_0,\Vec{\rmX}_1|\Vec{\rmX}_t)}\dif\Vec{\rmX}_0\dif\Vec{\rmX}_1\\
    &=\iint\exp(-k(\varepsilon(\Vec{\rmX}_0)+\varepsilon(\Vec{\rmX}_1)))q_t(\Vec{\rmX}_0,\Vec{\rmX}_1|\Vec{\rmX}_t)\nabla\log{q_t(\Vec{\rmX}_0,\Vec{\rmX}_1|\Vec{\rmX}_t)\dif\Vec{\rmX}_0\dif\Vec{\rmX}_1}\\
    &=\iint\exp(-k(\varepsilon(\Vec{\rmX}_0)+\varepsilon(\Vec{\rmX}_1)))q_t(\Vec{\rmX}_0,\Vec{\rmX}_1|\Vec{\rmX}_t)\nabla\log\frac{q_t(\Vec{\rmX}_t|\Vec{\rmX}_0,\Vec{\rmX}_1)q(\Vec{\rmX}_0,\Vec{\rmX}_1)}{q_t(\Vec{\rmX}_t)}\dif\Vec{\rmX}_0\dif\Vec{\rmX}_1\\
    &=\E_{q_t(\cdot,\cdot|\Vec{\rmX}_t)}[\exp(-k(\varepsilon(\Vec{\rmX}_0)+\varepsilon(\Vec{\rmX}_1)))(\nabla\log{q_t(\Vec{\rmX}_t|\Vec{\rmX}_0,\Vec{\rmX}_1)})-\nabla\log{q_t(\Vec{\rmX}_t)}].
\end{align}

Substitute the numerator back into \cref{eq:force-field-raw} and the conclusion holds.\qed

\textit{Proof of Proposition \ref{prop:vector-field}.} Similar to the definition of $v^*(\Vec{\rmX}_t,t)$ and $u^*(\Vec{\rmX}_t,t)$ under the probability path $q_t$, we define $v'(\Vec{\rmX}_t,t)$ and $u'(\Vec{\rmX}_t,t)$ under the probability path $p_t$ accordingly, which have the following form:
\begin{equation}
    v'(\Vec{\rmX}_t,t)=\E_{p_t(\cdot,\cdot|\Vec{\rmX}_t)}[\frac{\Vec{\rmX}_1-\Vec{\rmX}_t}{1-t}],~u'(\Vec{\rmX}_t,t)=\E_{p_t(\cdot,\cdot|\Vec{\rmX}_t)}[\frac{\Vec{\rmX}_t-\Vec{\rmX}_0}{t}].
\end{equation}

Further, the marginal scores of probability paths $q_t,p_t$ at diffusion time $t$ (termed as $s_t^*$ and $s_t'$ respectively) are connected based on \cref{eq:Boltzmann-constrained-dis} as follows:
\begin{equation}
    s_t'(\Vec{\rmX}_t)=s_t^*(\Vec{\rmX}_t)-k\nabla\varepsilon_t(\Vec{\rmX}_t).
\end{equation}

Considering the linkage between scores and vector fields in \cref{eq:marginal-score}, we obtain:
\begin{align}
    \label{eq:force-guided-score}
    -k\nabla\varepsilon_t(\Vec{\rmX}_t)&=\frac{v'(\Vec{\rmX}_t,t)-u'(\Vec{\rmX}_t,t)}{\sigma^2}-\frac{v^*(\Vec{\rmX}_t,t)-u^*(\Vec{\rmX}_t,t)}{\sigma^2}\\
    &=\frac{1}{\sigma^2}[(v'(\Vec{\rmX}_t,t)-v^*(\Vec{\rmX}_t,t))-(u'(\Vec{\rmX}_t,t)-u^*(\Vec{\rmX}_t,t))].
\end{align}


Now we expand the term $v'-v^*$ in the form of integral:
\begin{equation}
    \label{eq:delta-vector-field}
    v'(\Vec{\rmX}_t,t)-v^*(\Vec{\rmX}_t,t)=\frac{1}{1-t}\iint\Vec{\rmX}_1\left(p_t(\Vec{\rmX}_0,\Vec{\rmX}_1|\Vec{\rmX}_t)-q_t(\Vec{\rmX}_0,\Vec{\rmX}_1|\Vec{\rmX}_t)\right)\dif\Vec{\rmX}_0\dif\Vec{\rmX}_1.
\end{equation}
For convenience, we define:
\begin{align}
f(\Vec{\rmX}_0,\Vec{\rmX}_1,\Vec{\rmX}_t,t)&=p_t(\Vec{\rmX}_0,\Vec{\rmX}_1|\Vec{\rmX}_t)-q_t(\Vec{\rmX}_0,\Vec{\rmX}_1|\Vec{\rmX}_t)\\
&=\frac{p_t(\Vec{\rmX}_t|\Vec{\rmX}_0,\Vec{\rmX}_1)p(\Vec{\rmX}_0,\Vec{\rmX}_1)}{p_t(\Vec{\rmX}_t)}-\frac{q_t(\Vec{\rmX}_t|\Vec{\rmX}_0,\Vec{\rmX}_1)q(\Vec{\rmX}_0,\Vec{\rmX}_1)}{q_t(\Vec{\rmX}_t)}\\
\label{eq:f-def}
&=\frac{q_t(\Vec{\rmX}_t|\Vec{\rmX}_0,\Vec{\rmX}_1)q(\Vec{\rmX}_0,\Vec{\rmX}_1)}{q_t(\Vec{\rmX}_t)}[\frac{Z_t}{Z}\exp(-k(\varepsilon(\Vec{\rmX}_0)+\varepsilon(\Vec{\rmX}_1)-\varepsilon_t(\Vec{\rmX}_t)))-1],
\end{align}

Denote $g(\Vec{\rmX},\Vec{\rmX}_t,t)=\int{f(\Vec{\rmX}_0,\Vec{\rmX},\Vec{\rmX}_t,t)\dif\Vec{\rmX}_0}$, \cref{eq:delta-vector-field} can be rewritten by first integrating over $\Vec{\rmX}_0$:
\begin{equation}
    \label{eq:delta-vector-field-simplified-v}
    v'(\Vec{\rmX}_t,t)-v^*(\Vec{\rmX}_t,t)=\frac{1}{1-t}\int\Vec{\rmX}g(\Vec{\rmX},\Vec{\rmX}_t,t)\dif\Vec{\rmX}.
\end{equation}
Similarly, we can define $h(\Vec{\rmX},\Vec{\rmX}_t,t)=\int{f(\Vec{\rmX},\Vec{\rmX}_1,\Vec{\rmX}_t,t)\dif\Vec{\rmX}_1}$, then $u'-u^*$ admits the form:
\begin{equation}
    \label{eq:delta-vector-field-simplified-u}
    u'(\Vec{\rmX}_t,t)-u^*(\Vec{\rmX}_t,t)=-\frac{1}{t}\int\Vec{\rmX}h(\Vec{\rmX},\Vec{\rmX}_t,t)\dif\Vec{\rmX}.
\end{equation}

To establish the equality between $v'-v^*$ and $u'-u^*$ for solving \cref{eq:force-guided-score}, we further investigate the intrinsic relationship between functions $g$ and $h$.

Since Langevin dynamics~\citep{langevin1908theorie} in a conservative field can be considered to reach a stationary distribution after some time and satisfy the detailed balance~\citep{bussi2007accurate}, which means $\mu(\Vec{\rmX}_0)T(\Vec{\rmX}_1|\Vec{\rmX}_0)=\mu(\Vec{\rmX}_1)T(\Vec{\rmX}_0|\Vec{\rmX}_1)$ for any states $\Vec{\rmX}_0,\Vec{\rmX}_1$, where $\mu(\cdot)$ denotes the equilibrium probability and $T(\cdot|\cdot)$ denotes the Markov transition probability. In particular, we assume that with a sufficiently large number of non-redundant data pairs selected from long MD trajectories, our data distribution inherits the property, namely $q(\Vec{\rmX}_0,\Vec{\rmX}_1)=q(\Vec{\rmX}_1,\Vec{\rmX}_0)$. Therefore, we have:
\begin{align}
    q_t(\Vec{\rmX}_t)&=\iint{q_t(\Vec{\rmX}_t|\Vec{\rmX}_0,\Vec{\rmX}_1)q(\Vec{\rmX}_0,\Vec{\rmX}_1)}\dif\Vec{\rmX}_0\dif\Vec{\rmX}_1\\
    &=\iint{q_{1-t}(\Vec{\rmX}_t|\Vec{\rmX}_1,\Vec{\rmX}_0)q(\Vec{\rmX}_1,\Vec{\rmX}_0)\dif\Vec{\rmX}_1\dif\Vec{\rmX}_0}\\
    &=q_{1-t}(\Vec{\rmX}_t),
\end{align}
where in the second equality we use the property $q_t(\Vec{\rmX}_t|\Vec{\rmX}_0,\Vec{\rmX}_1)=q_{1-t}(\Vec{\rmX}_t|\Vec{\rmX}_1,\Vec{\rmX}_0)$, which can be verified based on \cref{eq:conditional}. Similarly, we can derive $\varepsilon_t=\varepsilon_{1-t}$ from \cref{eq:potential}, thereby implying $p_t=p_{1-t}$ and $Z_t=Z_{1-t}$. Then the following equation holds based on \cref{eq:f-def}:
\begin{equation}
\label{eq:f-sym}
f(\Vec{\rmX}_0,\Vec{\rmX}_1,\Vec{\rmX}_t,t)=f(\Vec{\rmX}_1,\Vec{\rmX}_0,\Vec{\rmX}_t,1-t).
\end{equation}
Taking all the above into consideration, an important observation is:
\begin{equation}
    \label{eq:hg-symmetry}
    h(\Vec{\rmX},\Vec{\rmX}_t,t)=g(\Vec{\rmX},\Vec{\rmX}_t,1-t).
\end{equation}

On the other hand, when $t$ approaches 1, the limit of function $g$ is given by:
\begin{align}
    &\lim_{t\to1^-}g(\Vec{\rmX},\Vec{\rmX}_t,t)\\
    &=\lim_{t\to1^-}\frac{1}{q_t(\Vec{\rmX}_t)}\int\delta(\Vec{\rmX}_t-\Vec{\rmX})q(\Vec{\rmX}_0,\Vec{\rmX})[\frac{Z_t}{Z}\exp(-k(\varepsilon(\Vec{\rmX}_0)+\varepsilon(\Vec{\rmX})-\varepsilon_t(\Vec{\rmX}_t)))-1]\dif\Vec{\rmX}_0\\
    &=\frac{1}{q_1(\Vec{\rmX})}\int{q(\Vec{\rmX}_0,\Vec{\rmX})}[\frac{Z_1}{Z}\exp(-k\varepsilon(\Vec{\rmX}_0))-1]\dif\Vec{\rmX}_0\\
    &=\frac{1}{q_1(\Vec{\rmX})}Z_1\exp(k\varepsilon(\Vec{\rmX}))\int{\frac{1}{Z}}q(\Vec{\rmX}_0,\Vec{\rmX})\exp(-k(\varepsilon(\Vec{\rmX}_0)+\varepsilon(\Vec{\rmX})))\dif\Vec{\rmX}_0-1\\
    &=\frac{1}{q_1(\Vec{\rmX})}Z_1\exp(k\varepsilon(\Vec{\rmX}))\int{p(\Vec{\rmX}_0,\Vec{\rmX})}\dif\Vec{\rmX}_0-1\\
    &=\frac{1}{q_1(\Vec{\rmX})}Z_1\exp(k\varepsilon(\Vec{\rmX}))p_1(\Vec{\rmX})-1=0,
\end{align}
where in the third and fourth equality we use the relationship between the marginal distribution and the joint distribution, and in the fifth equality we apply \cref{eq:Boltzmann-constrained-dis}.

Here we suppose the function $g$ is separable with respect to the time variable $t$. Formally, there exist functions $\iota$ and $\Gamma$ such that the following identity holds:
\begin{equation}
    g(\Vec{\rmX},\Vec{\rmX}_t,t)\equiv\iota(t)\Gamma(\Vec{\rmX},\Vec{\rmX}_t).
\end{equation}
Given $\lim_{t\to1^-}g(\Vec{\rmX},\Vec{\rmX}_t,t)=0$, we prescribe $\iota(t)=1-t$ for convenience, then we have $h(\Vec{\rmX},\Vec{\rmX}_t,t)=t\Gamma(\Vec{\rmX},\Vec{\rmX}_t)$ and subsequently we can derive the closed-form of $v'$ based on \cref{eq:delta-vector-field-simplified-v} and \ref{eq:delta-vector-field-simplified-u} by:
\begin{align}
    u'(\Vec{\rmX}_t,t)-u^*(\Vec{\rmX}_t,t)&=-\frac{1}{t}\int\Vec{\rmX}h(\Vec{\rmX},\Vec{\rmX}_t,t)\dif\Vec{\rmX}\\
    &=-\int\Vec{\rmX}\Gamma(\Vec{\rmX},\Vec{\rmX}_t)\dif\Vec{\rmX}\\
    &=-\frac{1}{1-t}\int\Vec{\rmX}(1-t)\Gamma(\Vec{\rmX},\Vec{\rmX}_t)\dif\Vec{\rmX}\\
    &=-(v'(\Vec{\rmX}_t,t)-v^*(\Vec{\rmX}_t,t)),
\end{align}
which implies $v'(\Vec{\rmX}_t,t)=v^*(\Vec{\rmX}_t,t)-\frac{\sigma^2}{2}k\nabla\varepsilon_t(\Vec{\rmX}_t)$ according to \cref{eq:force-guided-score}.\qed

\textit{Proof of Proposition \ref{prop:score}.} From the definition of $s_t^*$ in \cref{eq:marginal-score}, the following equation holds:
\begin{align}
    s_t^*(\Vec{\rmX}_t)&=\E_{q_t(\cdot,\cdot|\Vec{\rmX}_t)}[\nabla\log{q_t(\Vec{\rmX}_t|\Vec{\rmX}_0,\Vec{\rmX}_1)}]\\
    &=\iint\nabla\log{q_t(\Vec{\rmX}_t|\Vec{\rmX}_0,\Vec{\rmX}_1)q_t(\Vec{\rmX}_0,\Vec{\rmX}_1|\Vec{\rmX}_t)\dif\Vec{\rmX}_0\dif\Vec{\rmX}_1}\\
    &=\frac{1}{q_t(\Vec{\rmX}_t)}\iint\nabla{q_t(\Vec{\rmX}_t|\Vec{\rmX}_0,\Vec{\rmX}_1)}q(\Vec{\rmX}_0,\Vec{\rmX}_1)\dif\Vec{\rmX}_0\dif\Vec{\rmX}_1\\
    \label{eq:int-nabla-commute}
    &=\frac{1}{q_t(\Vec{\rmX}_t)}\nabla\iint{q_t(\Vec{\rmX}_t|\Vec{\rmX}_0,\Vec{\rmX}_1)}q(\Vec{\rmX}_0,\Vec{\rmX}_1)\dif\Vec{\rmX}_0\dif\Vec{\rmX}_1\\
    &=\frac{\nabla{q_t(\Vec{\rmX}_t)}}{q_t(\Vec{\rmX}_t)}=\nabla\log{q_t(\Vec{\rmX}_t)},
\end{align}
where in the second equality we use the Bayesian rule of probability, $q(\Vec{\rmX}_0,\Vec{\rmX}_1)$ denotes the joint distribution of random variables $\Vec{\rmX}_0,\Vec{\rmX}_1$ following $q_0,q_1$. Furthermore, the third equality is justified by assuming the integrands satisfy the regularity conditions of the Leibniz Rule.\qed

\textit{Proof of Proposition \ref{prop:boundary}.} We first check whether the continuity condition holds when $t\to0^+$. Note that under this condition, the density $q_t(\Vec{\rmX}_0,\Vec{\rmX}_1|\Vec{\rmX}_t)$ involves into the Dirac mass $\delta(\Vec{\rmX}_t-\Vec{\rmX}_0)$ at point $\Vec{\rmX}_t$, subsequently we have:
\begin{align}
    \lim_{t\to0^+}\nabla\varepsilon_t(\Vec{\rmX}_t)&=-\lim_{t\to0^+}\frac{\iint\exp(-k(\varepsilon(\Vec{\rmX}_0)+\varepsilon(\Vec{\rmX}_1)))\delta(\Vec{\rmX}_t-\Vec{\rmX}_0)\dif\Vec{\rmX}_0\dif\Vec{\rmX}_1}{k\iint\exp(-k(\varepsilon(\Vec{\rmX}_0)+\varepsilon(\Vec{\rmX}_1)))\delta(\Vec{\rmX}_t-\Vec{\rmX}_0)\dif\Vec{\rmX}_0\dif\Vec{\rmX}_1}\\
    &=-\lim_{t\to0^+}\frac{\nabla\iint\exp(-k(\varepsilon(\Vec{\rmX}_0)+\varepsilon(\Vec{\rmX}_1)))\delta(\Vec{\rmX}_t-\Vec{\rmX}_0)\dif\Vec{\rmX}_0\dif\Vec{\rmX}_1}{k\int\exp(-k(\varepsilon(\Vec{\rmX}_t)+\varepsilon(\Vec{\rmX}_1)))\dif\Vec{\rmX}_1}\\
    &=-\lim_{t\to0^+}\frac{\nabla\int\exp(-k(\varepsilon(\Vec{\rmX}_t)+\varepsilon(\Vec{\rmX}_1)))\dif\Vec{\rmX}_1}{k\int\exp(-k(\varepsilon(\Vec{\rmX}_t)+\varepsilon(\Vec{\rmX}_1)))\dif\Vec{\rmX}_1}\\
    &=-\lim_{t\to0^+}\frac{\nabla\exp(-k\varepsilon(\Vec{\rmX}_t))}{k\exp(-k\varepsilon(\Vec{\rmX}_t))}=\nabla\varepsilon(\Vec{\rmX}_0).
\end{align}
The case when $t$ approaches 1 is completely symmetrical and will not be elaborated further. Thus we have proven that the intermediate force field converges to the MD force field when $t$ approaches 0 and 1.

\section{Model Architecture}
\label{sec:arch}
In this work, we leverage the powerful \texttt{TorchMD-NET}~\citep{pelaez2024torchmdnet} as the backbone model to process molecular graphs, which intrinsically satisfies $\mathrm{SO}(3)$-equivariance with the \textit{equivariant transformer}~\citep{tholke2021equivariant} component. To adapt to our task setup, the inputs include not only the Cartesian coordinates $\Vec{\mX}$ and atom embeddings $\mZ$, but also a one-dimensional continuous diffusion time $t$. \texttt{TorchMD-NET} will then output $\mathrm{SO}(3)$-equivariant vectors $\Vec{\mV}\in\sR^{N\times{3}\times{H}}$ and node representations 
$\mH\in\sR^{N\times{H}}$. Formally, we have:
\begin{equation}
    \label{eq:torchmd-net}
    \Vec{\mV}, \mH=\texttt{TorchMD-NET}(\Vec{\mX}, \mZ, t).
\end{equation}
To streamline the model, we only add lightweight output heads to a single \texttt{TorchMD-NET} module for the baseline model as well as FBM. For the baseline model, we use two separate two-layer Feed-Forward Networks (FFN) with no shared weights to transform the node representations into weights of vectors, which are then multiplied by 
$\Vec{\mV}$ to obtain the final representation:
\begin{align}
    \label{eq:model-head}v_{\theta}(\Vec{\mX},t),u_{\theta}(\Vec{\mX},t)\coloneqq\Vec{\mV}\times\mathrm{FFN}(\mH)\in\sR^{N\times{3}},
\end{align}
where the dimensions of hidden and output layers of $\mathrm{FFN}$ are all $H$ and we use $\mathrm{SiLU}$~\citep{dong2017learning} for activation layers. Further, we construct the networks $\alpha_{\theta},\beta_{\theta},\gamma_{\theta}$ of FBM in the same way, while the only difference is that we add one $\mathrm{LayerNorm}$~\citep{ba2016layer} before the $\mathrm{FFN}$ layer due to the variance in scale of different targets.

\section{Training and Inference Details}
\label{sec:appendix-train-infer}
In this section, we provide additional details and pseudo codes for training and inference of the baseline bridge matching model FBM-{\small BASE} and the force-guided bridge matching model FBM.

\subsection{Normalization of Energies and Forces}
In practice, we found that the unnormalized potential function is numerically unstable and its variance is positively correlated with the number of atoms $N$. For stable training, we need to perform certain pre-processing steps. Specifically, for potentials in \texttt{kJ/mol}, we divide by $3N$ to obtain the average potential energy per degree of freedom for the entire molecule, where $N$ varies with different peptides. For force fields in \texttt{kJ/(mol$\cdot$nm)}, due to their relatively stable values across different molecular systems, we empirically multiply by the constant 0.002 for normalization.

\subsection{Guidance Strength $\eta$}
Similar to \citet{wang2024protein}, we introduce the \textit{guidance strength} $\eta$ for better approximation of the Boltzmann distribution. Formally, for any positive constant $\eta>0$, we can define a new probability path based on $q_t$:
\begin{equation}
    p_t(\Vec{\rmX}_t)=\frac{1}{Z_t}q_t(\Vec{\rmX}_t)\exp(-\frac{2\eta}{\sigma^2}k\varepsilon_t(\Vec{\rmX}_t)), \varepsilon_0=\varepsilon_1=\varepsilon.
\end{equation}
The only difference with \cref{eq:Boltzmann-constrained-dis} is the constant $\nicefrac{2\eta}{\sigma^2}$ in the exponential term, which can be interpreted as how well the probability path $p_t$ is guided by energies and forces. According to Proposition~\ref{prop:vector-field}, it can be easily deduced that the vector field $v'(\Vec{\rmX}_t,t)$ which generates $p_t$ has the following form:
\begin{equation}
    v'(\Vec{\rmX}_t,t)=v^*(\Vec{\rmX}_t,t)-\eta\cdot{k\varepsilon_t(\Vec{\rmX}_t)}.
\end{equation}
Thus practically, we regard $\eta$ as a hyperparameter during inference and enhance the similarity between $p_t$ and the Boltzmann distribution by selecting the proper guidance strength $\eta$.

\subsection{Refinement with Constrained Energy Minimization}
\label{sec:appedix-refinement}
We utilize the discrete form of the SDE process in \cref{eq:markov-diffusion} for inference with $T$ SDE steps, and the full conformation ensembles are generated in an autoregressive way, where the output from the previous step serves as the input for the next step. However, in the autoregressive fashion, errors at each inference step will accumulate, leading to out-of-distribution problem. Here we introduce an additional energy minimization procedure using \texttt{OpenMM}~\citep{eastman2017openmm} for refinement, which is performed for each generated conformation before sent to the next inference step. Note that we aim for the refinement to affect only the minor details (\emph{e.g.}, X-H bonds) without altering the overall conformation; therefore, independent harmonic constraints are further applied on all heavy atoms with spring constant of 10 kcal/mol$\cdot\mathring{\mathrm{A}}^2$ and the tolerance of 2.39 kcal/mol$\cdot\mathring{\mathrm{A}}^2$ without maximal step limits~\citep{wang2024protein}.

\subsection{Algorithms for Training and Inference} We provide pseudo codes for training and inference with our models FBM-{\small BASE} and FBM in Algorithm \ref{alg:base-train},\ref{alg:fbm-train},\ref{alg:infer} respectively.

\begin{algorithm}[ht]
   \caption{Training with FBM-{\small BASE}}
   \label{alg:base-train}
\begin{algorithmic}[1]
   \STATE {\bfseries Input:} peptide pairs $(\gG_0,\gG_1)$ in a batch $B$, vector field networks $u(\Vec{\rmX}_t,t),v(\Vec{\rmX}_t,t)$
   \FOR{training iterations}
   \STATE $t\sim\mathrm{Uni}(0,1)$
   \STATE $\Vec{\mX}_t\sim{q_t(\Vec{\rmX}_t|\Vec{\mX}_0,\Vec{\mX}_1)}$
   \STATE $\hat{\Vec{\mX}}_0 \gets \Vec{\mX}_t-tu_{\theta}(\Vec{\mX}_t, t),\hat{\Vec{\mX}}_1 \gets \Vec{\mX}_t+(1-t)v_{\theta}(\Vec{\mX}_t, t)$
   \STATE $(\mD_0,\mD_1,\hat{\mD}_0,\hat{\mD}_1) \gets$ pairwise interatomic distances of $(\Vec{\mX}_0,\Vec{\mX}_1,\hat{\Vec{\mX}}_0,\hat{\Vec{\mX}}_1)$
   \STATE $\gL_{\mathrm{fwd}} \gets \frac{1}{B}\sum_{\gG_0,\gG_1}||\nicefrac{(\Vec{\mX}_1-\Vec{\mX}_t)}{(1-t)}-v_{\theta}(\Vec{\mX}_t,t)||^2$
   \STATE $\gL_{\mathrm{rev}} \gets \frac{1}{B}\sum_{\gG_0,\gG_1}||\nicefrac{(\Vec{\mX}_t-\Vec{\mX}_0)}{t}-u_{\theta}(\Vec{\mX}_t,t)||^2$
   \STATE $\gL_{\mathrm{aux}} \gets \frac{1}{B}\sum_{\gG_0,\gG_1}(1-t)\cdot\frac{||\bm{1}_{\mD_0<6\mathring{\mathrm{A}}}(\mD_0-\hat{\mD}_0)||^2}{\sum{\bm{1}_{\mD_0<6\mathring{\mathrm{A}}}}-N}+t\cdot\frac{||\bm{1}_{\mD_1<6\mathring{\mathrm{A}}}(\mD_1-\hat{\mD}_1)||^2}{\sum{\bm{1}_{\mD_1<6\mathring{\mathrm{A}}}}-N}$
   \STATE $\gL_{\mathrm{base}} \gets \gL_{\mathrm{fwd}}+\gL_{\mathrm{rev}}+0.25\cdot\gL_{\mathrm{aux}}$
   \STATE $\min\gL_{\mathrm{base}}$
   \ENDFOR
\end{algorithmic}
\end{algorithm}

\begin{algorithm}[ht]
   \caption{Training with FBM}
   \label{alg:fbm-train}
\begin{algorithmic}[1]
   \STATE {\bfseries Input:} peptide pairs $(\gG_0,\gG_1)$ of \textbf{one molecular system} in a batch $B$, baseline model $v_{\theta}(\Vec{\rmX}_t,t),u_{\theta}(\Vec{\rmX}_t,t)$ in \S~\ref{sec:bbm} with frozen parameters, MD potentials $\varepsilon(\Vec{\mX}_0),\varepsilon(\Vec{\mX}_1)$, MD force fields $\nabla\varepsilon(\Vec{\mX}_0),\nabla\varepsilon(\Vec{\mX}_1)$, force field networks $\alpha_{\theta}(\Vec{\rmX}_t,t),\beta_{\theta}(\Vec{\rmX}_t,t),\gamma_{\theta}(\Vec{\rmX}_t,t)$
   \FOR{training iterations}
   \STATE $t\sim\mathrm{Uni}(0,1)$
   \STATE $\Vec{\mX}_t\sim{q_t(\Vec{\rmX}_t|\Vec{\mX}_0,\Vec{\mX}_1)}$
   \STATE $s_t^*(\Vec{\mX}_t) \gets \nicefrac{(v_{\theta}(\Vec{\mX}_t,t)-u_{\theta}(\Vec{\mX}_t,t))}{\sigma^2}$
   \STATE $\zeta(\Vec{\mX}_0,\Vec{\mX}_1,\Vec{\mX}_t) \gets s_t^*(\Vec{\mX}_t)-\nabla\log{q_t(\Vec{\mX}_t|\Vec{\mX}_0,\Vec{\mX}_1)}$
   \STATE $M \gets \frac{1}{B}\sum_{\gG_0,\gG_1}q_t(\Vec{\mX}_t|\Vec{\mX}_0,\Vec{\mX}_1)\exp(-k(\varepsilon(\Vec{\mX}_0)+\varepsilon(\Vec{\mX}_1)))$
   \STATE $w(\Vec{\mX}_t,t) \gets (1-t)\cdot\texttt{detach}(\alpha_{\theta}(\Vec{\mX}_t,t))+t\cdot\texttt{detach}(\beta_{\theta}(\Vec{\mX}_t,t))+t(1-t)\cdot\gamma_{\theta}(\Vec{\mX}_t,t)$
   \STATE $\gL_{\mathrm{iff}} \gets \frac{1}{B}\sum_{\gG_0,\gG_1}||\nicefrac{\exp(-k(\varepsilon(\Vec{\mX}_0)+\varepsilon(\Vec{\mX}_1)))\zeta(\Vec{\mX}_0,\Vec{\mX}_1,\Vec{\mX}_t)}{kM}-w_{\theta}(\Vec{\mX}_t,t)||^2$
   \STATE $\gL_{\mathrm{bnd}} \gets \frac{1}{B}\sum_{\gG_0,\gG_1}||\nabla\varepsilon(\Vec{\mX}_0)-\alpha_{\theta}(\Vec{\mX}_t,t)||^2+||\nabla\varepsilon(\Vec{\mX}_1)-\beta_{\theta}(\Vec{\mX}_t,t)||^2$
   \STATE $\gL_{\mathrm{FBM}} \gets \gL_{\mathrm{iff}}+\gL_{\mathrm{bnd}}$
   \STATE $\min\gL_{\mathrm{FBM}}$
   \ENDFOR
\end{algorithmic}
\end{algorithm}

\begin{algorithm}[ht]
   \caption{Autoregressive inference with FBM/FBM-{\small BASE}}
   \label{alg:infer}
\begin{algorithmic}[1]
   \STATE {\bfseries Input:} Initial state $\gG_0$, chain length $L$, discrete SDE step $T$, guidance strength $\eta$, baseline model $v(\Vec{\rmX}_t,t)$ in \S~\ref{sec:bbm}, FBM model $w(\Vec{\rmX}_t,t)$ in \S~\ref{sec:fbm}, model type $c\in$\{FBM-{\small BASE}, FBM\}
   \STATE $C \gets$ []
   \STATE $\Delta \gets \nicefrac{1}{T}$
   \FOR{$l \gets 1$ {\bfseries to} $L$}
       \FOR{$t$ {\bfseries in} $\mathrm{linspace}(0,1-\Delta,T)$}
       \STATE $\epsilon\sim\gN(\bm{0},\bm{I})$
       \IF{$c=\mathrm{FBM}$}
       \STATE $v'(\Vec{\mX}_t,t) \gets v_{\theta}(\Vec{\mX}_t,t)-\eta\cdot{kw_{\theta}(\Vec{\mX}_t,t)}$
       \ELSE
       \STATE $v'(\Vec{\mX}_t,t) \gets v_{\theta}(\Vec{\mX}_t,t)$
       \ENDIF
       \STATE $\Vec{\mX}_{t+\Delta} \gets \Vec{\mX}_t+v'(\Vec{\mX}_t,t)\Delta+\sqrt{t}\sigma\epsilon$
       \ENDFOR
   \STATE $\Vec{\mX}_1' \gets \mathrm{energy\_minim}(\Vec{\mX}_1)$
   \STATE $\Vec{\mX}_0 \gets \Vec{\mX}_1'$
   \STATE $C \gets C \cup \Vec{\mX}_1$  
   \ENDFOR
   \STATE {\bfseries Output} $C$
\end{algorithmic}
\end{algorithm}

\subsection{Hyperparameters} The hyperparameters we choose are listed in Table \ref{tab:hyper}.

\begin{table}[ht]
\centering
\caption{Hyperparameter choice of FBM-{\small BASE} and FBM.}
\label{tab:hyper}
\begin{tabular}{ll}
\hline
Hyperparameters                           & Values                                                   \\ \hline
\multicolumn{2}{c}{\textbf{Network}}                                                                 \\ \hline
Hidden dimension $H$ of FBM-{\small BASE} & 128                                                      \\
Hidden dimension $H$ of FBM               & 176                                                      \\
RBF dimension                             & 32                                                       \\
Number of attention heads                 & 8                                                        \\
Number of layers                          & 6                                                        \\
Cutoff threshold $r_{\mathrm{cut}}$       & 5.0$\mathring{\mathrm{A}}$                               \\ \hline
\multicolumn{2}{c}{\textbf{Training}}                                                                \\ \hline
Learning rate                             & 5e-4                                                     \\
Optimizer                                 & Adam                                                     \\
Warm up steps                             & 1,000                                                    \\
Warm up scheduler                         & LamdaLR                                                  \\
Training scheduler                        & ReduceLRonPlateau(factor=0.8, patience=5, min\_lr=1e-7) \\
Batch size of FBM-{\small BASE}           & 16                                                       \\
Batch size of FBM                         & 10                                                       \\
SDE noise scale $\sigma$                  & 0.2                                                      \\ \hline
\multicolumn{2}{c}{\textbf{Inference}}                                                               \\ \hline
SDE steps $T$                             & [25,30]                                                  \\
Guidance strength $\eta$ of FBM           & [0.04,0.05,0.06,0.07,0.08]                               \\ \hline
\end{tabular}
\end{table}

\section{Experimental Details}
\label{sec:appendix-experimental-detail}

\subsection{Dataset Details}
\label{sec:appendix-dataset-detail}
As mentioned in \S~\ref{sec:experiment-setup}, all peptides of PepMD are simulated using \texttt{OpenMM}~\citep{eastman2017openmm}. The parameters we used for MD simulations are listed in Table \ref{tab:simulation-setup} and the statistical information of PepMD is shown in Table \ref{tab:dataset-stats}.

Additionally, all 14 peptides of our test set are listed below with the format \{pdb-id\}:\{chain-id\}, including 1hhg:C, 1k8d:P, 1k83:M, 1bz9:C, 1i7u:C, 1gxc:B, 1ar8:0, 2xa7:P, 1e28:C, 1gy3:F, 1n73:I, 1fpr:B, 1aze:B, 1qj6:I.

\begin{table}[ht]
\centering
\caption{MD simulation setups using $\texttt{OpenMM}$.}
\label{tab:simulation-setup}
\begin{tabular}{ll}
\hline
Property              & Value                    \\ \hline
Forcefield            & AMBER-14                 \\
Integrator            & LangevinMiddleIntegrator \\
Integration time step & 1fs                      \\
Frame spacing         & 1ps                      \\
Friction coefficient  & 1.0$\text{ps}^{-1}$      \\
Temperature           & 300K                     \\
Electrostatics        & NoCutoff                 \\
Constraints           & HBonds                   \\ \hline
\end{tabular}
\end{table}

\begin{table}[ht]
\centering
\caption{Dataset statistics.}
\label{tab:dataset-stats}
\begin{tabular}{ll}
\hline
Dataset name                         & PepMD               \\ \hline
Training set simulation time         & 100ns               \\
Test set simulation time             & 100ns               \\
MD integration time step $\Delta{t}$ & 1fs                 \\
coarsened predition time $\tau$      & $0.5\times{10^6}$fs \\
\# Clusters                          & 2480                \\
\# Training peptides                 & 136                 \\
\# Training pairs per peptide        & $2\times{10^3}$     \\
\# Validation pairs per peptide      & $4\times{10^2}$     \\
\# Test peptides                     & 14                  \\ \hline
\end{tabular}
\end{table}

\subsection{Details on Evaluation Metrics}
\label{sec:appendix-metric-detail}
In this part, we provide details for computing the evaluation metrics in \S~\ref{sec:experiment-setup}.

\paragraph{Flexibility}
Following \citet{janson2023direct}, we compute the contact rates between residues as a measure of structural flexibility. For each residue pair $i,j$ ($1\leq{i}<j\leq{R}$) of a peptide with $R$ residues, the contact rate $r(i,j)$ of residue $i,j$ is defined as follows:
\begin{equation}
    r(i,j)=\frac{1}{L}\sum_{l=1}^L\bm{1}_{d_l(i,j)<10\mathring{\mathrm{A}}},
\end{equation}
where $d_l(i,j)$ denotes the Euclidean distance between $\alpha$-carbons of residue $i,j$ of conformation $l$. Now we compute the root mean square error of contact maps between generated conformation ensembles and reference MD trajectories:
\begin{equation}
    \text{C{\small ONTACT}}=\sqrt{\frac{2}{R(R-1)}\sum_{1\leq{i}<j\leq{R}}(r(i,j)-r_{\mathrm{ref}}(i,j))^2}.
\end{equation}

\paragraph{Validity}
We assess the structural validity by checking for bond breaks between adjacent residues and bond clashes between any residue pairs. The same as in \citet{wang2024protein}, \textit{bond clash} occurs when the distance between $\alpha$-carbons of any residue pair is less than the threshold $\delta_{\mathrm{clash}}=3.0\mathring{\mathrm{A}}$, and \textit{bond break} occurs when the distance between adjacent $\alpha$-carbons is greater than the threshold $\delta_{\mathrm{break}}=4.19\mathring{\mathrm{A}}$. Then the metric V{\small AL}-CA is assessed by the fraction of conformations without bond break and bond clash.

\paragraph{Distributional Similarity}
Similar to \citet{lu2023str2str}, we project peptide conformations onto the following three low-dimensional feature space: (i) Pairwise Distance (P{\small W}D) between $\alpha$-carbons excluding residue pairs within an offset of 3. (ii) Radius of gyration (R{\small G}) which computes the geometric mean of the distances from $\alpha$-carbons to the center-of-mass. (iii) Time-lagged Independent Components (TIC), where we featurize structures using backbone dihedrals $\psi,\phi,\omega$ and pairwise distances between $\alpha$-carbons~\citep{klein2024timewarp}, then TIC analysis is performed using \texttt{Deeptime}~\citep{hoffmann2021deeptime}. Only the slowest components, TIC 0 and TIC 1, are taken for further evaluation~\citep{perez2013identification}. (iv) the joint distribution of TIC 0 and TIC 1, termed as TIC-2D. (v) Specifically for the evaluation on AD, the joint distribution of backbone dihedrals $\psi$ and $\phi$, namely the \textit{Ramachandran plot}~\citep{1963stereochemistry}, is taken into consideration (R{\small AM}).

Afterwards we compute the Jensen-Shannon (JS) distance between generated samples and reference MD trajectories on the projection space. Features are discretized with 50 bins based on the reference ensembles, and a pseudo count 1e-6 is added for numerical stability. For each feature space, we report the mean distance along all dimensions.

\section{Additional Experimental Results}
\label{sec:appendix-additional-experimental-result}

\subsection{Ablation Study}
In Table \ref{tab:appendix-ablation-result} we provide ablation results of SDE steps $T$ and the guidance strength $\eta$ on test peptides of PepMD. A clear pattern is that when $T$ is fixed, the greater the guidance strength $\eta$, the more likely it is to generate reasonable conformations, which demonstrates a strong correlation between the intermediate force field and real interatomic constraints of molecular systems.

\begin{table}[H]
\caption{Ablation results of SDE steps $T$ and the guidance strength $\eta$ on the test set of PepMD. Values of each metric are first averaged over 3 independent runs for each peptide and then shown in mean/std of all 14 test peptides.}
\label{tab:appendix-ablation-result}
\resizebox{\textwidth}{!}{%
\begin{tabular}{lcccccc}
\hline
\multirow{2}{*}{Hyperparameters} & \multicolumn{4}{c}{JS {\small DISTANCE} ($\downarrow$)} & \multirow{2}{*}{V{\small AL}-CA ($\uparrow$)} & \multirow{2}{*}{C{\small ONTACT} ($\downarrow$)} \\ \cline{2-5}
                                 & P{\small W}D  & R{\small G} & TIC         & TIC-2D      &                                               &                                                  \\ \hline
$T=25,\eta=0.04$                 & 0.586/0.059   & 0.540/0.143 & 0.640/0.056 & 0.809/0.020 & 0.559/0.190                                   & 0.195/0.115                                      \\
$T=25,\eta=0.05$                 & 0.578/0.069   & 0.550/0.151 & 0.639/0.075 & 0.805/0.027 & 0.606/0.192                                   & 0.179/0.132                                      \\
$T=25,\eta=0.06$                 & 0.573/0.064   & 0.542/0.140 & 0.631/0.077 & 0.801/0.032 & 0.616/0.188                                   & 0.188/0.127                                      \\
$T=25,\eta=0.07$                 & 0.585/0.084   & 0.548/0.177 & 0.644/0.089 & 0.800/0.044 & 0.647/0.182                                   & 0.201/0.135                                      \\
$T=25,\eta=0.08$                 & 0.573/0.064   & 0.577/0.144 & 0.638/0.079 & 0.804/0.033 & 0.675/0.169                                   & 0.200/0.111                                      \\ \hline
$T=30,\eta=0.04$                 & 0.585/0.074   & 0.587/0.151 & 0.644/0.077 & 0.804/0.029 & 0.553/0.208                                   & 0.208/0.131                                      \\
$T=30,\eta=0.05$                 & 0.582/0.066   & 0.542/0.162 & 0.640/0.088 & 0.806/0.035 & 0.615/0.183                                   & 0.217/0.130                                      \\
$T=30,\eta=0.06$                 & 0.596/0.085   & 0.591/0.139 & 0.637/0.089 & 0.803/0.041 & 0.604/0.205                                   & 0.231/0.116                                      \\
$T=30,\eta=0.07$                 & 0.576/0.059   & 0.590/0.120 & 0.649/0.074 & 0.806/0.023 & 0.655/0.180                                   & 0.182/0.116                                      \\
$T=30,\eta=0.08$                 & 0.590/0.077   & 0.575/0.151 & 0.614/0.101 & 0.789/0.065 & 0.661/0.176                                   & 0.208/0.116                                      \\ \hline
\end{tabular}%
}
\end{table}

\subsection{PepMD Additional Results}

\begin{figure}[h]
    \centering
    \includegraphics[width=\textwidth]{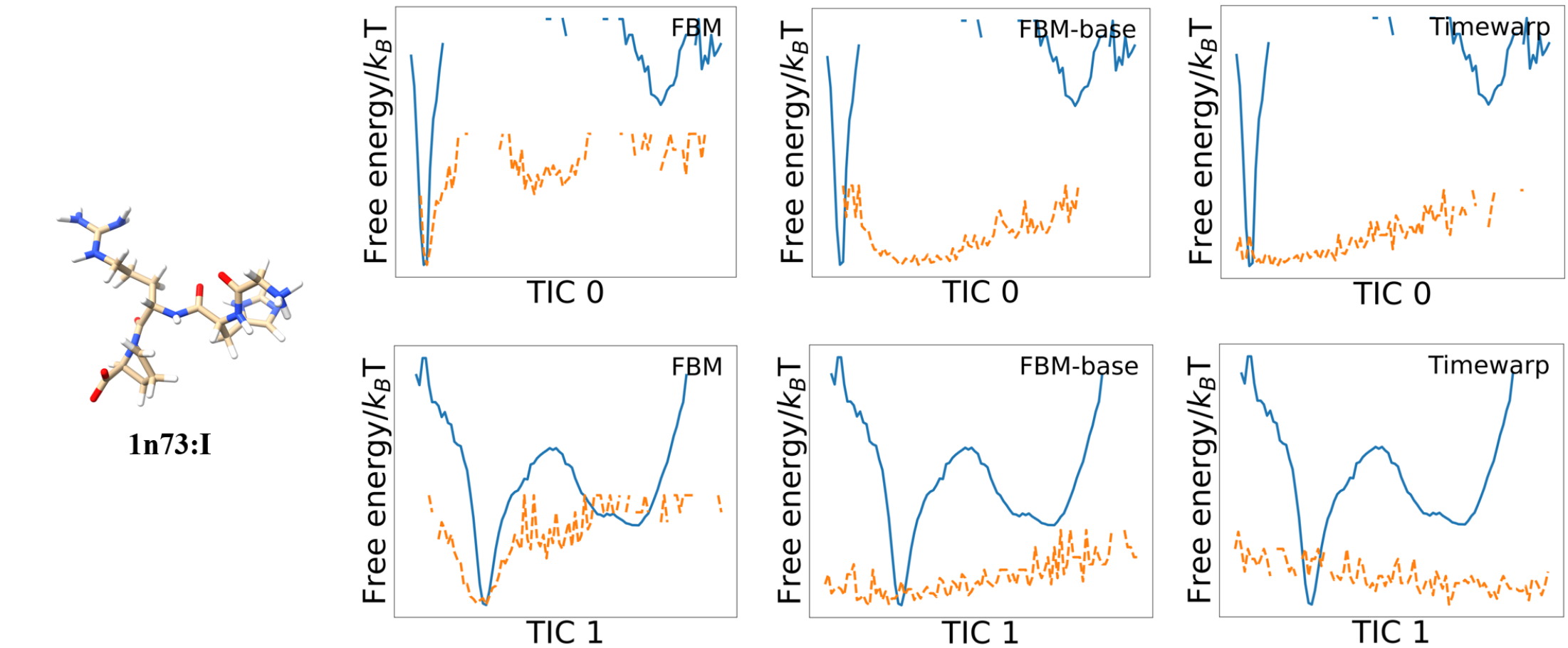}
    \includegraphics[width=\textwidth]{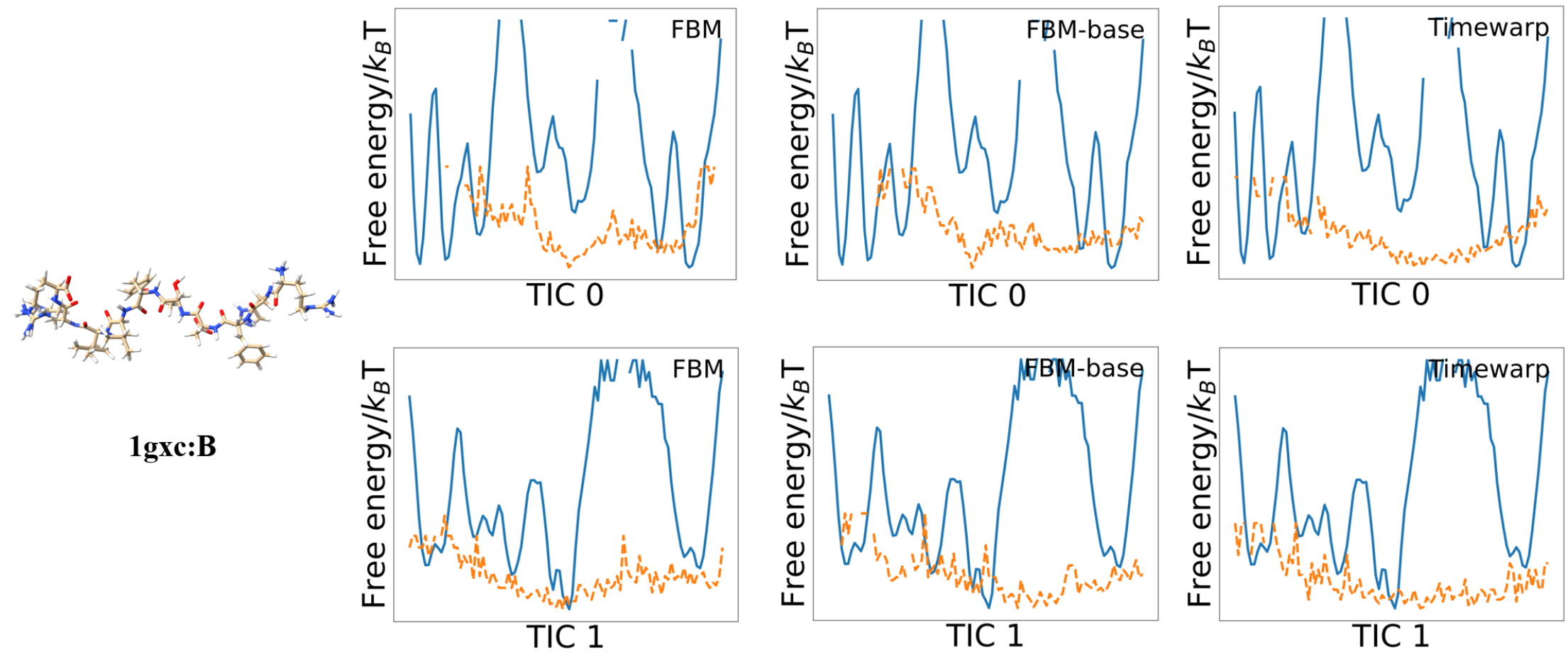}
    \includegraphics[width=\textwidth]{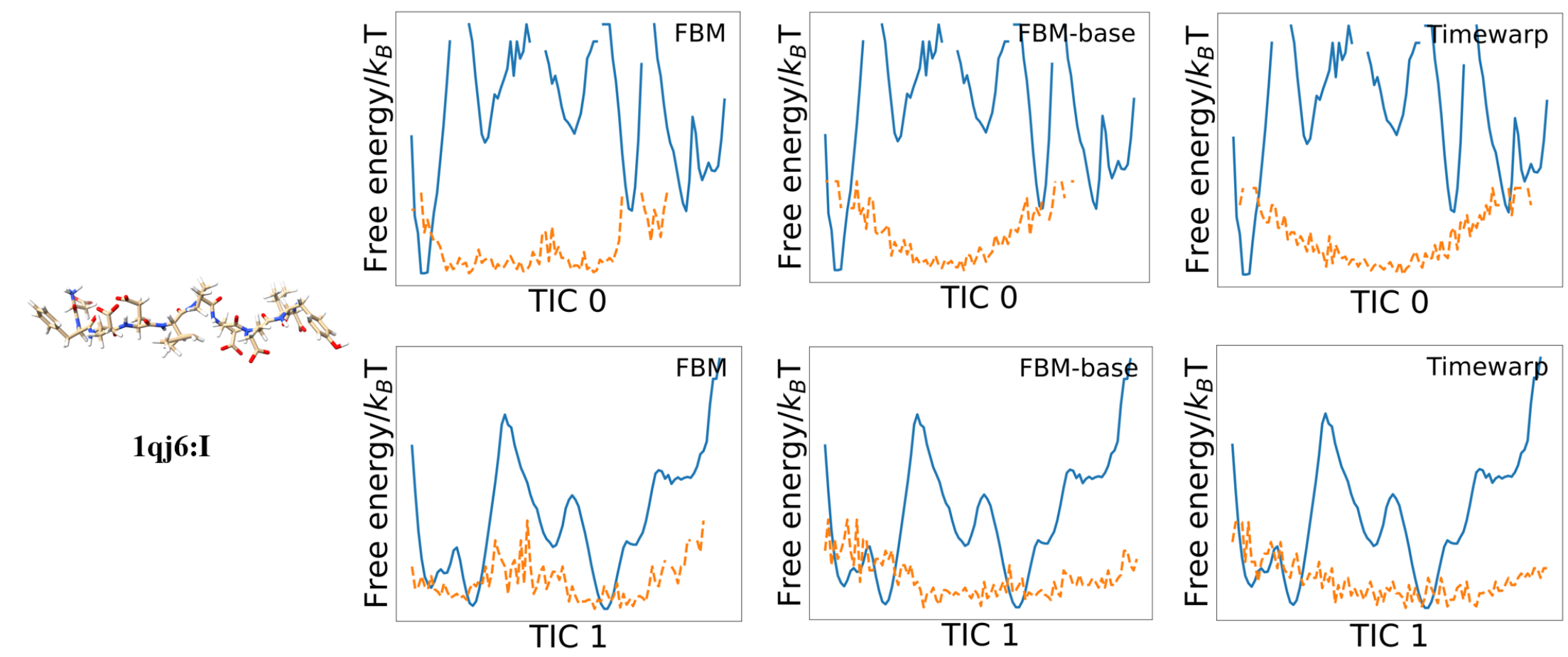}
    \caption{Experiments on PepMD test peptides 1n73:I, 1gxc:B and 1qj6:I (top, middle and bottom). Samples were generated in the time-coarsened manner for a chain length of $10^3$. Free energies (\emph{i.e.}, the relative log probability) along the first two TIC components of FBM, FBM-{\small BASE} and Timewarp are displayed in the left, middle, and right columns, respectively. The blue solid line represents the full MD trajectories, while the yellow dashed line represents model-generated samples.}
    \label{fig:appendix-pepmd-tic}
\end{figure}

In Figure~\ref{fig:appendix-pepmd-tic}, we provide the comparison of different methods for free energy projections on the slowest two TIC components for test peptides 1n73:I (GHRP), 1gxc:B (RHFDTYLIRR) and 1qj6:I (DFEEIPEEYL) from PepMD. Note that, due to the discrepancies in the prediction time interval and the number of samples, there may be certain systematic errors compared with the MD simulation data. Compared to the baselines, FBM presents more accurate depictions of the molecular free energy landscape in most cases. We also found that for peptides with fewer residues (\emph{e.g.}, the tetrapeptide 1n73:I), FBM often achieves higher accuracy. This condition aligns with the scaling law, suggesting that accurate molecular dynamics simulations for more complex molecules may require larger training datasets and more parameters.

\begin{figure}[htbp]
    \centering

    \begin{subfigure}[b]{0.9\textwidth}
        \centering
        \begin{minipage}{0.3\textwidth}
            \includegraphics[width=\textwidth]{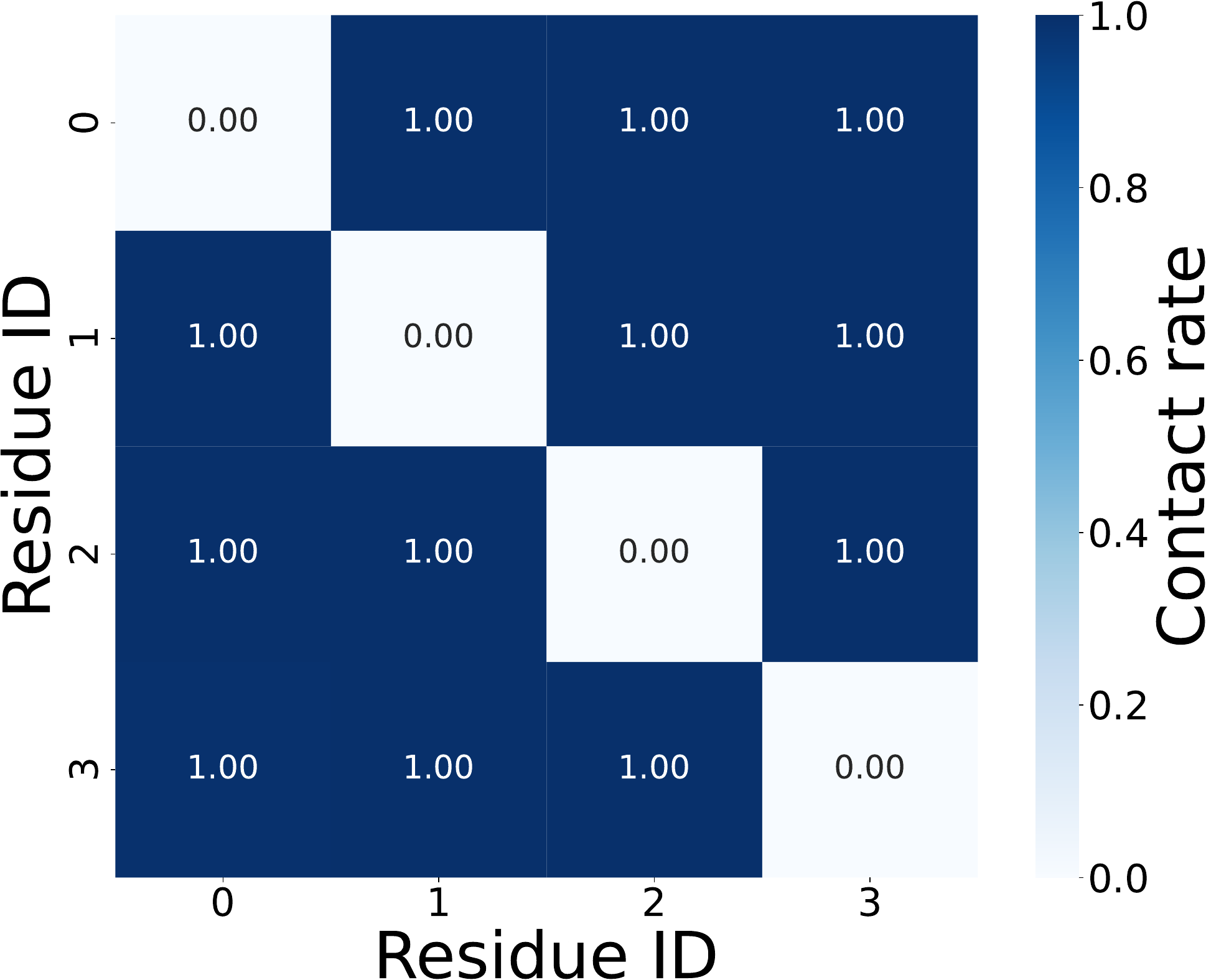}
        \end{minipage}
        \begin{minipage}{0.3\textwidth}
            \includegraphics[width=\textwidth]{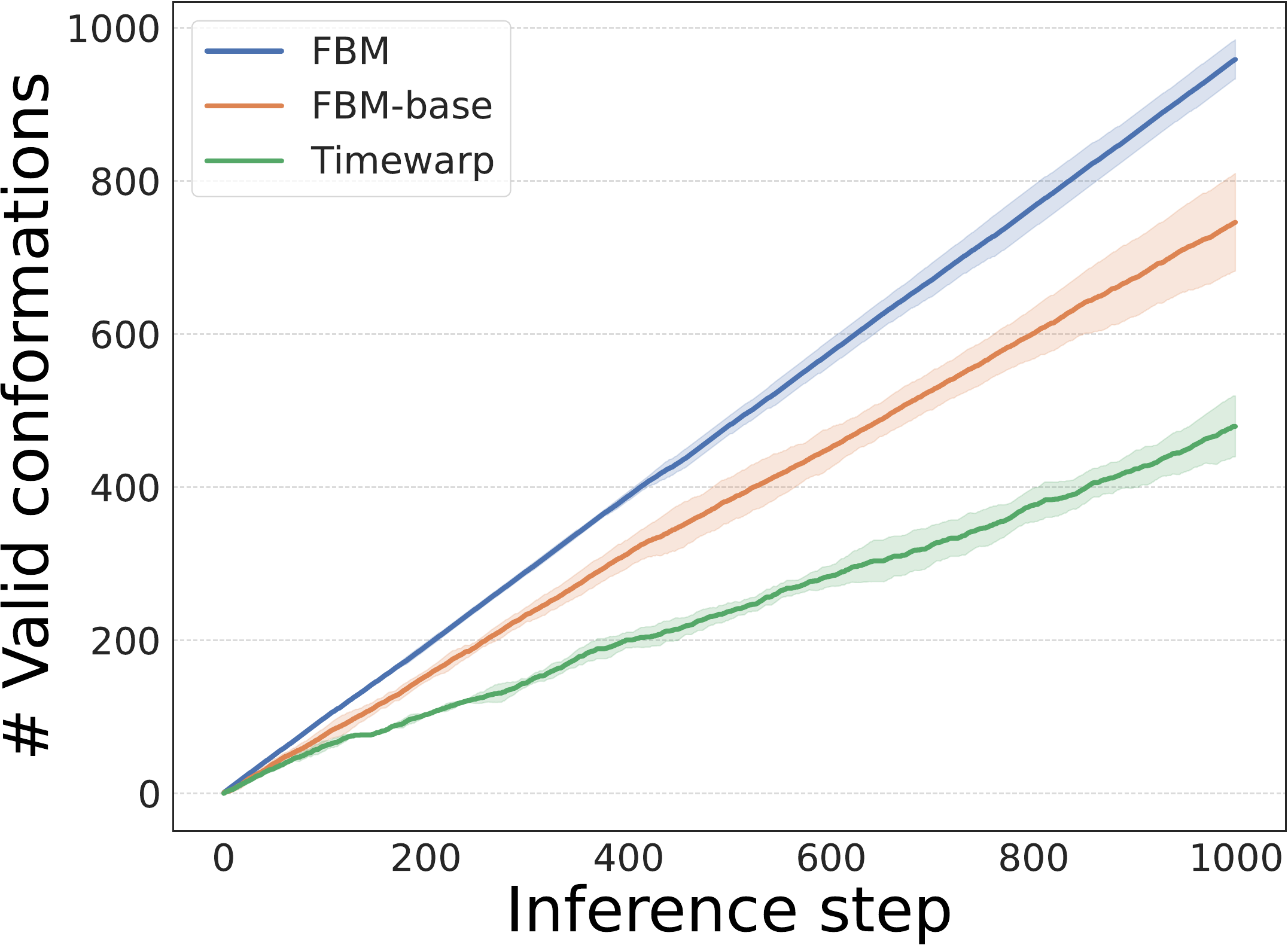}
        \end{minipage}
        \\
        \begin{minipage}{0.3\textwidth}
            \includegraphics[width=\textwidth]{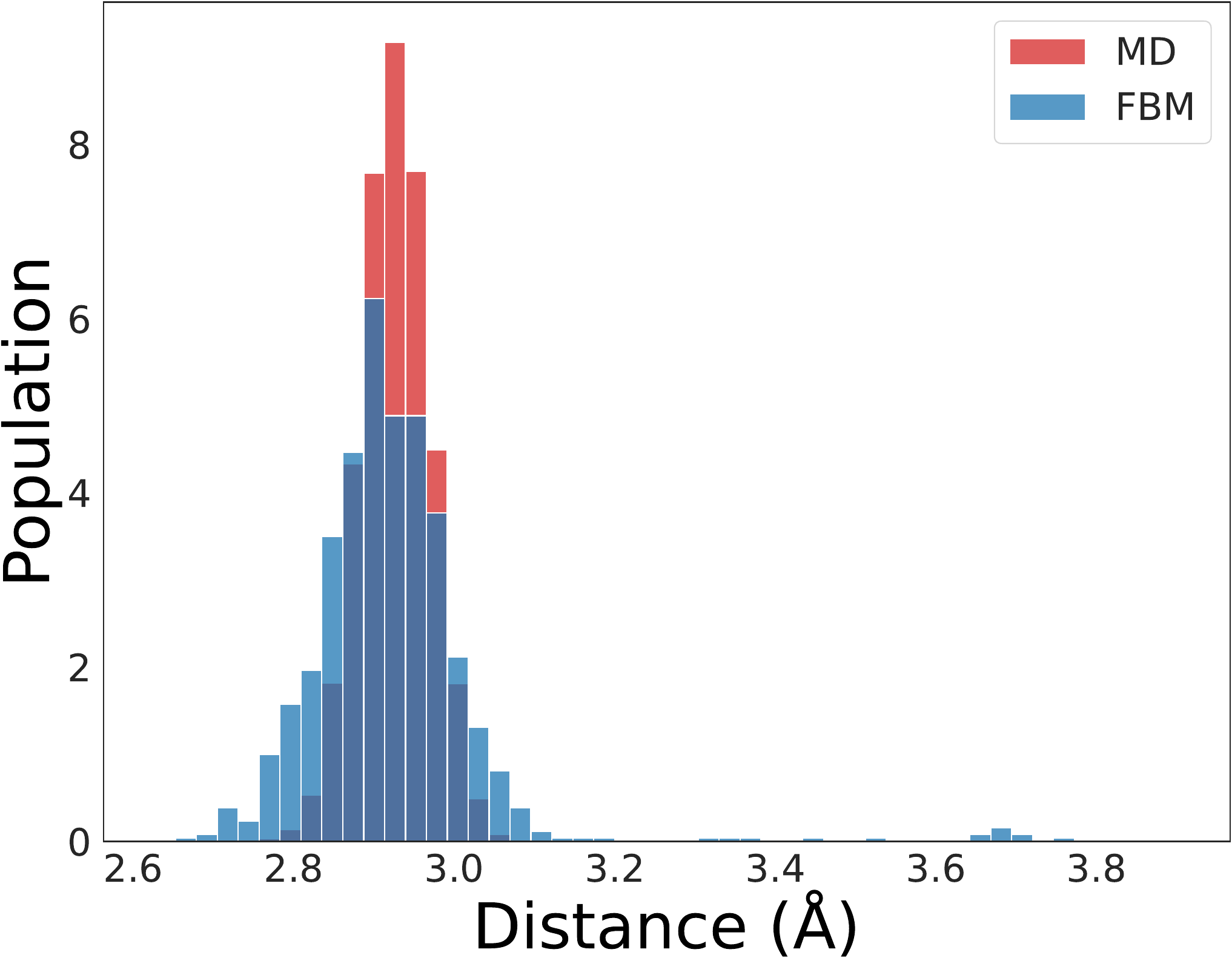}
        \end{minipage}
        \begin{minipage}{0.3\textwidth}
            \includegraphics[width=\textwidth]{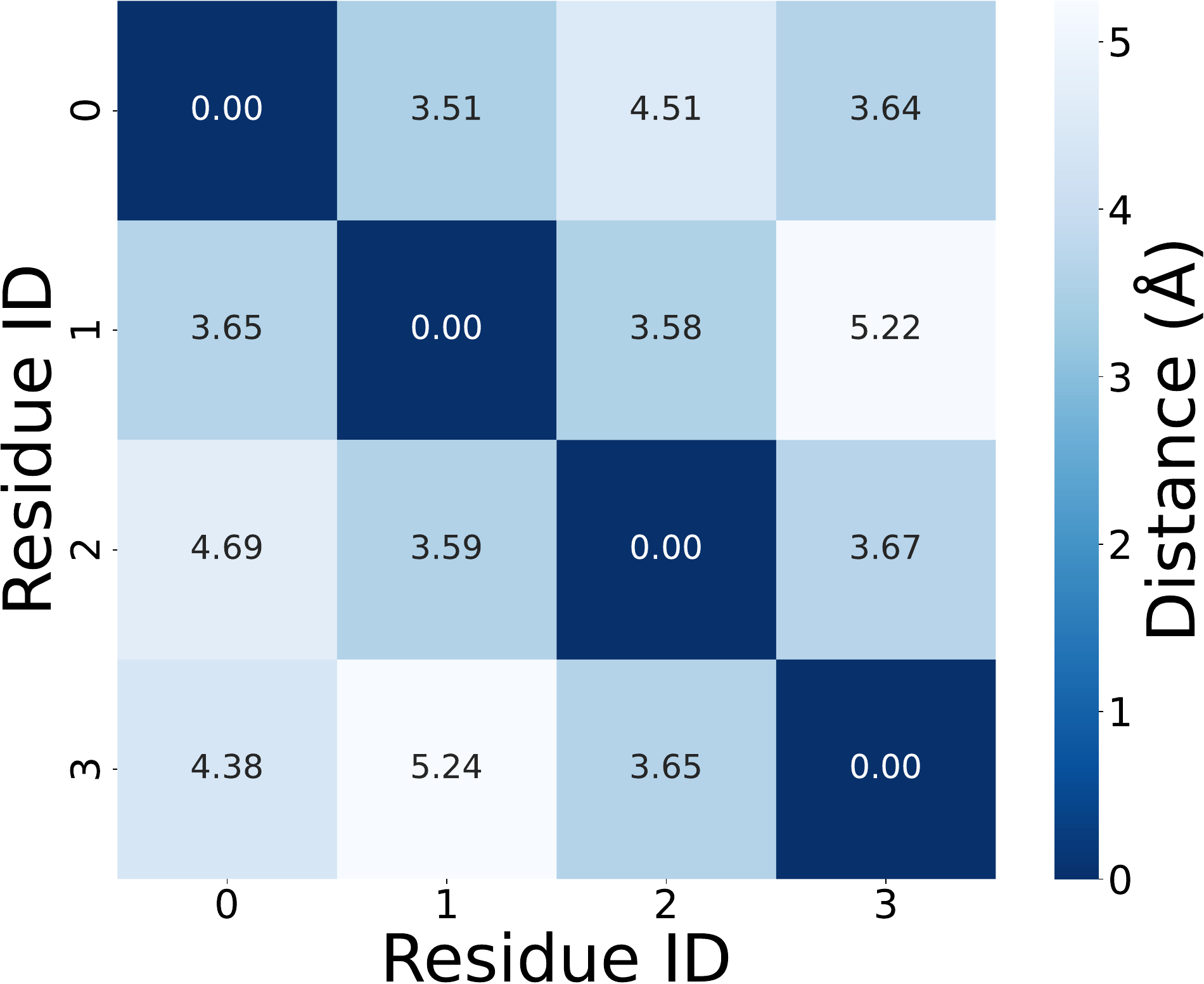}
        \end{minipage}
        \begin{minipage}{0.3\textwidth}
            \includegraphics[width=\textwidth]{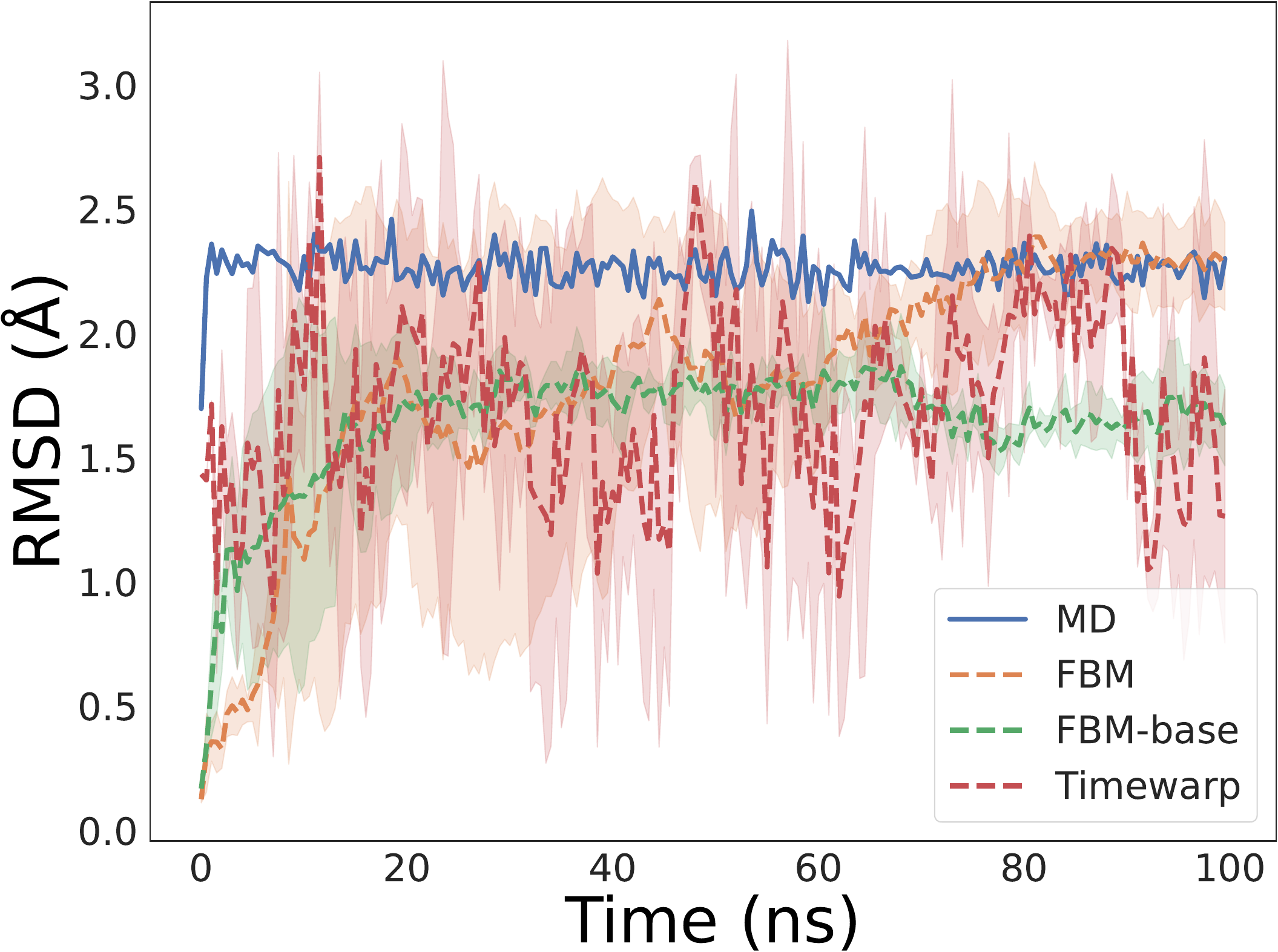}
        \end{minipage}
        \caption{Visualization of peptide 1n73:I (GHRP).}
    \end{subfigure}

    \begin{subfigure}[b]{0.9\textwidth}
        \centering
        \begin{minipage}{0.3\textwidth}
            \includegraphics[width=\textwidth]{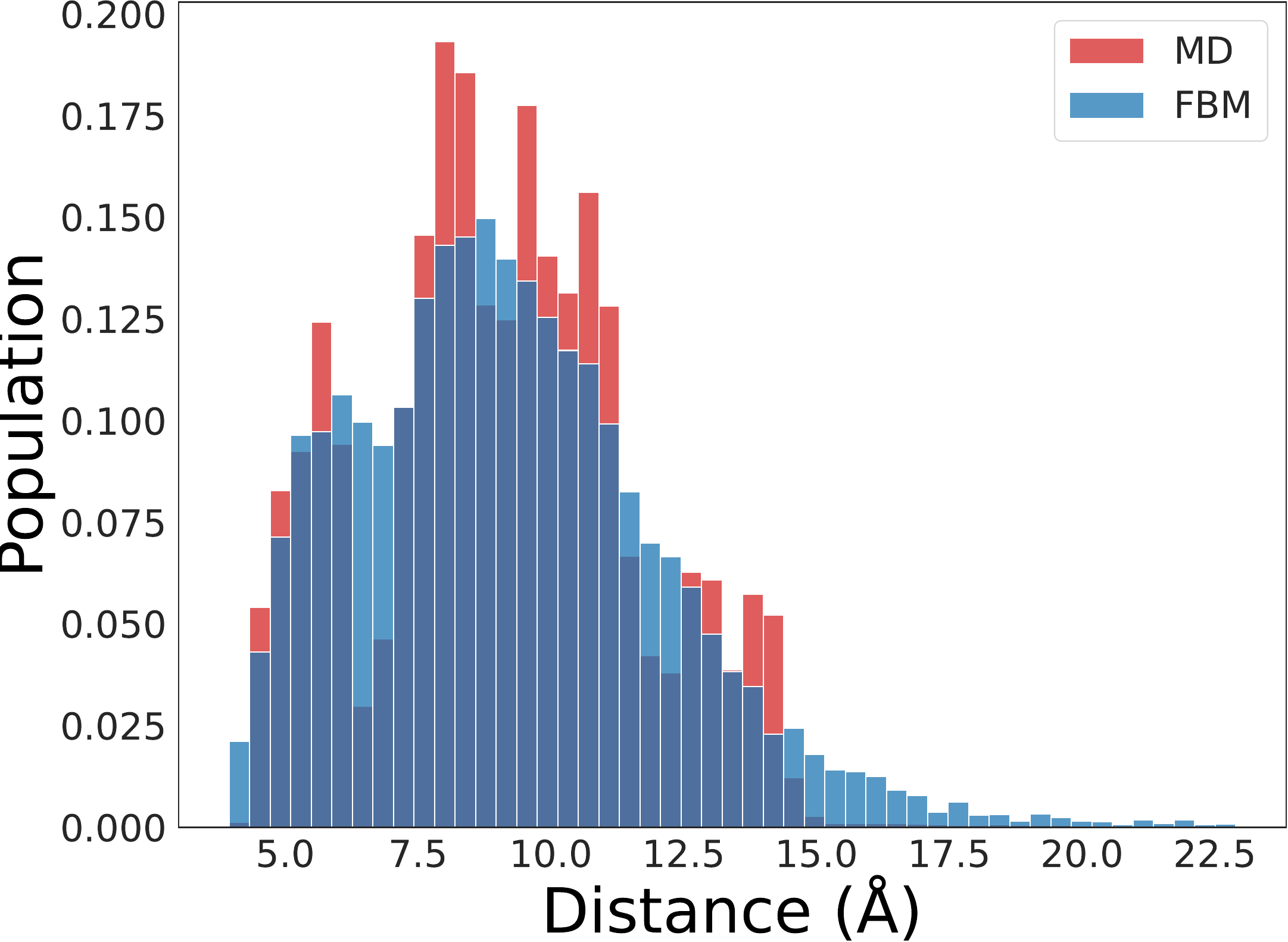}
        \end{minipage}
        \begin{minipage}{0.3\textwidth}
            \includegraphics[width=\textwidth]{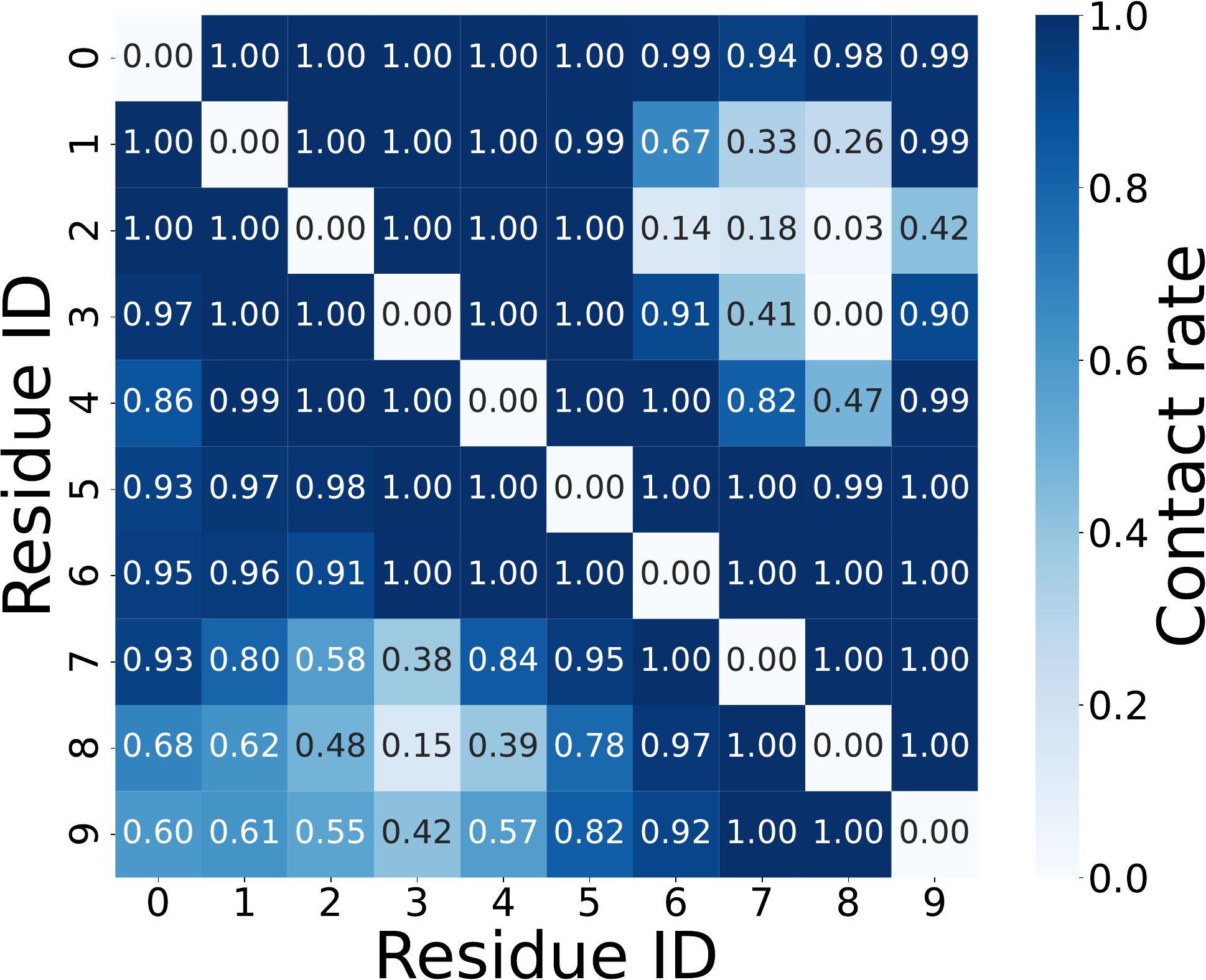}
        \end{minipage}
        \begin{minipage}{0.3\textwidth}
            \includegraphics[width=\textwidth]{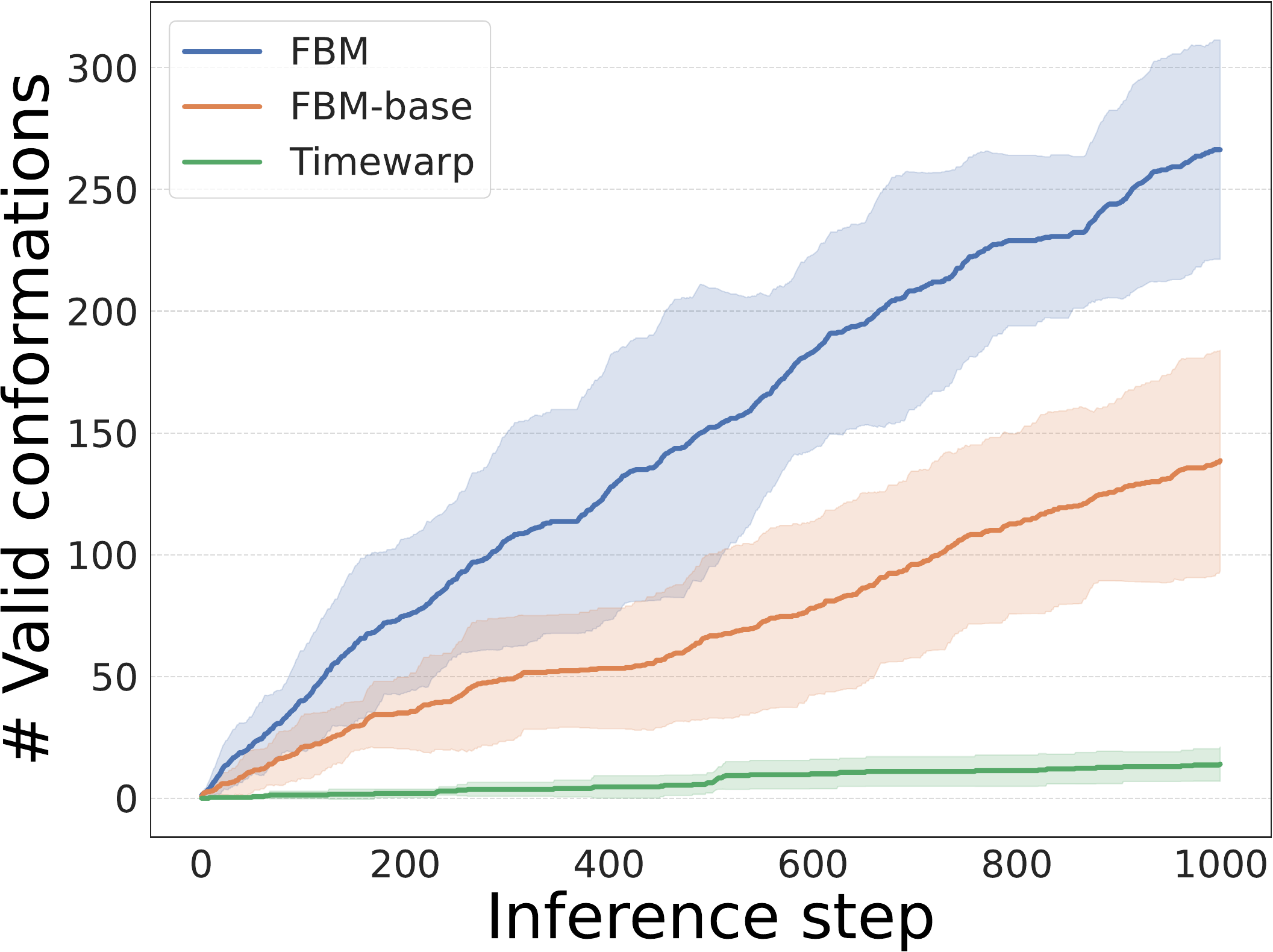}
        \end{minipage}
        \\
        \begin{minipage}{0.3\textwidth}
            \includegraphics[width=\textwidth]{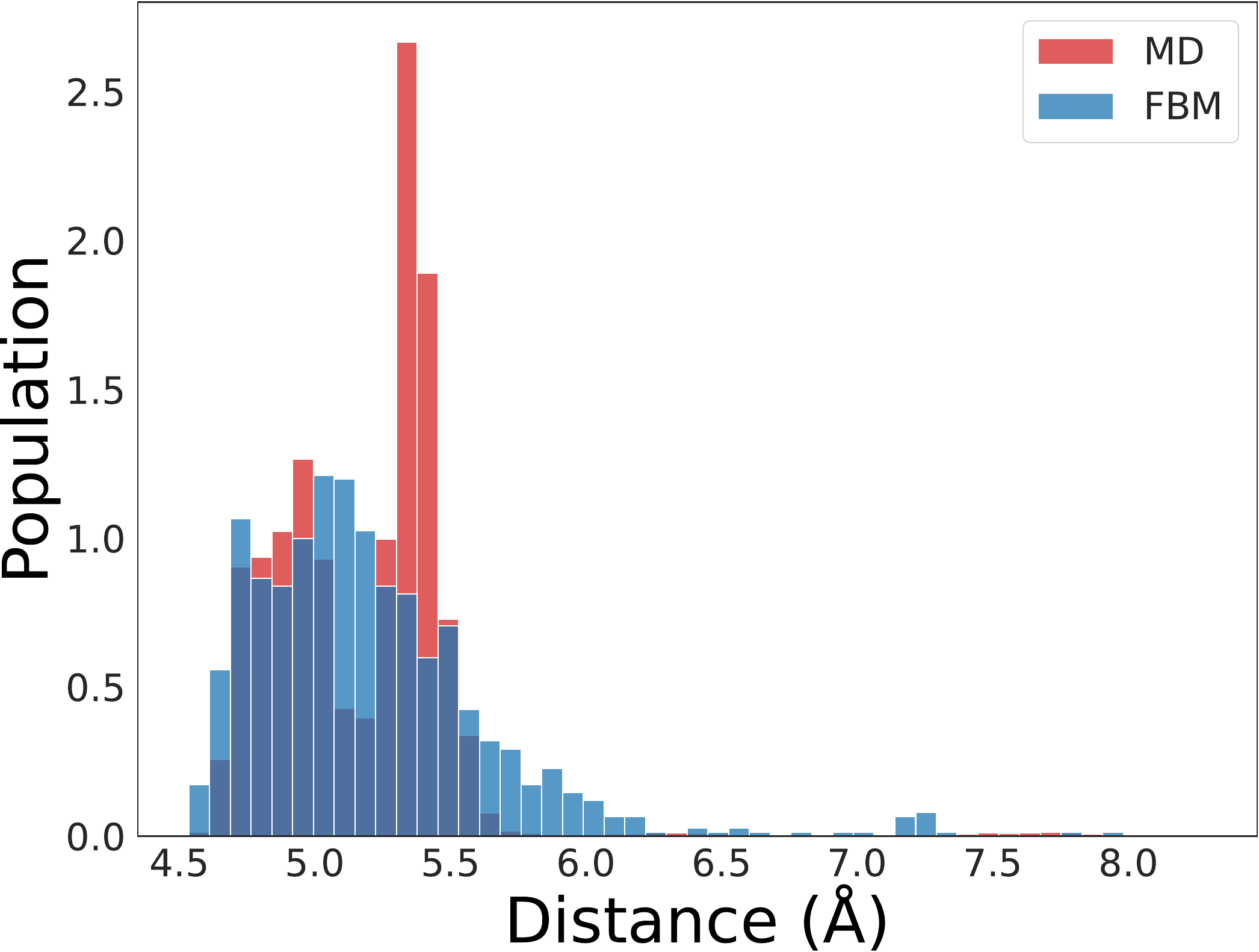}
        \end{minipage}
        \begin{minipage}{0.3\textwidth}
            \includegraphics[width=\textwidth]{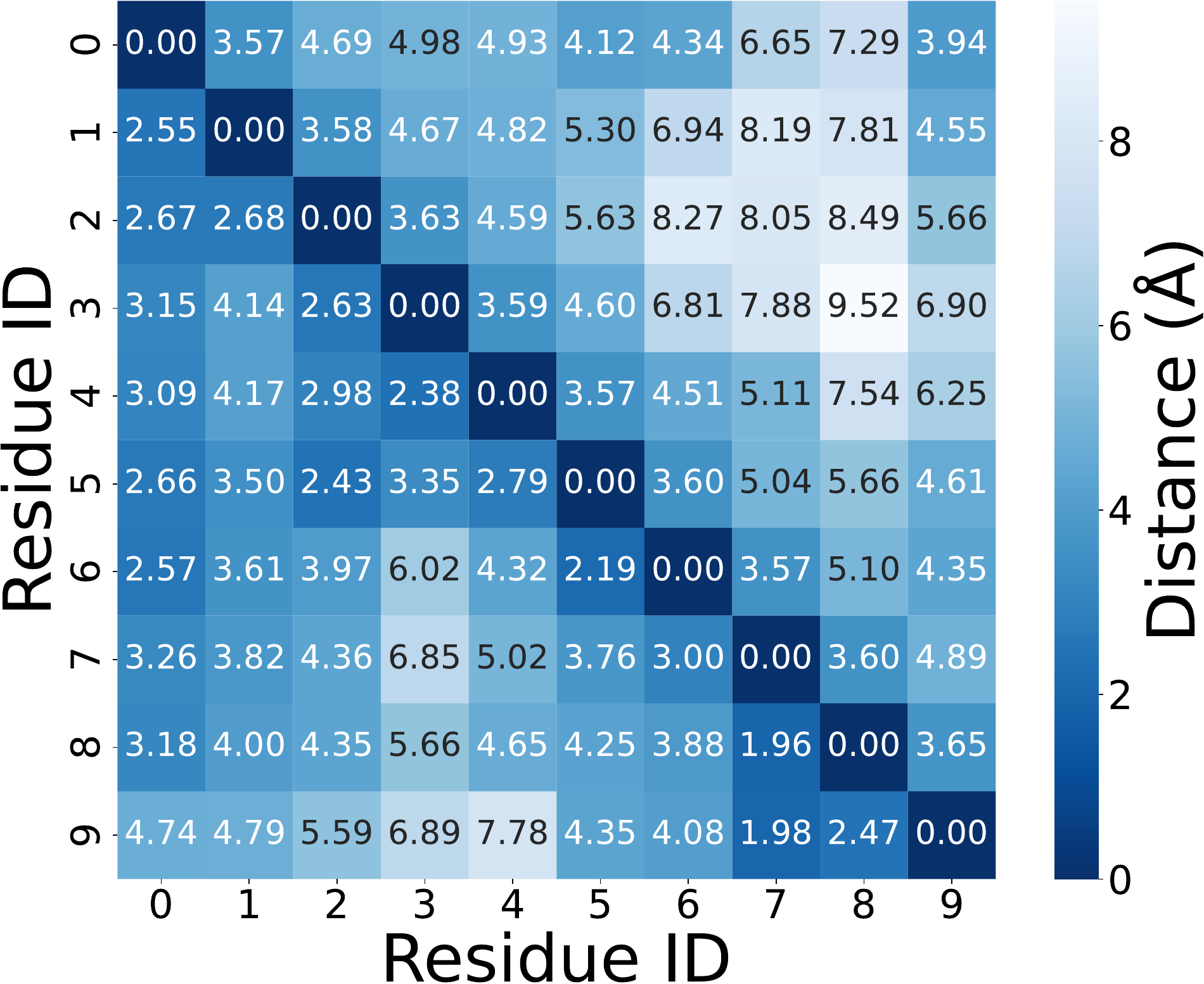}
        \end{minipage}
        \begin{minipage}{0.3\textwidth}
            \includegraphics[width=\textwidth]{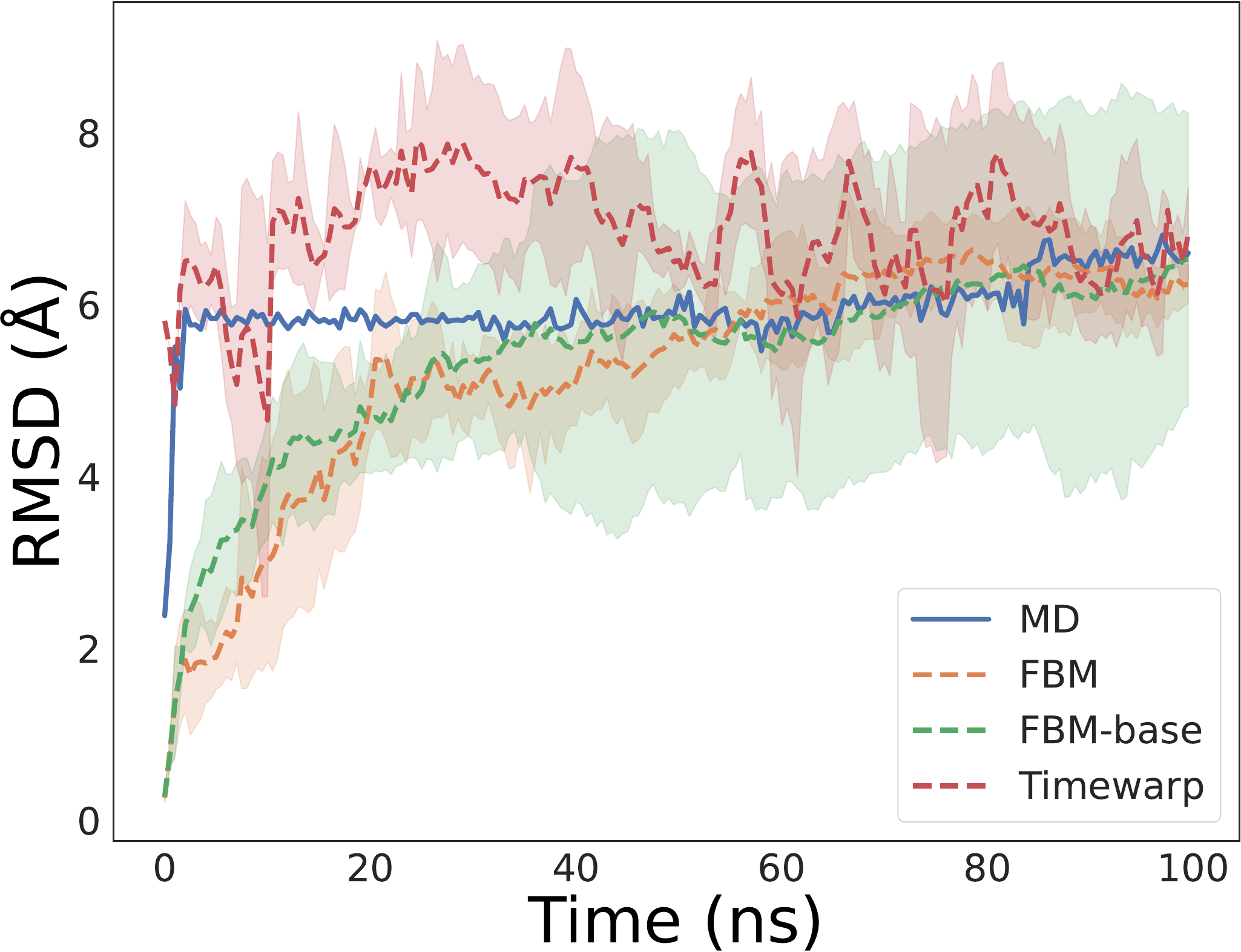}
        \end{minipage}
        \caption{Visualization of peptide 1gxc:B (RHFDTYLIRR).}
    \end{subfigure}

    \begin{subfigure}[b]{0.9\textwidth}
        \centering
        \begin{minipage}{0.3\textwidth}
            \includegraphics[width=\textwidth]{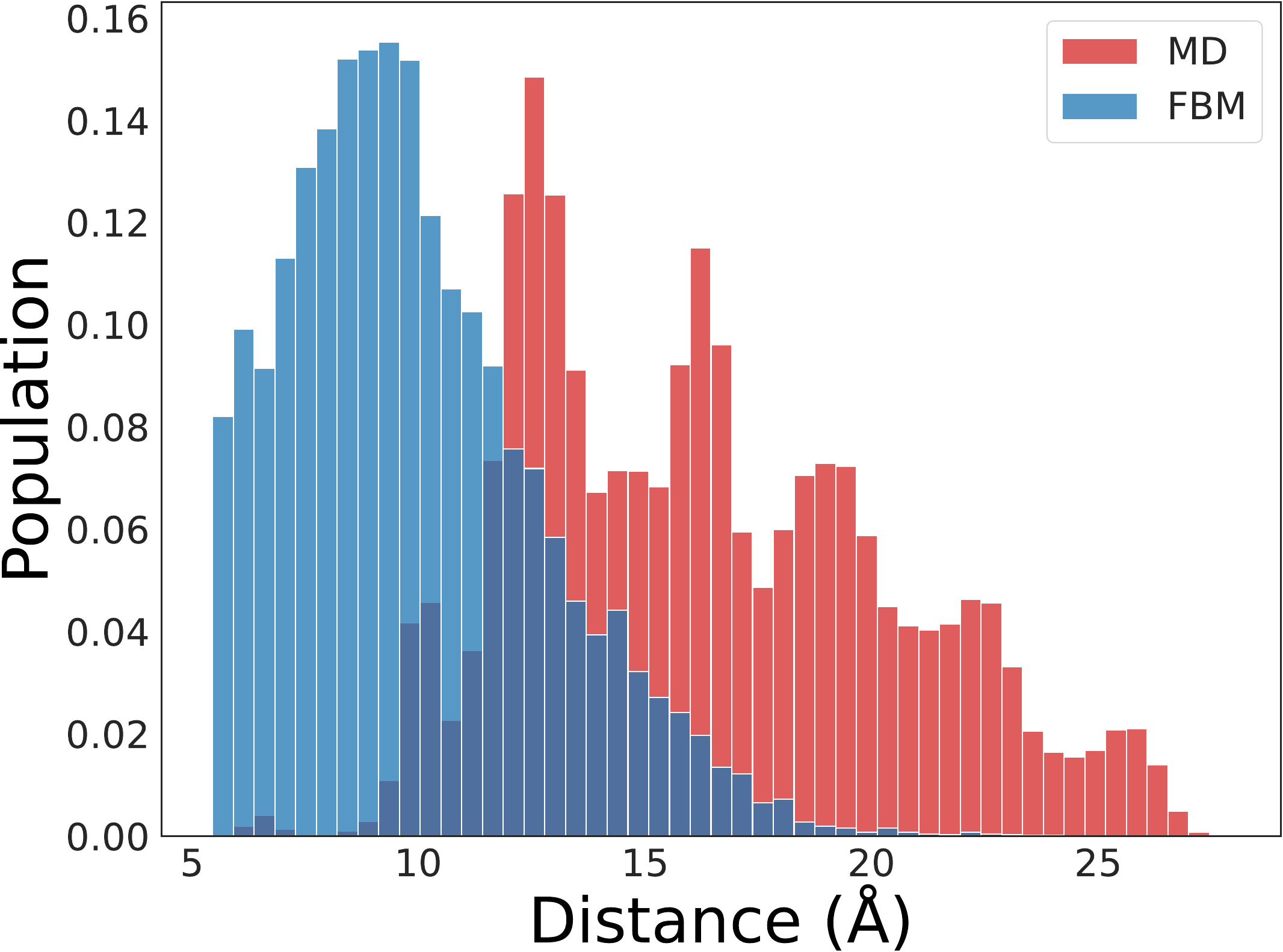}
        \end{minipage}
        \begin{minipage}{0.3\textwidth}
            \includegraphics[width=\textwidth]{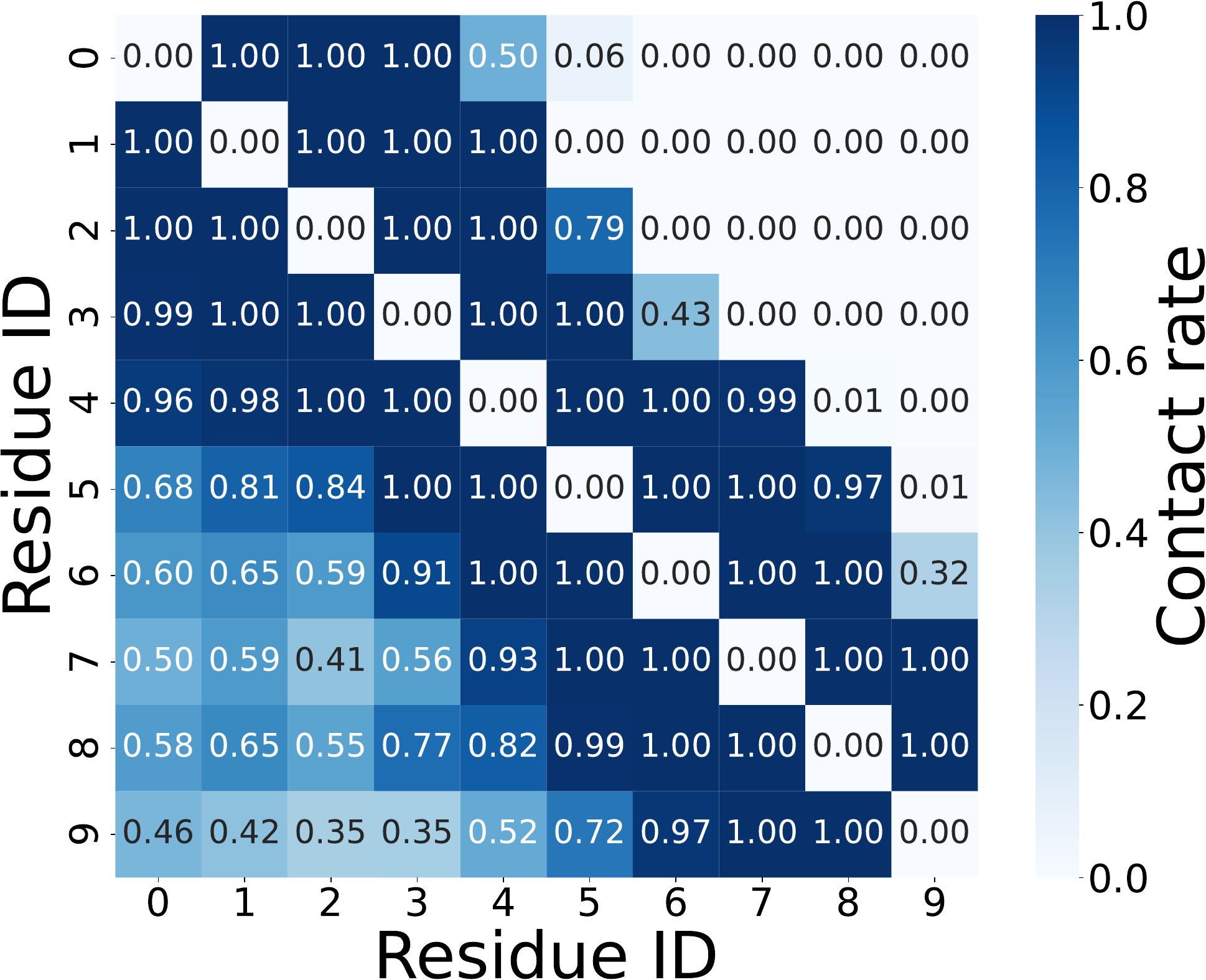}
        \end{minipage}
        \begin{minipage}{0.3\textwidth}
            \includegraphics[width=\textwidth]{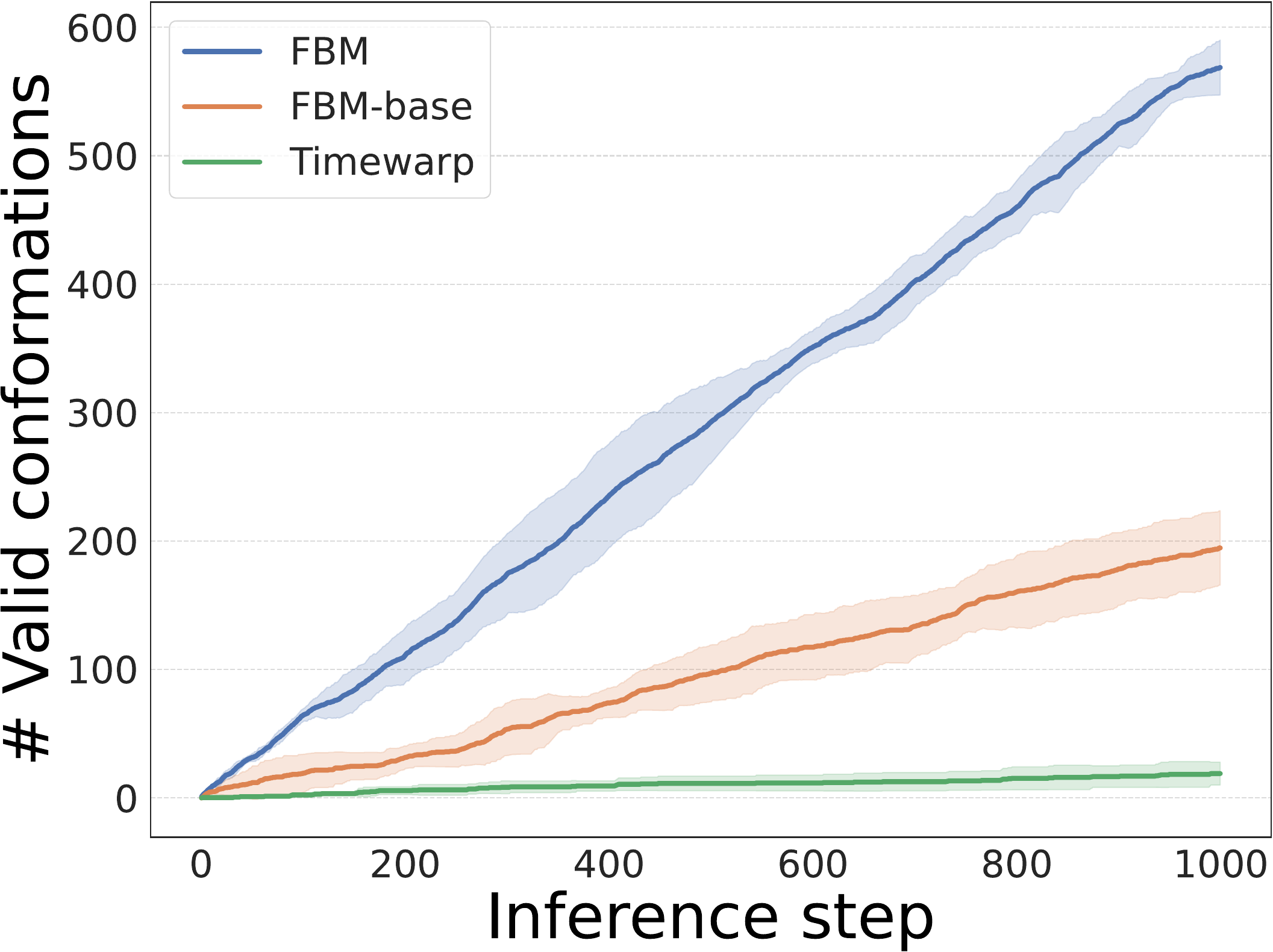}
        \end{minipage}
        \\
        \begin{minipage}{0.3\textwidth}
            \includegraphics[width=\textwidth]{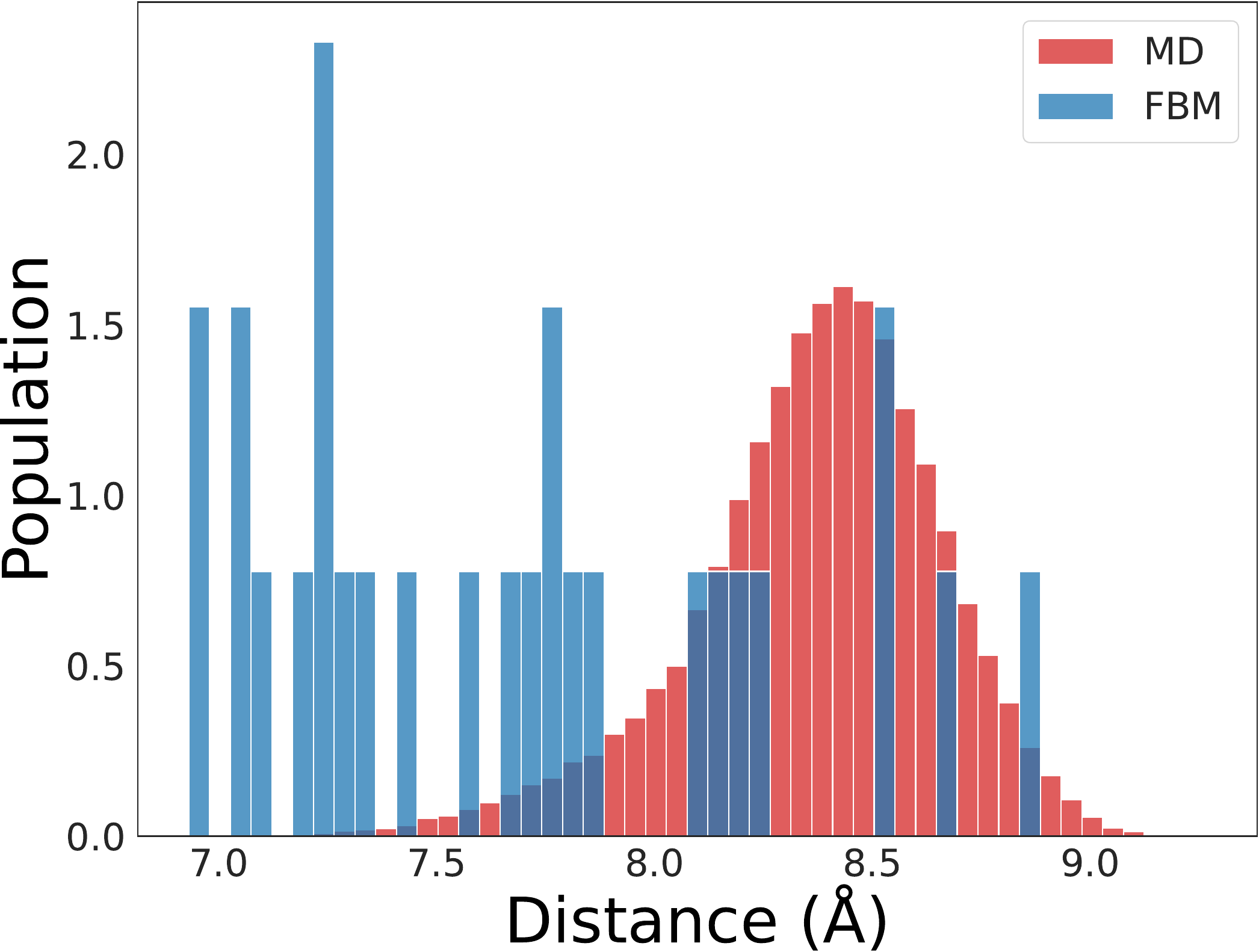}
        \end{minipage}
        \begin{minipage}{0.3\textwidth}
            \includegraphics[width=\textwidth]{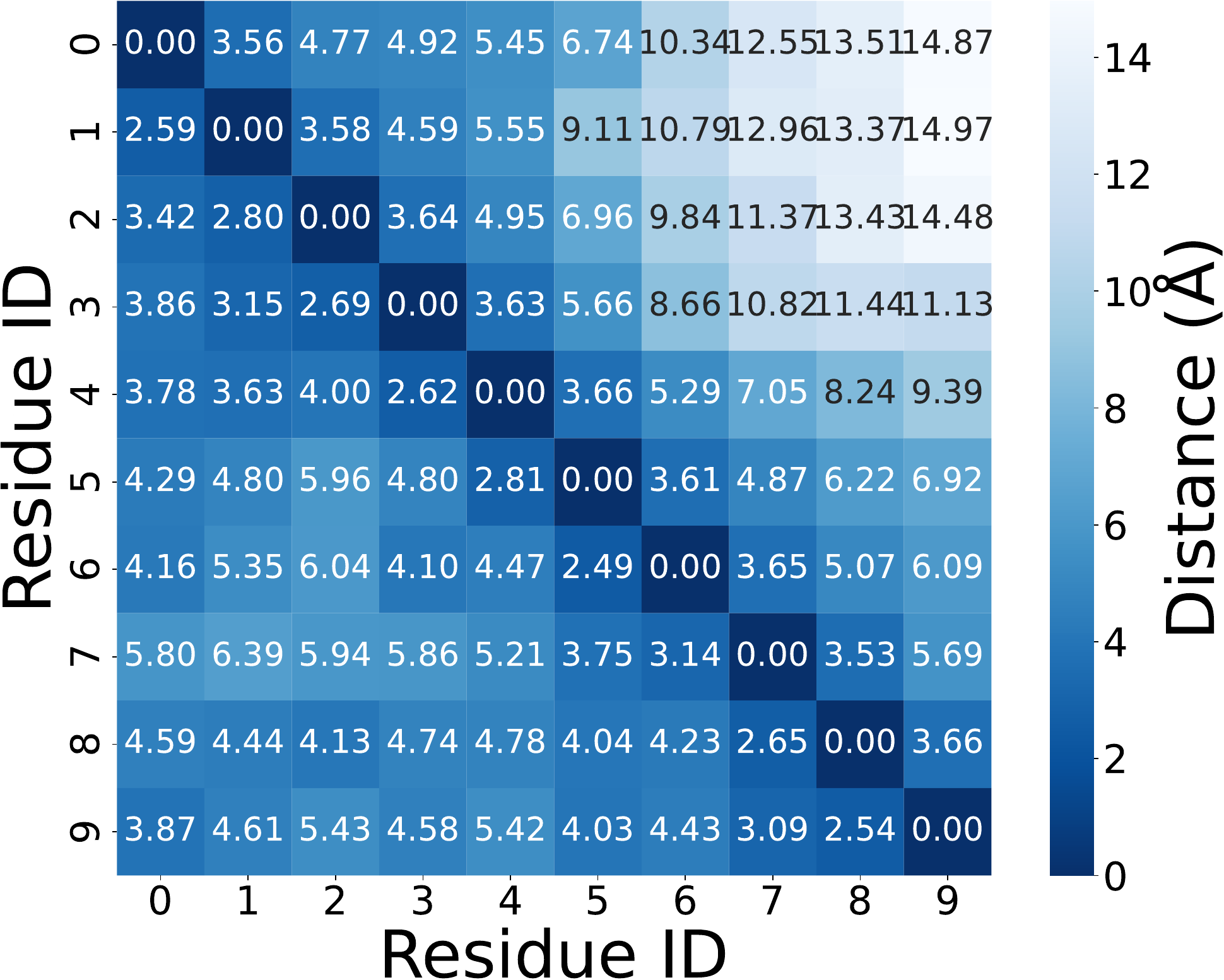}
        \end{minipage}
        \begin{minipage}{0.3\textwidth}
            \includegraphics[width=\textwidth]{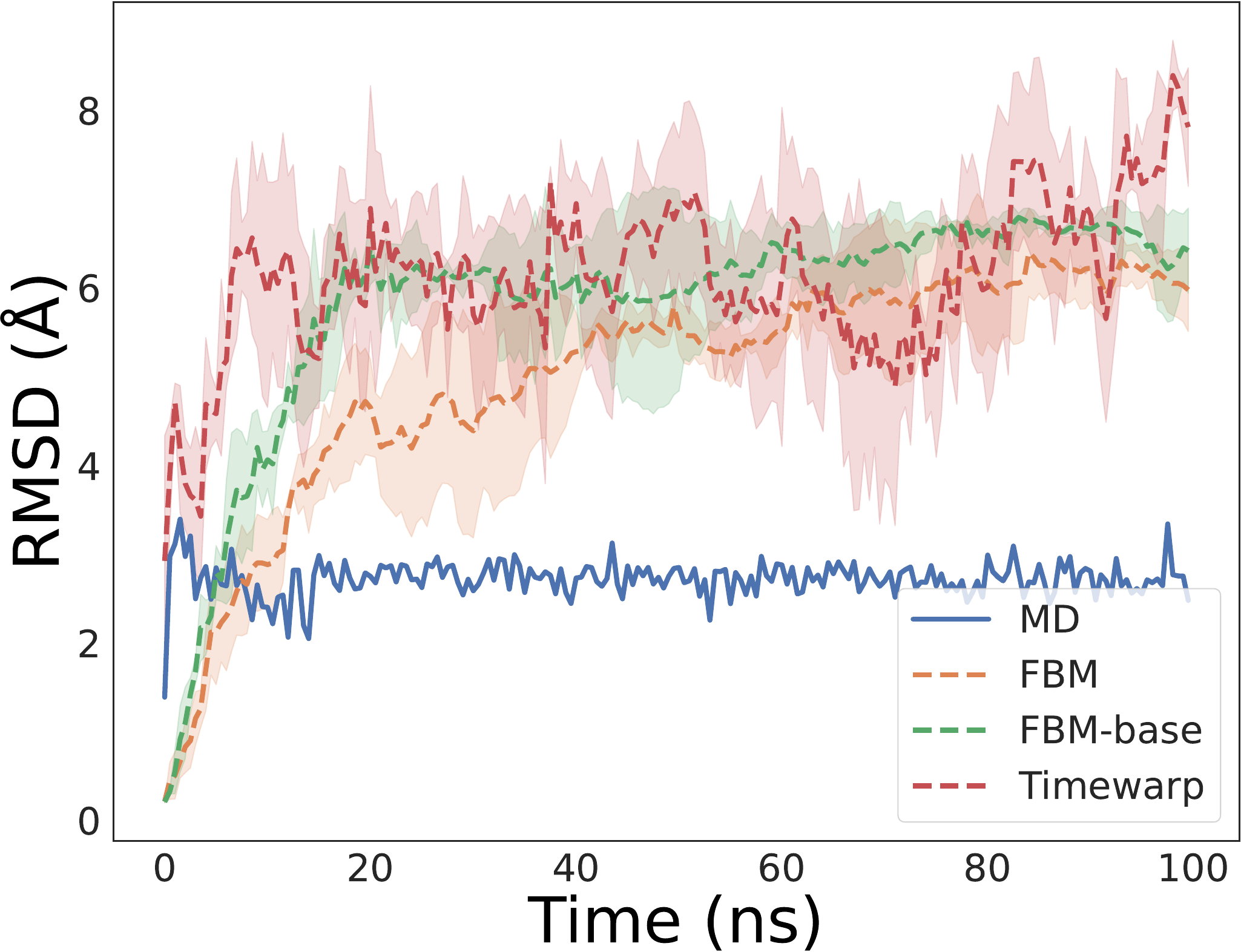}
        \end{minipage}
        \caption{Visualization of peptide 1qj6:I (DFEEIPEEYL).}
    \end{subfigure}

    \caption{The visualization of comprehensive metrics on test peptides 1n73:I, 1gxc:B and 1qj6:I (top, middle and bottom). For each sub-figure of the corresponding peptide: \textbf{1.} The top-left and bottom-left plots show the joint distribution of pairwise distances between residues and the distribution of the radius of gyration, respectively. \textbf{2.} The top-middle and bottom-middle plots demonstrate the residue contact map and residue minimum-distance map, respectively. \textbf{3.} The top-right plot compares the cumulative valid conformations of different methods during inference, with each method undergoing three independent runs. \textbf{4.} The bottom-right plot shows the $C_{\alpha}$-RMSD relative to the initial state along trajectories.}
    \label{fig:appendix-pepmd-metric}

\end{figure}

To provide a more comprehensive evaluation, we present the visualization of comprehensive metrics for the three peptides in Figure~\ref{fig:appendix-pepmd-metric} similar to that in Figure~\ref{fig:1e28-all}. Note that, the P{\small W}D distribution plot of peptide 1n73:I is not displayed since its residue length is too short for an offset of 3~\citep{wang2024protein}. FBM undoubtedly outperforms all other baselines in stably generating valid conformations of all three peptides. For the tetrapeptide 1n73:I and decapeptide 1gxc:B, FBM achieves a close match with MD trajectories in terms of distributions on projected features, residue contact rates and inter-residue distances. Thrillingly, FBM consistently achieves equilibrium distributions during the inference of both peptides and aligns quite well with MD trajectories under the $C_{\alpha}$-RMSD evaluation, showing a good transferability to Out-Of-Distribution (OOD) peptides with different residue lengths. In contrast, other methods either deviate from the real distribution or exhibit excessive fluctuations during the generation process.

Meanwhile, we also provide a \textbf{failure case} of FBM, \emph{i.e.}, peptide 1qj6:I in Figure~\ref{fig:appendix-pepmd-metric}(c). For this peptide, the trajectories generated by FBM show significant deviation from the reference distribution. Based on the residue contact map, we observe that the residues in the peptide are spatially dispersed, which may hinder the graph neural network from efficiently capturing global dynamics and interactions of the molecule. In such cases, increasing the model parameter size, stacking more layers, and expanding the dataset are likely to help generate more accurate trajectories.

\subsection{A Tiny Experiment on Chignolin}

To extend FBM on more complex molecular systems, we perform a tiny experiment on Chignolin, a small protein consisting of 10 residues and 175 atoms. The trajectory data of Chignolin is downloaded from \texttt{figshare}\footnote{https://figshare.com/articles/dataset/Chignolin\_Simulations/13858898}, which was curated by \cite{PérezCulubret2021}. MD simulations were performed with ACEMD, using CHARMM22 force field~\citep{mackerell1998all} and TIP3P water model~\citep{jorgensen1983comparison} at 350K temperature, which contains 1,881 water molecules and two Na$^+$ ions to neutralize the peptide's negative charge. The dataset consists of 3,744 independent product simulations of 50 ns, for a total aggregate time of 187.2 µs. Considering the difference of MD simulation setups from ours, we need to fine-tune our model on the dataset to align with its simulation environment.

To be specific, we first randomly select 500/50/3 \textbf{independent} trajectories from the dataset for training/validation/test and select 200 data pairs for each trajectory with $\tau=$100 ps. The trained FBM-{\small BASE} model will then be fine-tuned on the training data for 20 epochs, with a relatively low learning rate of 2e-4. Further, to obtain atomic forces and potentials of molecules for training with FBM, we use CHARMM36 force field~\citep{best2012optimization} and implicit solvation of GB-OBC I parameters~\citep{onufriev2004exploring} on the \texttt{OpenMM} platform. Afterwards, a new FBM model will be trained for 100 epochs, with the fine-tuned FBM-{\small BASE} as the baseline model.

\begin{figure}[htbp]
    \centering

    \begin{subfigure}[b]{0.9\textwidth}
        \centering
        \begin{minipage}{0.3\textwidth}
            \includegraphics[width=\textwidth]{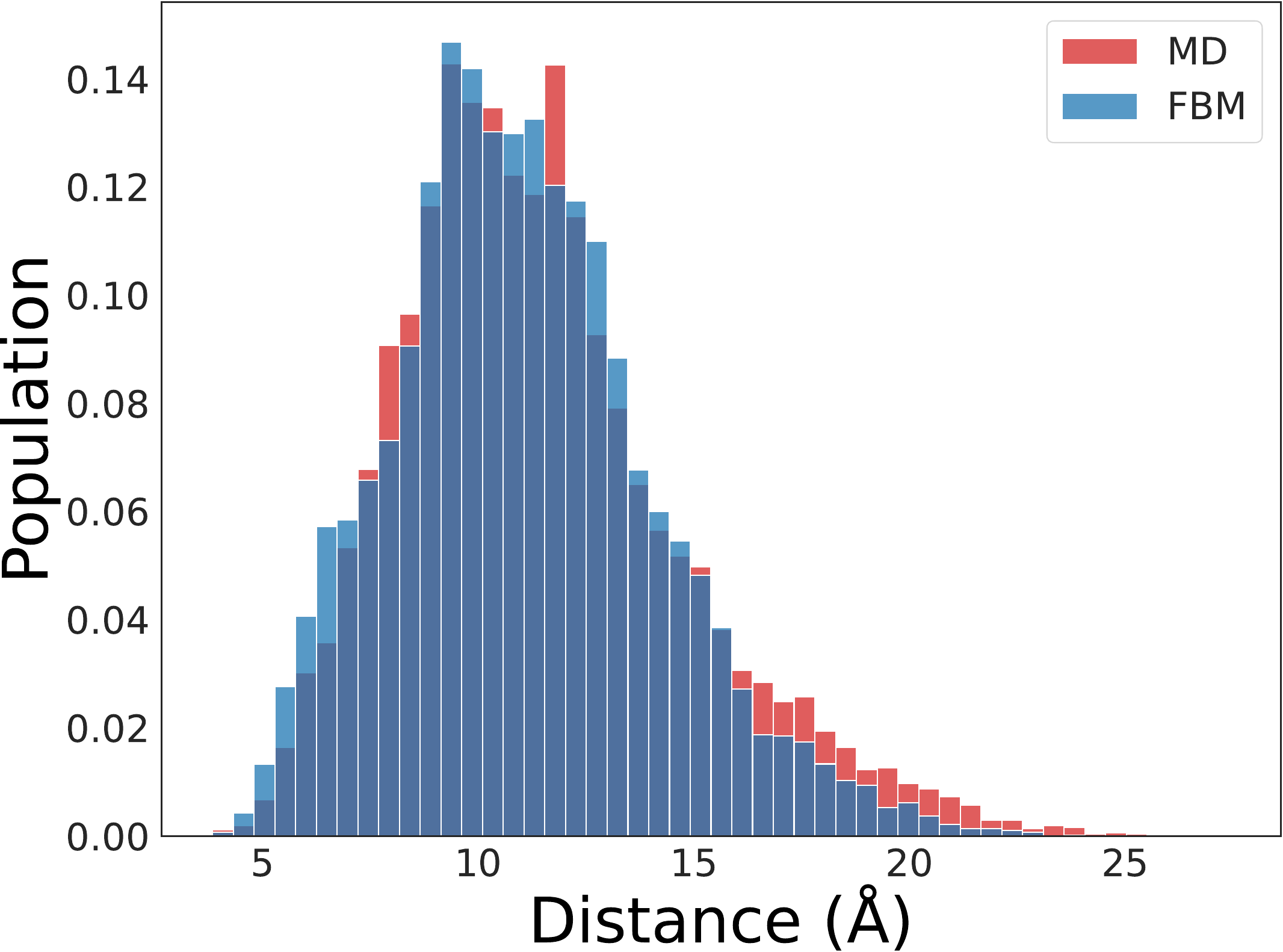}
        \end{minipage}
        \begin{minipage}{0.3\textwidth}
            \includegraphics[width=\textwidth]{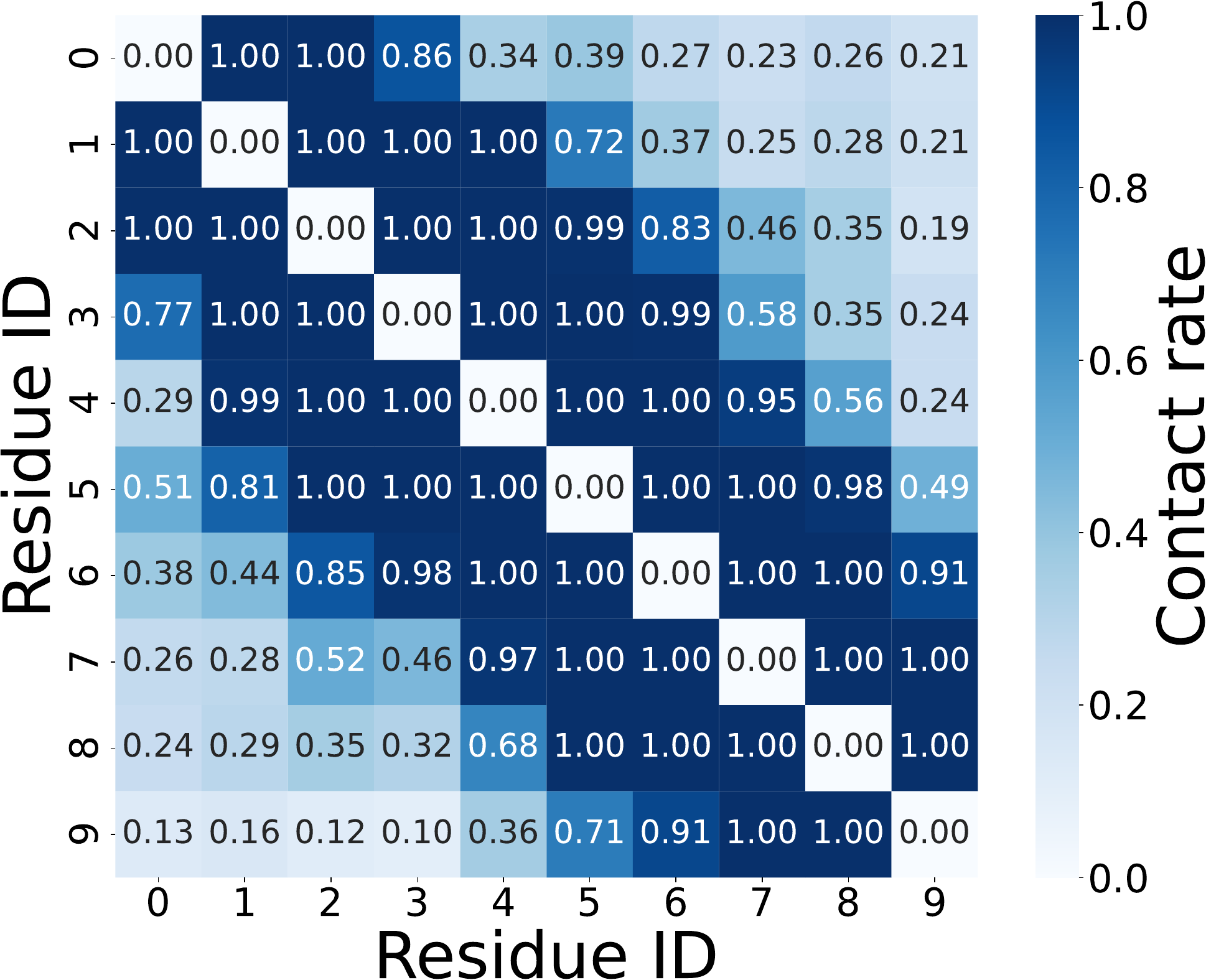}
        \end{minipage}
        \begin{minipage}{0.3\textwidth}
            \includegraphics[width=\textwidth]{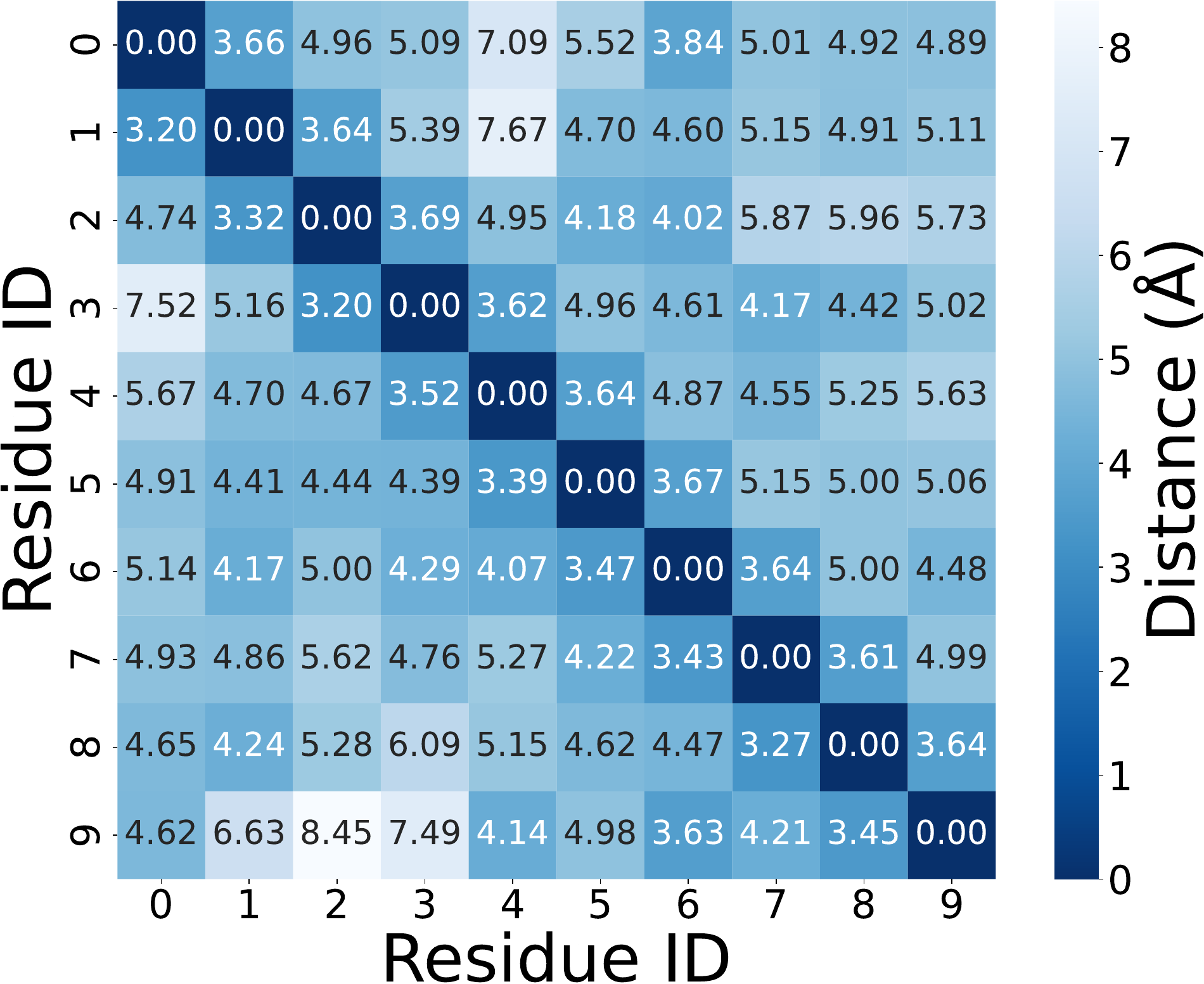}
        \end{minipage}
        \\
        \begin{minipage}{0.3\textwidth}
            \includegraphics[width=\textwidth]{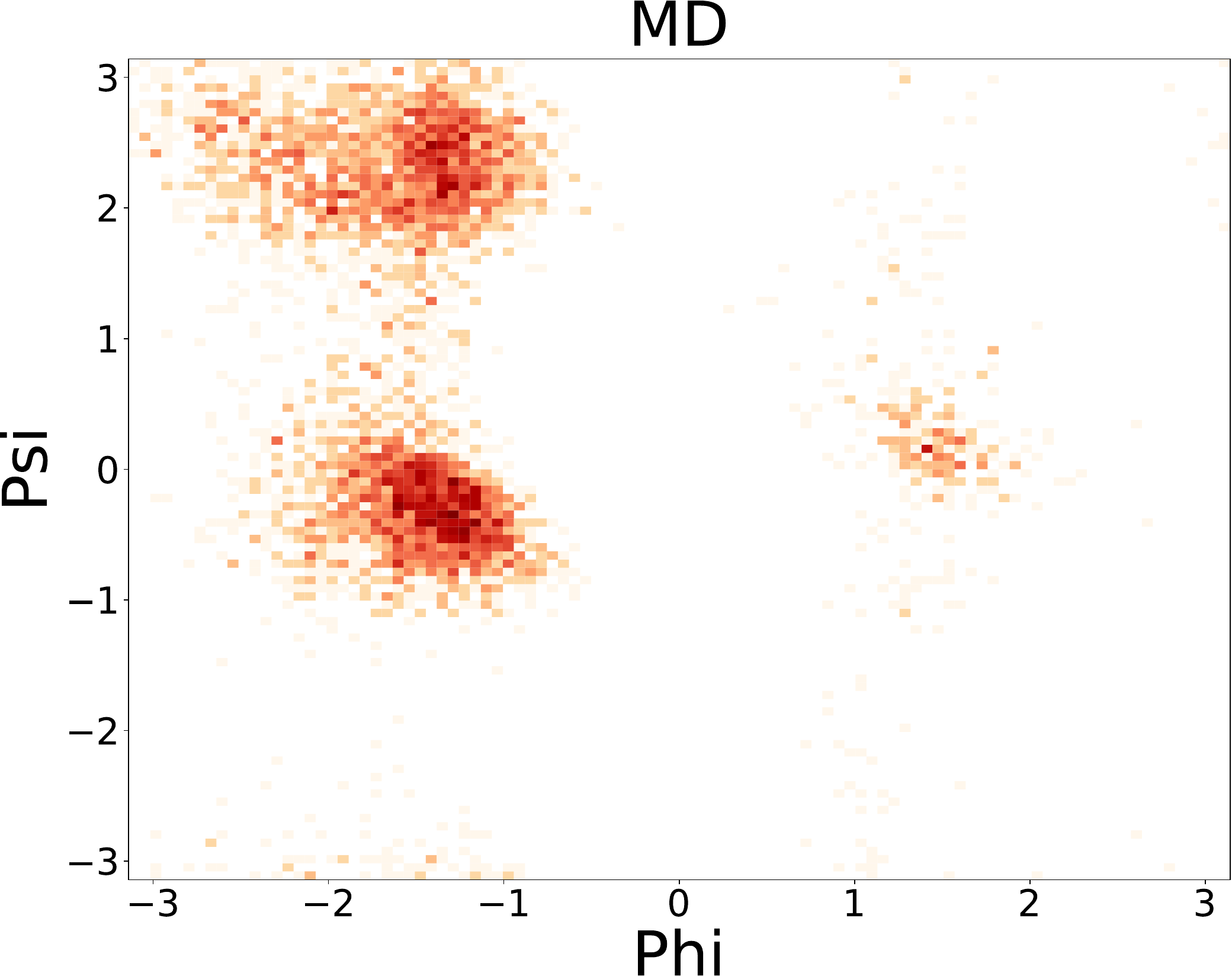}
        \end{minipage}
        \begin{minipage}{0.3\textwidth}
            \includegraphics[width=\textwidth]{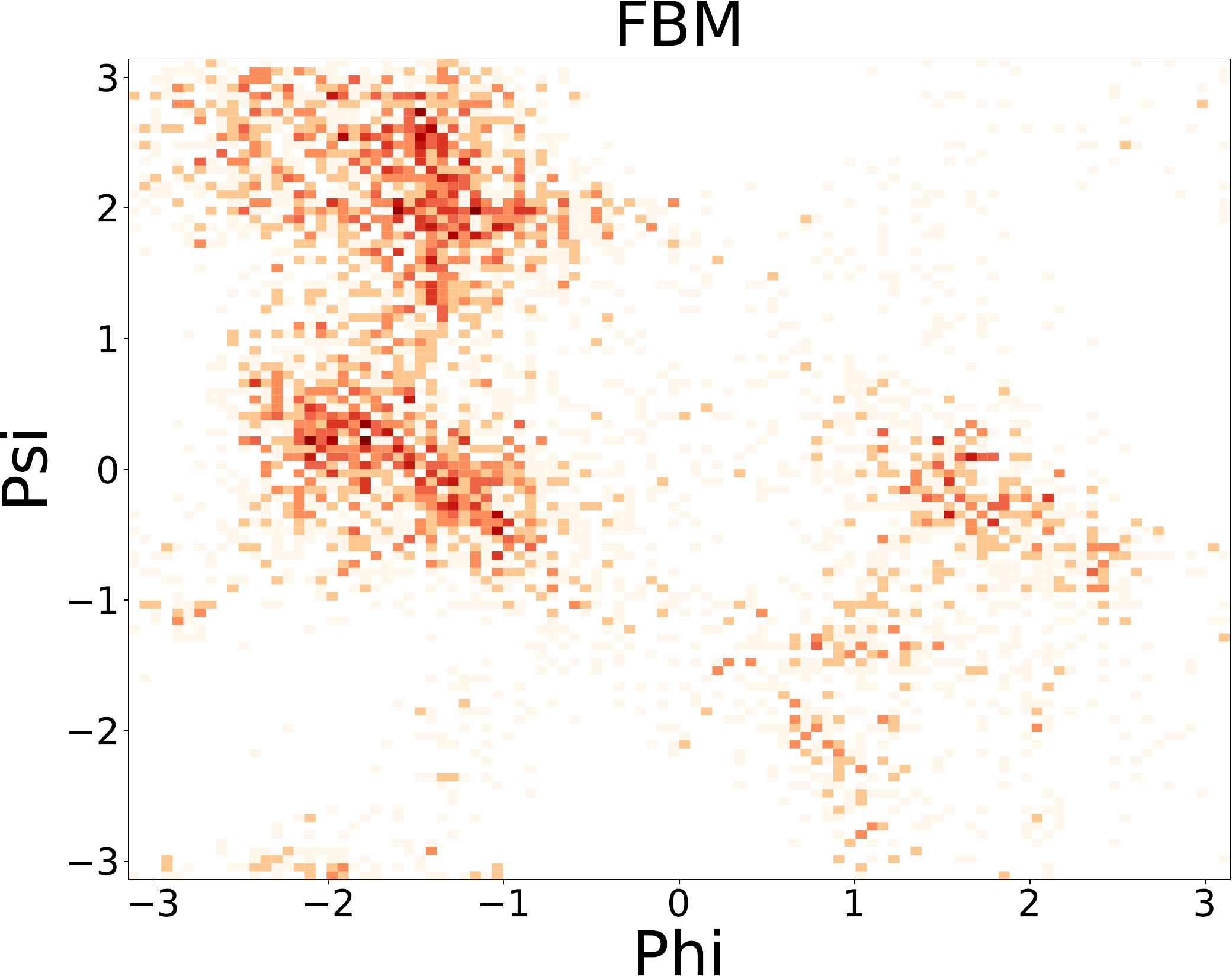}
        \end{minipage}
        \begin{minipage}{0.3\textwidth}
            \includegraphics[width=\textwidth]{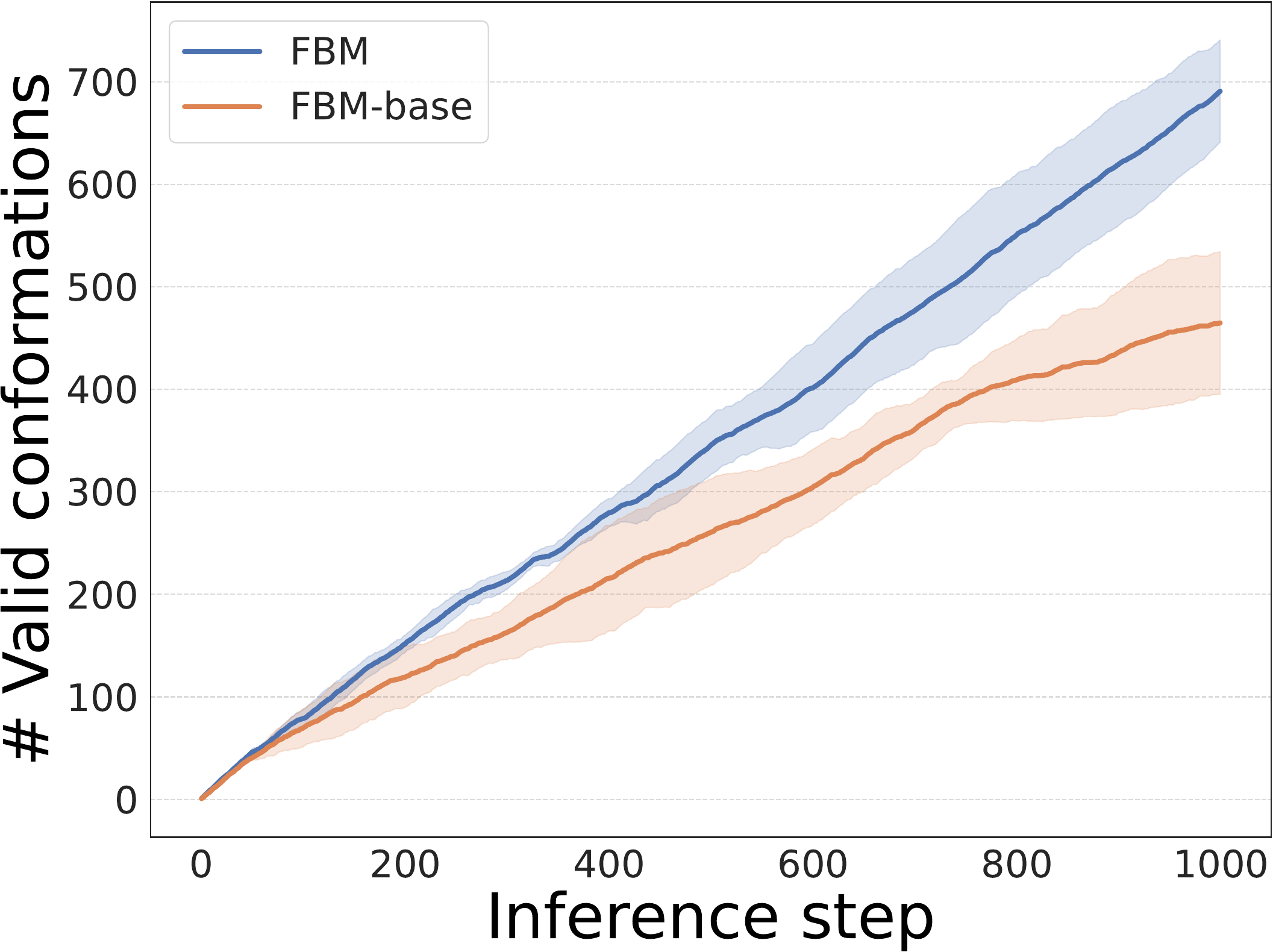}
        \end{minipage}
        \caption{Visualization of evaluations on trajectory e1s44.}
    \end{subfigure}

    \begin{subfigure}[b]{0.9\textwidth}
        \centering
        \begin{minipage}{0.3\textwidth}
            \includegraphics[width=\textwidth]{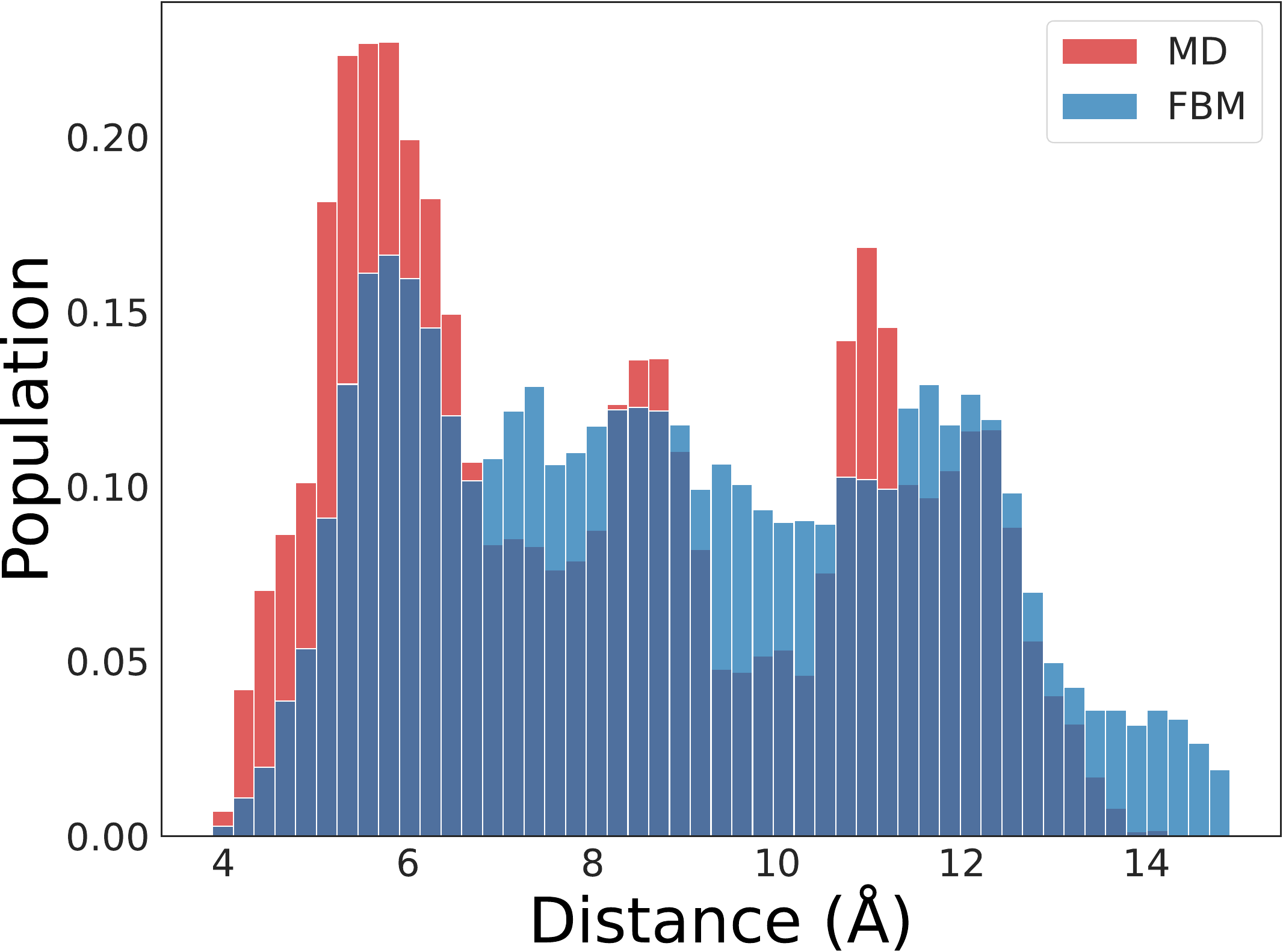}
        \end{minipage}
        \begin{minipage}{0.3\textwidth}
            \includegraphics[width=\textwidth]{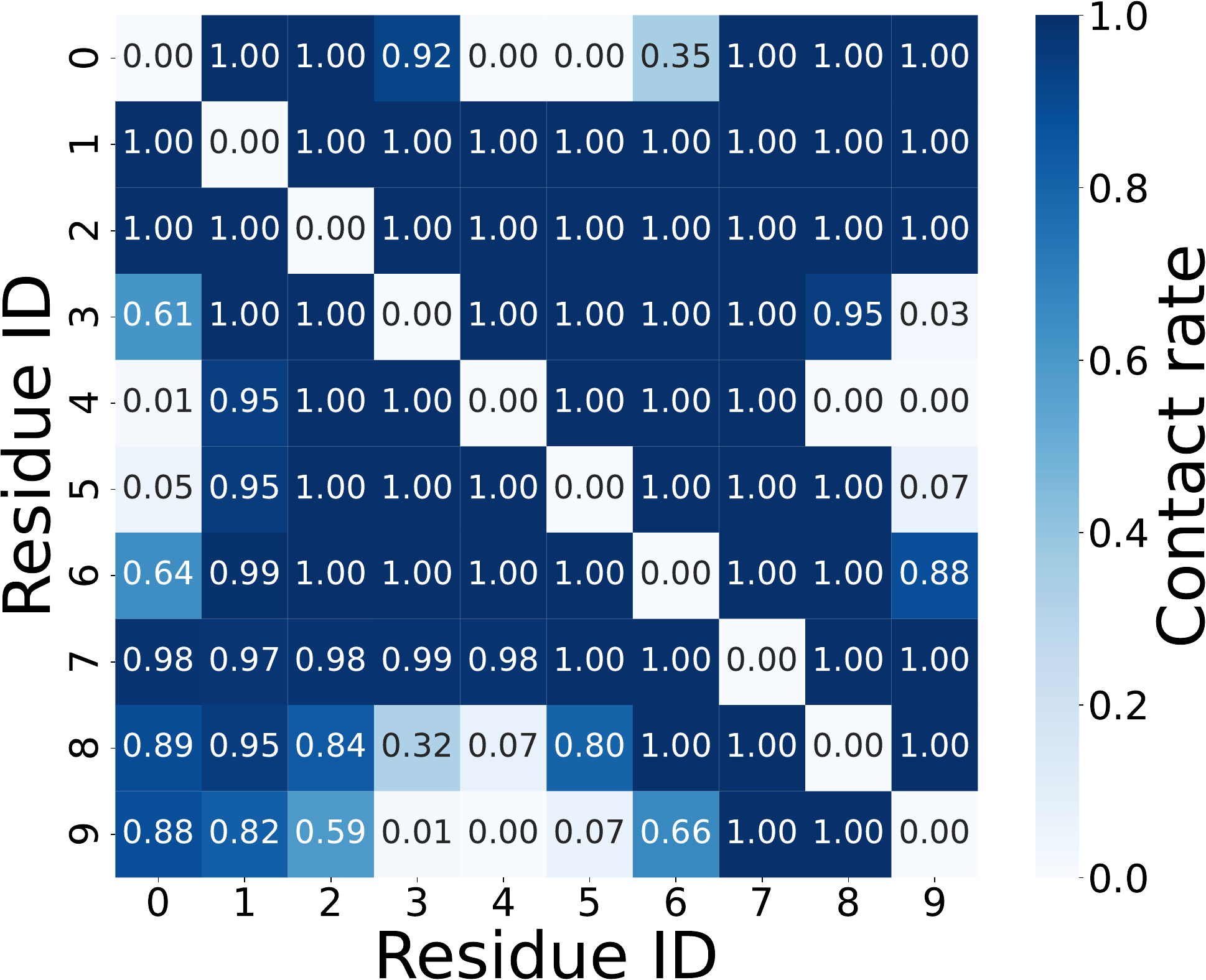}
        \end{minipage}
        \begin{minipage}{0.3\textwidth}
            \includegraphics[width=\textwidth]{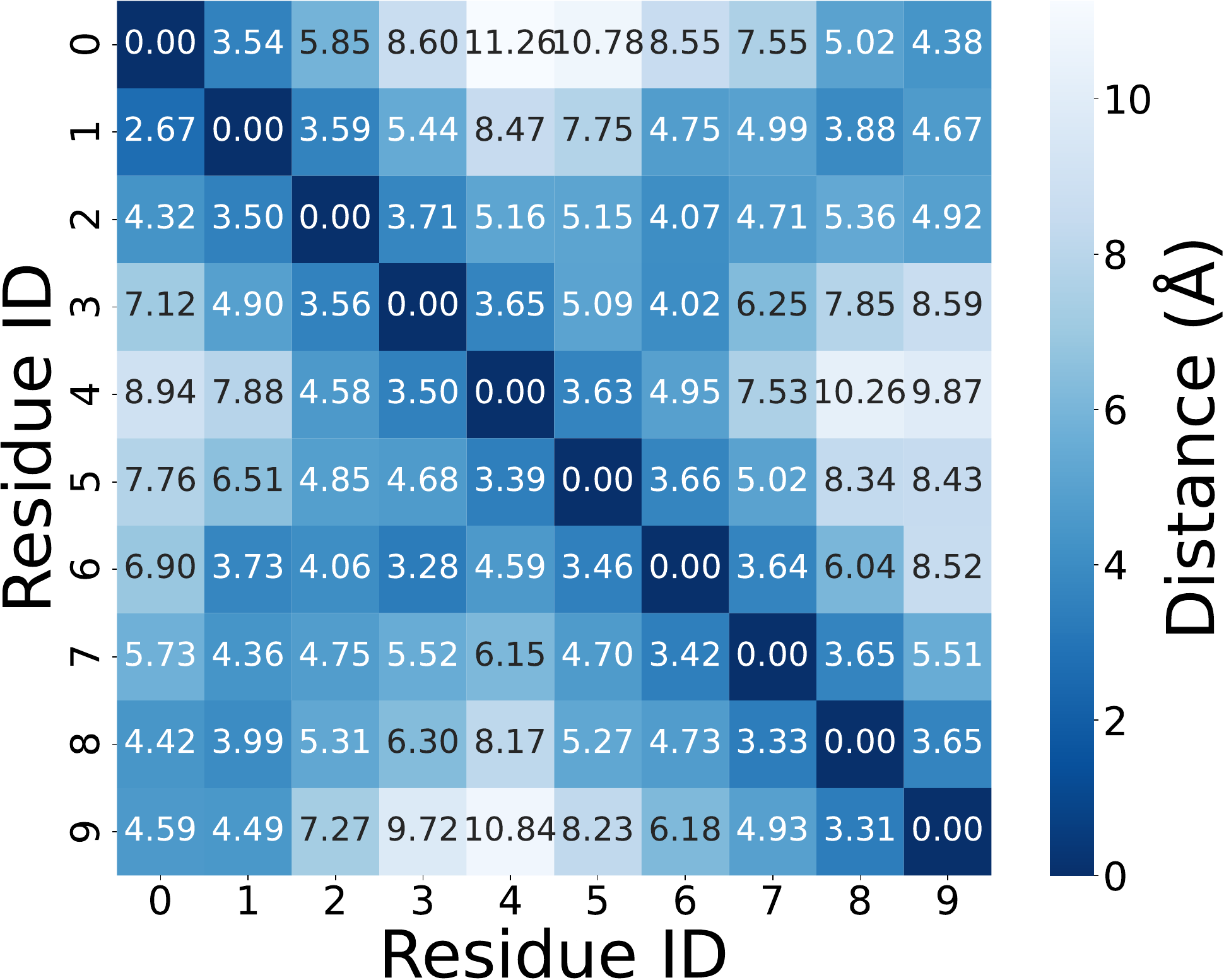}
        \end{minipage}
        \\
        \begin{minipage}{0.3\textwidth}
            \includegraphics[width=\textwidth]{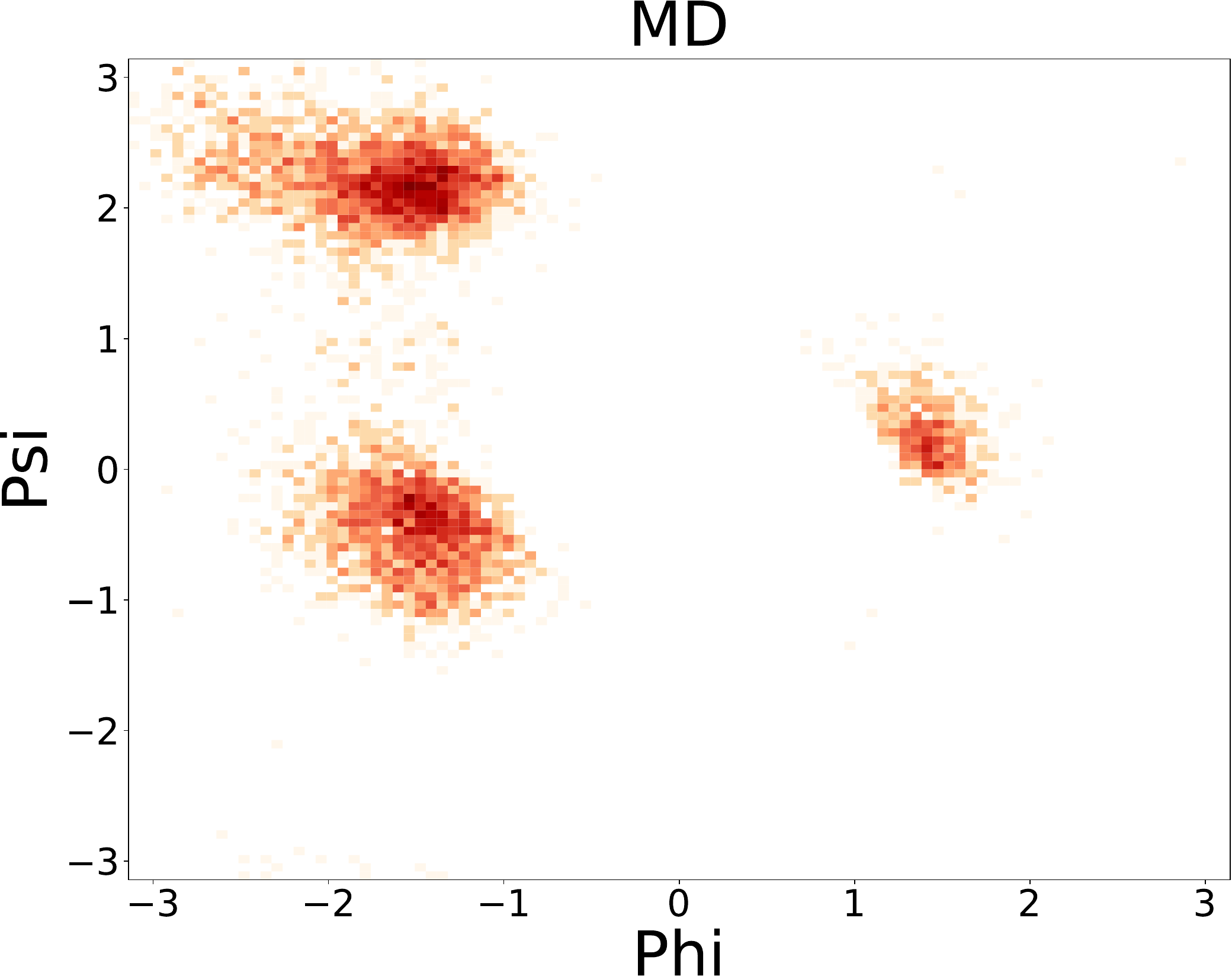}
        \end{minipage}
        \begin{minipage}{0.3\textwidth}
            \includegraphics[width=\textwidth]{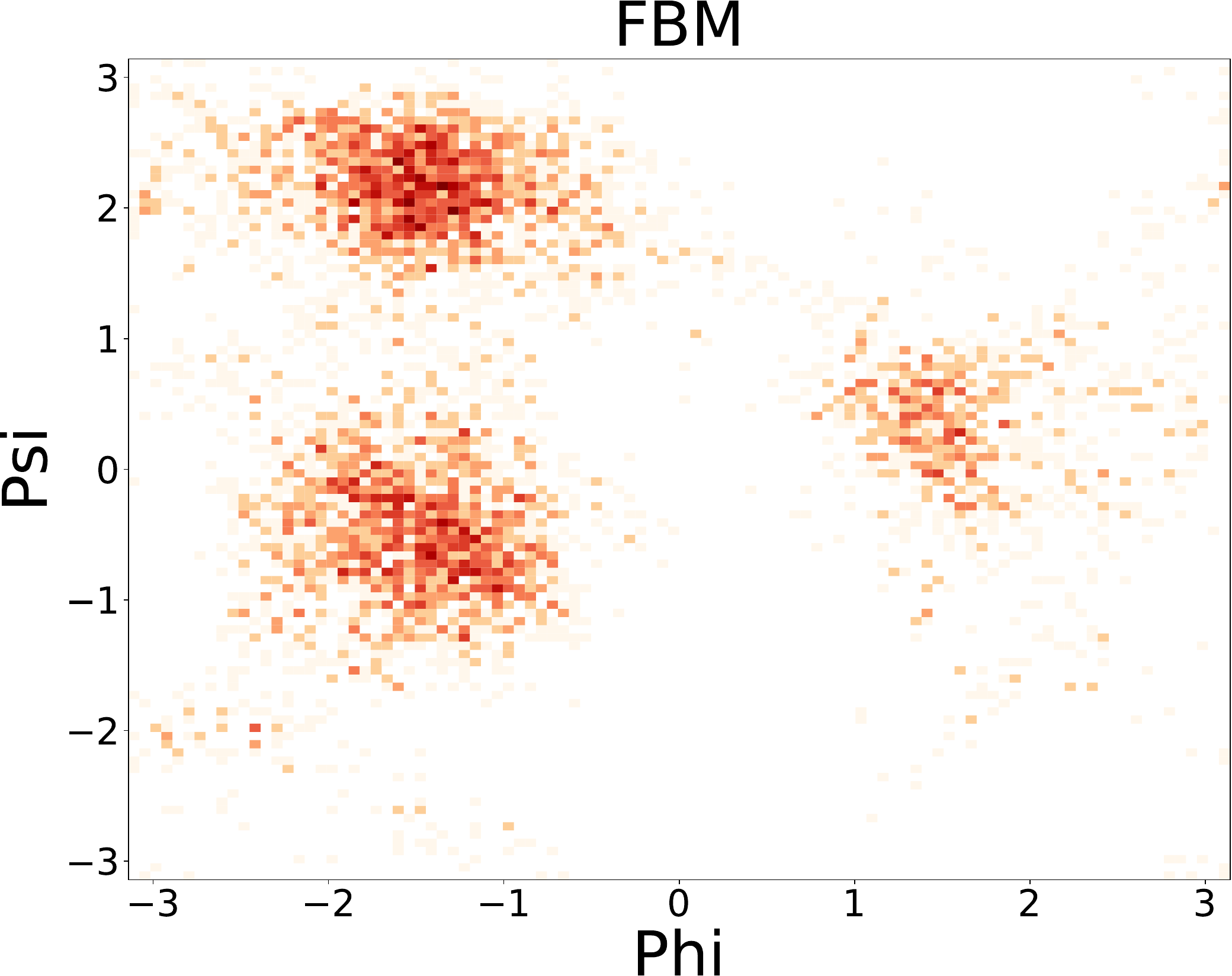}
        \end{minipage}
        \begin{minipage}{0.3\textwidth}
            \includegraphics[width=\textwidth]{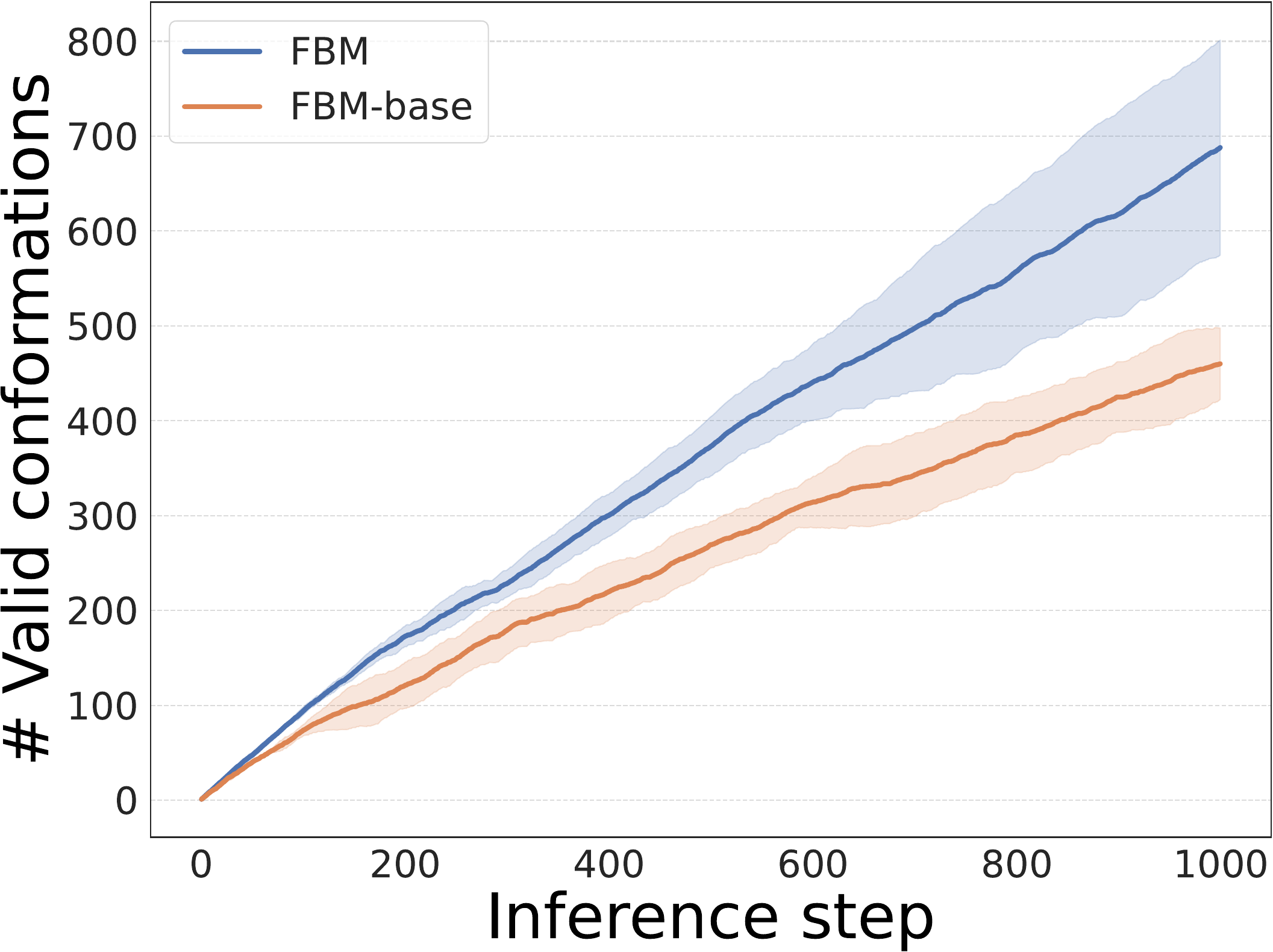}
        \end{minipage}
        \caption{Visualization of evaluations on trajectory e59s7.}
    \end{subfigure}

    \begin{subfigure}[b]{0.9\textwidth}
        \centering
        \begin{minipage}{0.3\textwidth}
            \includegraphics[width=\textwidth]{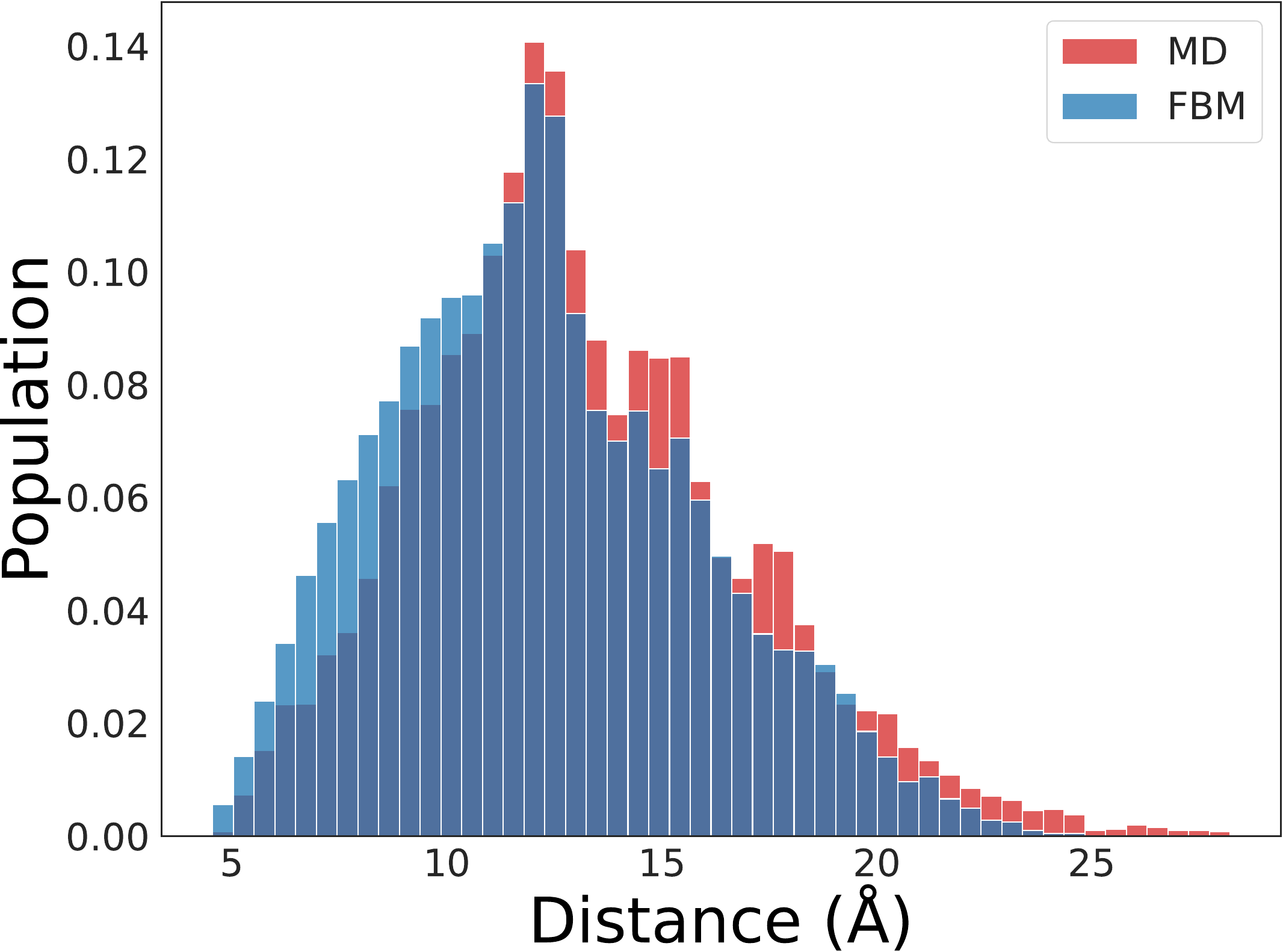}
        \end{minipage}
        \begin{minipage}{0.3\textwidth}
            \includegraphics[width=\textwidth]{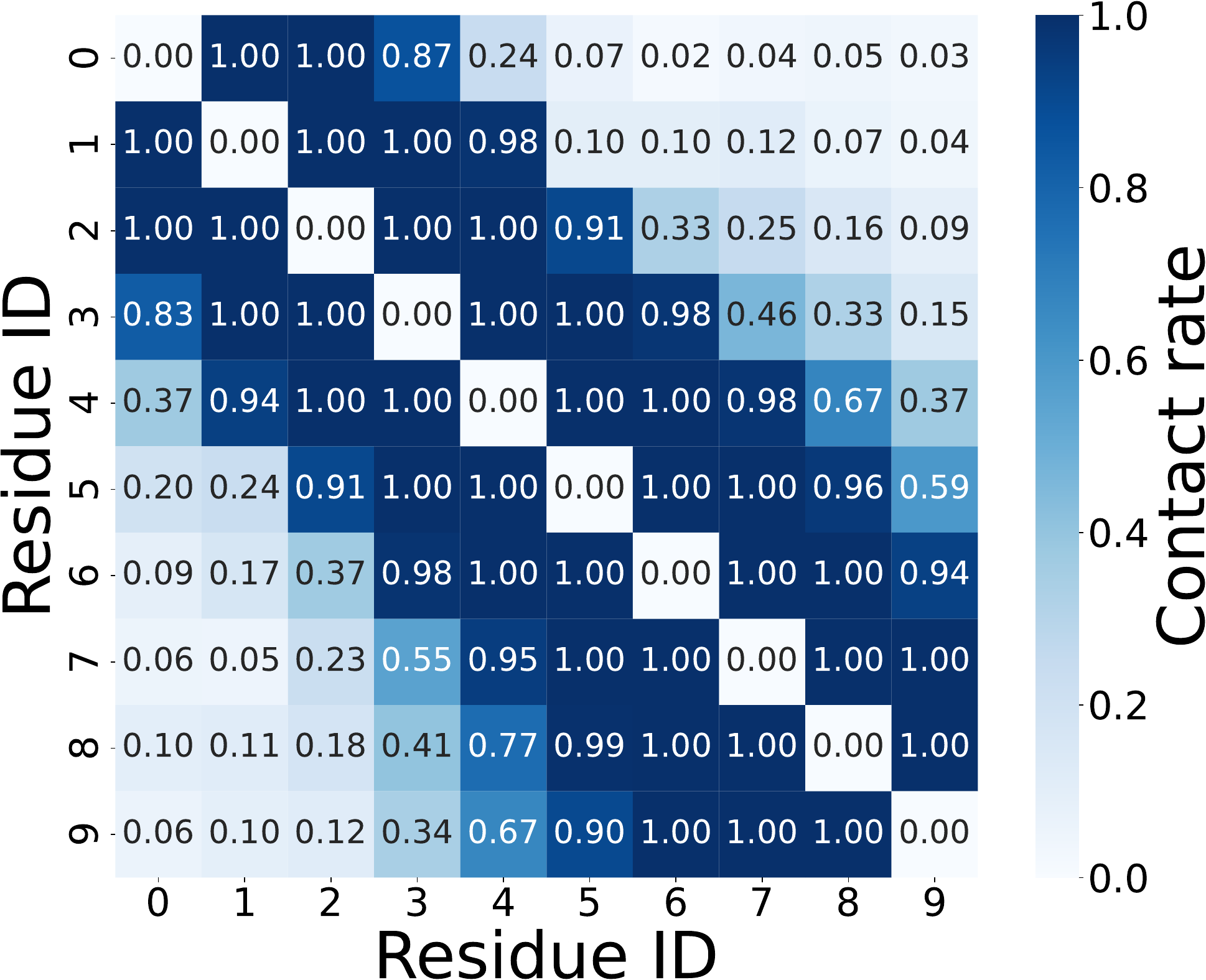}
        \end{minipage}
        \begin{minipage}{0.3\textwidth}
            \includegraphics[width=\textwidth]{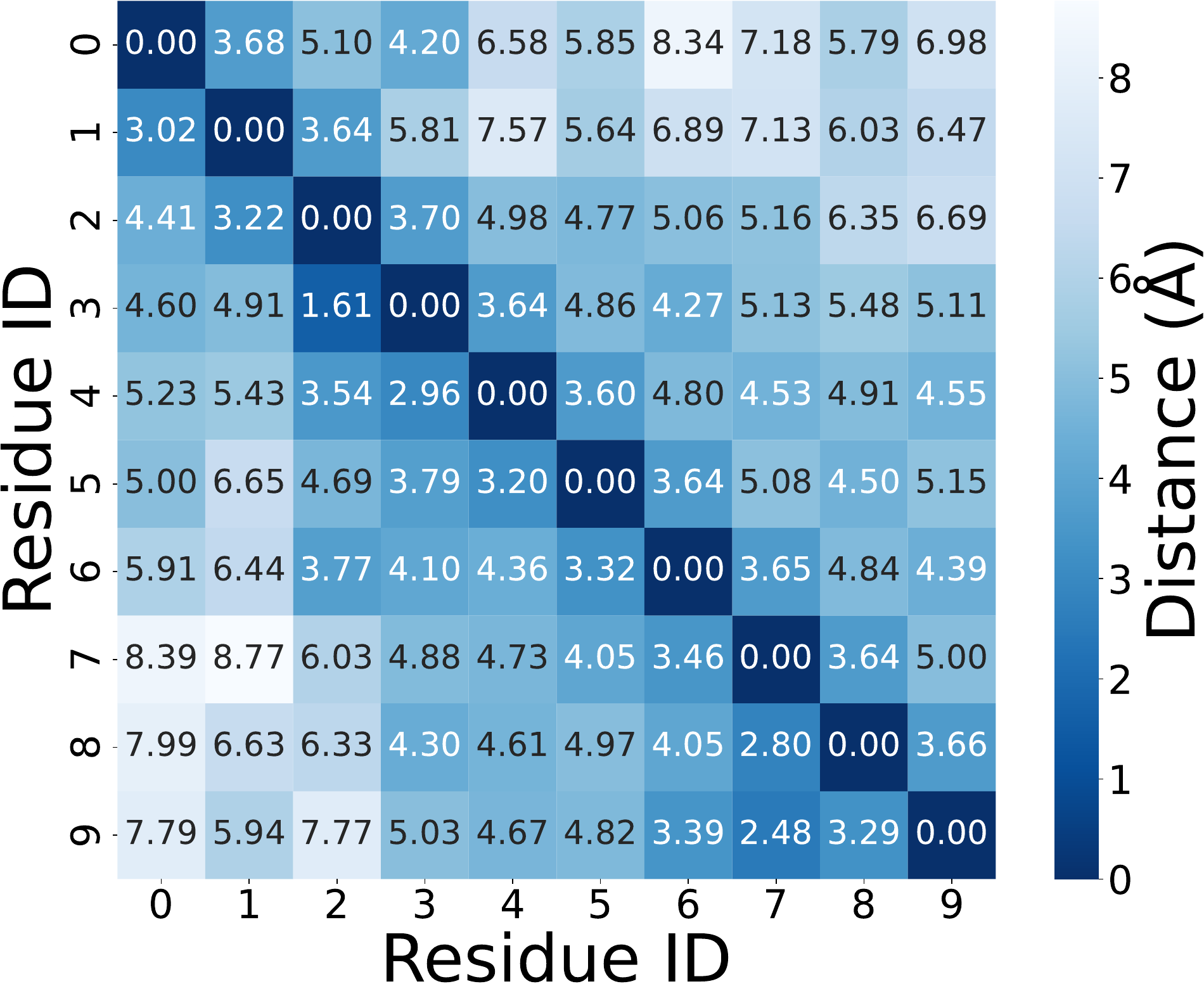}
        \end{minipage}
        \\
        \begin{minipage}{0.3\textwidth}
            \includegraphics[width=\textwidth]{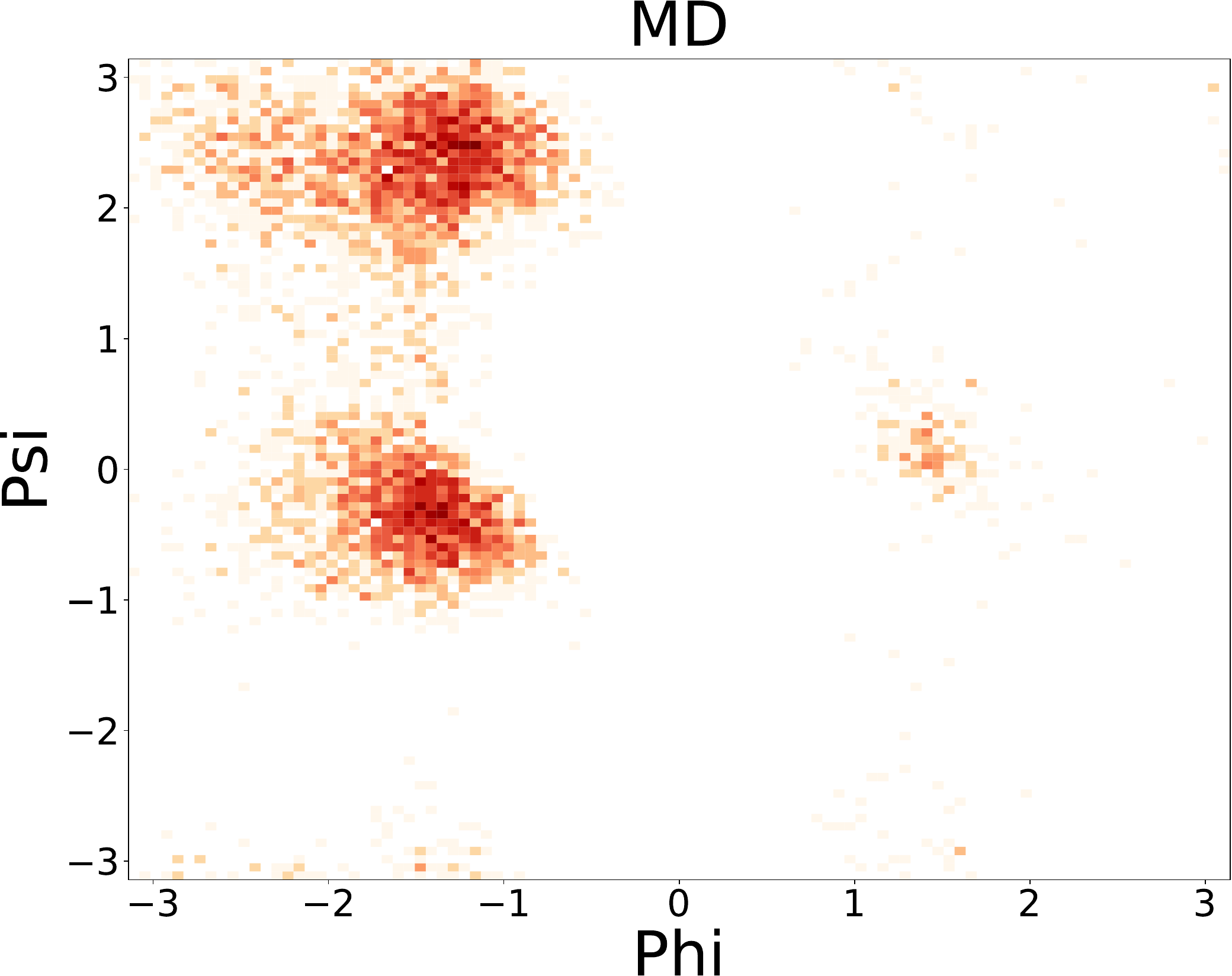}
        \end{minipage}
        \begin{minipage}{0.3\textwidth}
            \includegraphics[width=\textwidth]{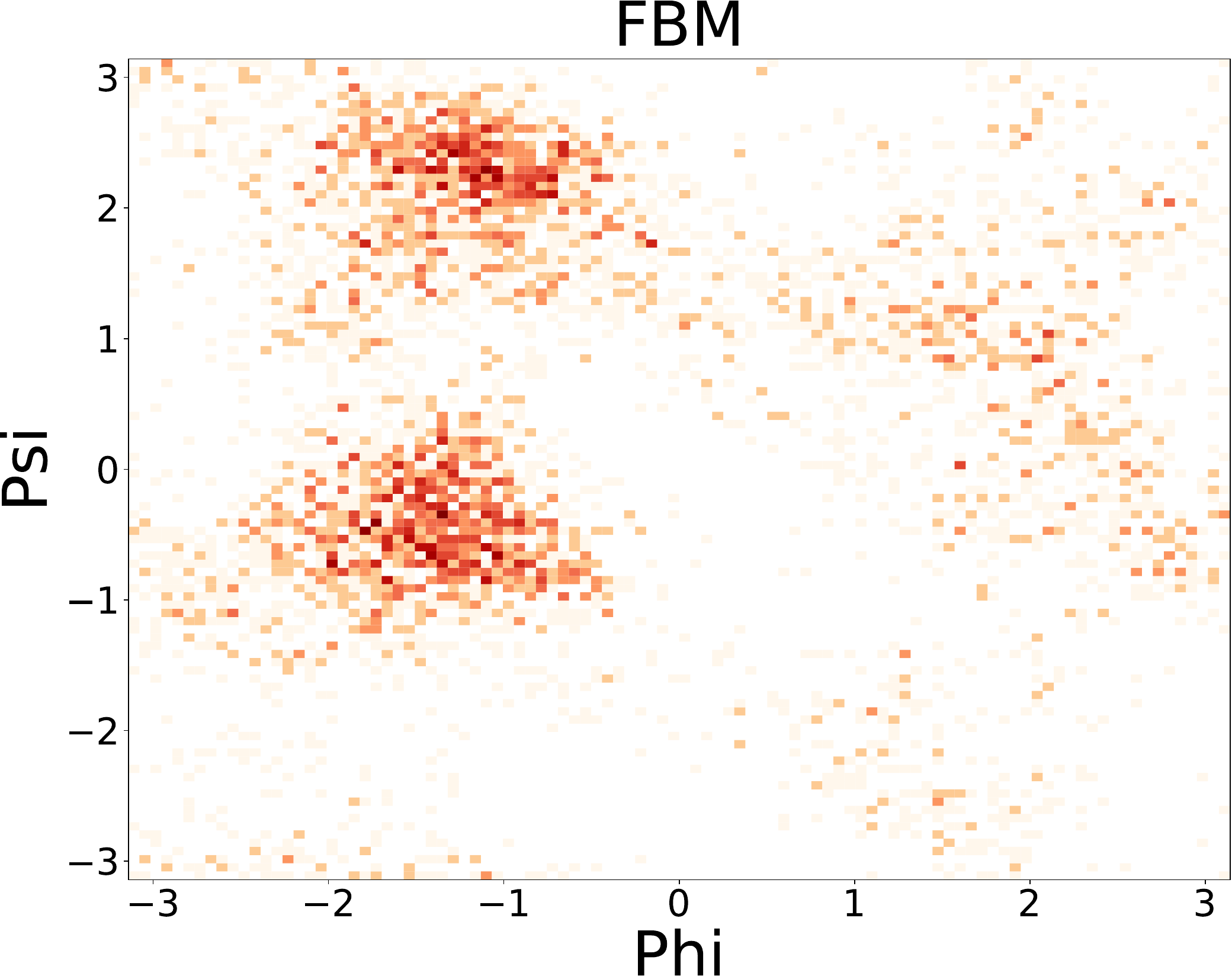}
        \end{minipage}
        \begin{minipage}{0.3\textwidth}
            \includegraphics[width=\textwidth]{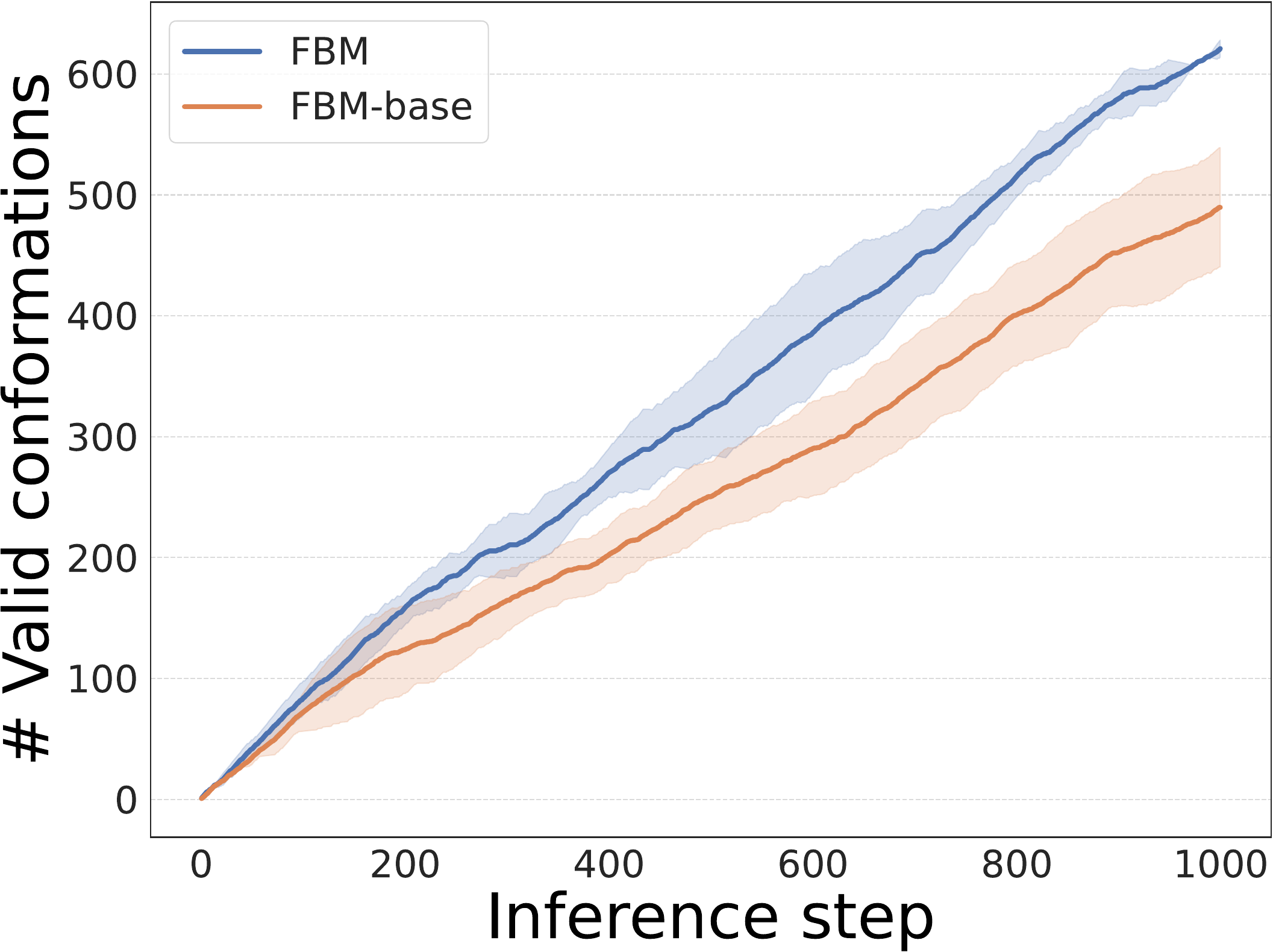}
        \end{minipage}
        \caption{Visualization of evaluations on trajectory e3s24.}
    \end{subfigure}

    \caption{The visualization of comprehensive metrics on three test trajectories of Chignolin. For each sub-figure of the corresponding peptide: \textbf{1.} The top-left plot shows the joint distribution of pairwise distances (P{\small W}D). \textbf{2.} The top-middle and top-right plots demonstrate the residue contact map and the residue minimum-distance map, where the data in the lower and upper triangle are obtained from FBM and MD, respectively. \textbf{3.} The bottom-left and bottom-middle are Ramachandran plots of MD and FBM, respectively. \textbf{4.} The bottom-right plot compares the cumulative valid conformations of different methods during inference, with each method undergoing three independent runs.}
    \label{fig:appendix-chignolin}

\end{figure}

The evaluation results on three test trajectories are shown in Table~\ref{tab:appendix-result-chig}, where the identifiers of these trajectories in the original dataset are labeled as e1s44, e59s7 and e3s24, respectively. We find that FBM significantly outperforms FBM-{\small BASE} across multiple comprehensive metrics, showcasing a strong and stable generation ability for molecular dynamics. Moreover, we provide the visualizations of various metrics on the test set in Figure~\ref{fig:appendix-chignolin}, where the generated trajectories of FBM shows a close match to those of MD in most cases, demonstrating its usefulness and scalability to more complex molecular systems.

\begin{table}[H]
\centering
\caption{Evaluation results of FBM on three test trajectories of Chignolin. Values of each metric are first averaged over 3 independent runs for each peptide and then shown in mean/std of all 14 test peptides. The best result for each metric is shown in \textbf{bold} and the second best is {\ul underlined}.}
\label{tab:appendix-result-chig}
\resizebox{\textwidth}{!}{%
\begin{tabular}{llcccccc}
\hline
\multirow{2}{*}{Index} & \multirow{2}{*}{Model} & \multicolumn{4}{c}{JS {\small DISTANCE} ($\downarrow$)}                                   & \multirow{2}{*}{V{\small AL}-CA ($\uparrow$)} & \multirow{2}{*}{C{\small ONTACT} ($\downarrow$)} \\ \cline{3-6}
                       &                        & P{\small W}D         & R{\small G}          & TIC                  & R{\small AM}         &                                               &                                                  \\ \hline
\multirow{2}{*}{e1s44} & FBM                    & \textbf{0.315}/0.056 & \textbf{0.285}/0.080 & \textbf{0.544}/0.011 & 0.509/0.029          & \textbf{0.691}/0.041                          & \textbf{0.161}/0.069                             \\
                       & FBM-{\small BASE}      & 0.417/0.069          & 0.467/0.148          & 0.554/0.023          & \textbf{0.480}/0.013 & 0.465/0.057                                   & 0.249/0.066                                      \\ \hline
\multirow{2}{*}{e59s7} & FBM                    & \textbf{0.395}/0.017 & \textbf{0.400}/0.034 & \textbf{0.522}/0.015 & \textbf{0.443}/0.020 & \textbf{0.780}/0.012                          & \textbf{0.184}/0.029                             \\
                       & FBM-{\small BASE}      & 0.456/0.026          & 0.407/0.016          & 0.526/0.002          & 0.490/0.017          & 0.460/0.031                                   & 0.219/0.056                                      \\ \hline
\multirow{2}{*}{e3s24} & FBM                    & \textbf{0.305}/0.046 & \textbf{0.332}/0.089 & \textbf{0.524}/0.004 & 0.519/0.011          & \textbf{0.628}/0.010                          & \textbf{0.123}/0.032                             \\
                       & FBM-{\small BASE}      & 0.450/0.086          & 0.549/0.149          & 0.527/0.015          & \textbf{0.502}/0.019 & 0.490/0.040                                   & 0.278/0.104                                      \\ \hline
\end{tabular}%
}
\end{table}

\section{Computing Infrastructure}
Our models, FBM-{\small BASE} and FBM, were trained on 4 NVIDIA GeForce RTX 3090 GPUs within a week. The inference procedure with baselines and our model were all performed on one single NVIDIA GeForce RTX 3090 GPU.

\end{document}